\documentclass{article}
\usepackage{graphicx}

\title{Surface Effects On Wetting and Layering Transitions}
\author{Hamid Ez-Zahraouy\footnote{Corresponding author : ezahamid@fsr.ac.ma} \footnote{Editor},
Abdelilah Benyoussef, Lahoucine Bahmad}
\begin{document}
\maketitle
\begin{center}
{Facult\'e des Sciences, D\'epartement de Physique, Laboratoire de Magn\'etisme et Physique des Hautes Energies, B.P. 1014, Rabat, Morocco.\\
}
\end{center}
\begin{abstract}
Wetting phenomena plays  an interesting role in the technological development  of materials. Recently, much attention has been directed to the study of magnetic solid films. To understand, theoretically, the effect of surface on  wetting and layering transitions in these systems, we will give in this chapter a presentation of several relevant recent theoretical works realized on wetting and layering transitions in which we explain how the nature of surface can affect the behavior of wetting and layering transitions. Using  different spin systems models and different  numerical and approximate methods such as mean field and effective field theories, real space renormalization group technique, transfer matrix method and Monte-Carlo simulations, it is found that the  wetting and layering transitions depend on the nature of the surface magnetic field, the surface coupling strength, the surface crystal field, the geometry of surface, the in-homogeneity of substrate and the quantum fluctuations at the surface�. Indeed the layering transition temperature increases or decreases as a function of the thickness of the film depending on which the surface coupling is smaller or greater than a critical value. Thus the wetting transition temperature can be greater or smaller than the layering temperature. In the case of  variable surface crystal field an order-disorder layering reentrance appears with  new critical exponents  and multi-critical  behaviors. Wetting phenomena is also influenced by the curvature of the surface. In fact the intra-layers transitions appear under the effect  of edge on the surface and/or corrugated surface. However, there exist an intra-layering  surface temperature above which the surface and the intra-layer surface  transitions occur . While the bulk layering and intra-layering transitions above an other finite temperature which is greater than the former one. Such behavior is not seen in the case of a perfect surface. Beside this and under the corrugation effect , the pre-wetting phenomena and layering transitions occur at zero temperature, even in the case of a uniform magnetic field. In this case the wetting temperature depends strongly on the width of the corrugated-steps but weakly on its number. The quantum fluctuations introduce a new wetting transitions, namely the wetting, layering and roughening transverse fields. These fluctuations  lead to the appearance of a sequence of  layering sublimations transitions which occur before bulk melting above a critical transverse field  which depends on temperature and/or chemical potential. Moreover, the wetting temperature depends strongly on the randomness of the quantum fluctuations. The chapter is organized as follows: Section {\bf 1} is reserved to the Introduction and definitions. In section {\bf 2} we present the relation between the nature of surface and wetting phenomena. The effect of the surface geometry  on wetting and layering transitions is presented in section {\bf 3}. While quantum fluctuations effects are given in section {\bf 4}. References are cited in section {\bf 5}.
\end{abstract}

\newpage
\tableofcontents


\newpage
\section{Introduction}
Recently, the wetting and layering transitions of magnetic Ising systems have been studied by several authors.
Multilayer films adsorbed on attractive substrates may exhibit a variety of possible phase transitions, as has been reviewed by Pandit {\it et al.} [1], Pandit and Wortis [2], Nightingale {\it et al.} [3] and Ebner {\it et al.} [4-7]. A variation of phase diagrams with the strength of the substrate potential in lattice gas model for multilayer adsorption has been studied by Patrykiejew {\it et al.} [8] using Monte Carlo simulations and molecular field approximation.
One type of transitions is the layering transitions, in which the thickness of a solid film increases discontinuously by one layer as the pressure is increased. Such transitions have been observed in a variety of systems including for example $^{4}He$ [9,10] and ethylene [11,12] adsorbed on graphite. 

\mbox{  }Many experiments works have for subject the study of wetting transitions. However, the effect of substrate roughness on wetting has been studied [13,14] and the thickness profile of the drop of a nematic liquid crystal on a rough surface [15]. The effect of the nature of the adsorbate has been studied by several authors namely [16] the critical wetting of pentane on water. the critical adsorption and the layering transitions, at the free surface,  of a smectic liquid crystal [17],  the effect of segregation on wetting in films of a partially miscible polymer blend[18], the effect of the chain length of alkanes on water [19], the critical adsorption and dimensional crossover in epitaxial FeCo films [20], the heterogeneous hole nucleation in electron-charged ${ }^{4}He$ wetting films [21], the existence of tricritical wetting in liquid mixtures of methanol and the n-alkanes [22], the microscopic structure of the wetting film at the surface of liquid Ga-Bi alloys [23], the tetra point wetting at the free surface of liquid Ga-Bi, in which a continuous surface wetting transition, pinned to a solid-liquid-liquid-vapor tetra coexistence point occurs [24], the short range wetting at liquid (Ga-Bi) alloy surfaces [25], the critical Casimir effect and wetting by Helium mixtures [26]. The dynamic aspect has been studied by several authors; namely, the dynamic properties of a fluid interface of nanometric curvature formed by capillary condensation between solid substrates [27], the dynamic of wetting layer [28], and the dynamic of layering transition in confined liquids [29]. The nature of the transitions has attracted attention like; the wetting near the triple point [30], and the sequence of two wetting transitions induced by tuning the Hamaker constant [31].
However,  a simple lattice gas model with layering transitions and critical points has
been introduced and studied in the mean field approximation by de Oliveira
and Griffiths [32]. These experimental studies have motivated many theoretical works. However several methods have been used to study wetting and layering transitions. The transfer matrix and scaling theory have been used to study the curvature controlled wetting in two dimensions [33], the adsorption of polyelectrolytes at an oppositely charged surface [34]. Using Monte Carlo simulations the effects of van der Waals surface interactions on wetting transitions in polymer blends [35], the wetting properties of neon on heterogeneous surfaces [36], the wetting transitions of hydrogen and deuterium on the surface of alkali [37], the multilayer adsorption of binary mixtures [38], the competition between short-ranged attractions and long-ranged repulsions in systems like langmuir monolayers, magnetic films, and adsorbed monolayers using numerical simulations and analytic theory [39], the double wedge wetting [40], and the wetting behavior of associating binary mixtures at attractive walls [41]. The variational and mean field approximation have been used to investigate many interest problems such as; the dewetting, partial wetting and spreading of a two-dimensional monolayer on solid surface [42], the line tension between fluid phases and a substrate [43], the effect of confinement and surface enhancement on superconductivity [44], the symmetric polymer blend confined into a film with antisymmetric surfaces [45], the wetting of potassium surfaces by superfluid ${ }^{4}He$ [46], the filling transition for edge  [47], the wetting of film on chemically heterogeneous substrates [48], the wetting and the interaction potential between spherical particles  [49], the phase diagram for morphological transitions of wetting films on chemically structured substrates [50], the phase behavior of colloidal binary hard-platelet and hard-rod fluids [51], the influence of random self-affine and mound substrate roughness on the wetting of adsorbed van der Waals films [52], the wetting of van der Waals solid films on self-affine rough surfaces [53], the preroughening and reentrant layering transitions on triangular lattice substrates [54], the layering transitions, disordered flat phases, reconstruction, and roughening of a semi-infinite crystal [55].
Other method have been used to study  wetting phenomena such as; the renormalization group of the capillary parameter at complete wetting [56], the surface deconstruction and roughening in the multiziggurat model of wetting [57], the morphological transitions of wetting layers on structured surfaces [58], the limiting height at which a solid can approach an air-water surface without becoming wet [59], the surface roughness and hydrodynamic boundary slip of a newtonian fluid in completely wetting system [60], the modeling at low temperature of a ${ }^{4}He$ film adsorbed on a Li substrate [61] and the prewetting of ${ }^{4}He$ on rubidium [62], the critical adsorption on  defects in Ising magnets and binary alloys  [63], the exact study of edge filling transition for Ising corners [64], the wetting hysteresis at the molecular scale [65], the wetting transition in a two-dimensional Ising model with a free rough surface [66]. Ebner and Saam [67] carried out Monte Carlo simulations of such a
lattice gas model. Huse [68] applied renormalizaion group technique to this
model. It allowed the study of the effects on an atomic scale or order
disorder transitions in the adsorbed layers, which may have considerable
influence on the layering transitions and tracing back macroscopic phenomena
on inter-atomic potentials.
Benyoussef and Ez-Zahraouy have studied the layering transitions of Ising model thin films using a real space renormalization group [69], and transfer matrix methods [70].
Using the mean field theory, Hong [71], have found that depending on the values of the exchange integrals near the surface region, the film critical temperature may be lower, higher than, or equal to that of the bulk. 

On the other hand, the development of molecular-beam epitaxy has allowed the fabrication of high quality thin films [72]. Among them, the magnetic film consisting on a few atomic layers is of particular importance, as it has been used as a convenient model for theoretical and experimental analysis of a variety of magnetic phenomena. One other problem investigated is the critical temperature at which the system undergoes a magnetic-nonmagnetic phase transition. Of particular interest is the change of this temperature when the magnetic system becomes finite in a given direction.

Briefly, depending on the values of
magnetic coupling constants near the surface region, two different behaviors
at the surface can be demonstrated [71]: (i) the surface critical temperature
is the same as of the bulk one when the coupling constant on the surface is
smaller than a critical value "ordinary transition". (ii) The surface
critical temperature may be higher than the corresponding bulk one when
the surface coupling is larger than the critical value "extraordinary
transition". At the critical ratio, one encounters the "special transition
point", with critical properties of the surface transition deviating from
those at the ordinary or the distinct surface transition. As for the film system, which is finite in one direction, it has been established that its magnetic properties can differ greatly from those of the corresponding bulk [73-77].
Experimental results [78] showed that the critical temperature of a vanadium film depends on the film thickness and its critical behavior is like that of the two-dimensional system rather than that of the three-dimensional bulk.

Since the theory of surface critical phenomena started developing, much attention
has been devoted to the study of the Blume Capel (BC) model over semi-infinite lattices, with modified surface couplings. Phase diagrams in the mean field approximation reporting
four possible topologies at fixed bulk/surface coupling ratios have been determined [79-83]. An analogous analysis have also been done using a real space renormalization group transformation [84]. Some other works referring to particular regions of the
phase space are, for example, those using: the mean field approximation [85], the
effective field approximation [86] and the low temperature expansion [87]. These works show that it is possible to have a phase with ordered surface and disordered bulk, which is separated from the completely ordered phase by the so-called {\it extraordinary} transition and from the completely disordered phase by the {\it surface} transition. In the absence of this phase, the transition between the completely ordered and the completely disordered phase is called {\it ordinary}. These three kinds of phase transitions have a meeting point named {\it special}
which is generally a multi-critical point.
{\bf }The discussion presented in Ref. [88] shows that the strong interest in these
models arises partly from the unusually rich phase transition behavior they
display as their interaction parameters are varied, and partly from their many
possible applications. The bilinear interaction considered in most of these cases is ferromagnetic.
The spin-$1$ Ising systems are used to describe both the order-disorder
transition and the crystallization of the binary allow, and it was solved
for the anti-ferromagnetic case in the mean field approach [89].
The decomposition of a line of tri-critical points into a line of critical
end points and one of double critical points is one of the most interesting
features of the mean field phase diagram for the anti-ferromagnetic
spin-$1$ Blume-Capel model in an external magnetic field [90].
The transfer-matrix and Monte Carlo finite-size-scaling methods [91],
are also applied to study this model but such decomposition does not occur
in the two dimensional case.
The finite cluster approximation has been applied by Benyoussef {\it et al.}
[92] in order to study the spin-$1$ Ising model with a random crystal field.
On the other hand, the transverse field or crystal field effects
of spin-$1$ Ising model has been studied by several authors [93-96].

The experimental measurements of layer-by-layer ordering phenomena have been
established on free-standing liquid crystals films such as ${\it nm}OBC$
$(n-alkyl-4'-n-alkyloxybiphenyl-4-carboxylate)$ [97,98] and $54COOBC$ (
n-pentyl-4'-n-pentanoyloxy-biphenyl-4-carboxylate) [99] for several
molecular layers. More recently, Lin {\it el al.} [100] have used the
three-level Potts model to show the existence of layer-by-layer ordering
of ultra thin liquid crystal films of free-standing $54COOBC$ films, by
adjusting the interlayer and intra layer couplings between nearest-
neighboring molecules.

Depending on the value of the crystal field at the surface
some systems undergo a phase transition from a completely
disordered state, at high temperatures,
to a state with the first $k$ layers disordered and the rest of the
layers ordered at low very temperatures.
One can then observe a particular behavior when the transition to
the order  state of the
$k^{th}$ layer occurs not only with the decrease but also with the  increase
of the temperature. Such phase transitions are called
the  reentrant phenomena.

The reentrant first-order layering transitions have been observed
experimentally in multilayer argon films on graphite by Youn and Hess [101].
Interesting results were obtained within MFA such [102]
as order-disorder layering transitions and reentrant phenomena.
But it is well known that the MFA approximation does not take into account
the thermal fluctuations and neglects correlations between spins.
This can lead to incorrect results.

\mbox{  } The Blume-Capel model (BC) was originally proposed to study the first-order magnetic phase transitions in spin$-1$ Ising systems [103].
This model was generalized to the Blume-Emery-Griffiths (BEG) to study phase separation and superfluity in $^3$He-$^4$He mixtures [104]. Later it has been applied to describe properties of multicomponent fluids [105], semiconductor alloys [106] and electronic conduction models [107]. The (BC) model is not exactly solvable in more than one dimension, but it has been studied over infinite $d$-dimensional lattices by means of many different approximate techniques and its phase diagram is well known.

On the other hand, ferroelectric films can be described by an Ising model and when the film becomes very thick, its properties are those of the semi-infinite Ising system [108-110]. From the experimental point of view, the most commonly studied magnetic multilayers are those of ferromagnetic transition metal such as Fe/Ni, where the coupling can exist between magnetic layers [111-113]. The discovery of enormous values of magnetoresistance in magnetic multilayers are far exceeding those found in single layer films and so exceeds the discovery of oscillatory interlayer coupling in transition metal multilayers. These experimental studies have motivated much theoretical works to study magnetic thin films as well as critical phenomena [114-117]. This is partly motivated by the development of new growth and characterization techniques, but perhaps more so by the discovery of many exciting new properties, some quite unanticipated.
The effect of the surface and bulk transverse fields on the phase diagrams of a semi infinite spin-$1$ ferromagnetic Ising model with a crystal field was investigated [118-119] within a finite cluster approximation with an expansion technique for cluster identities of spin-$1$ localized spin systems.

Substantial research efforts have been devoted to
the fabrication of quantum-dot structures, due to the interesting electronic
and optical properties that can result from quantum confinement.
Several novel applications for such structures have been envisaged such as
the single-electron transistor and intersubband quantum-dot infrared
photodetectors.
The well studied technique of creating quantum dots is the use of $Si$
deposition on $Ge$ [120,121].
A significant number of studies have been performed with the aim of
understanding the structural and compositional evolution of $Ge$ islands
growing.
The role played by different growth parameters such as temperature,
pressure and others have been studied for different growth techniques.
The evolution of self assembled $Ge$ islands in low pressure chemical vapor
deposition of $Ge$ on $Si$, using high growth rates, has been
investigated by atomic force microscopy and Rutherford backscattering
spectrometry [122].

On the other hand, the understanding of the nature and consequences of
adsorbate-adsorbate interactions at solid surfaces is of great interest in
engineering properties of interfacial materials.
Many recent studies show
that indirect interactions, mediated by the structure, can be significant
enough to influence the formation of nano-structures at surfaces [123].
Indeed, much progress has been made recently in understanding
adatom-pair interactions mediated by Shockley surface state electrons and
the oscillatory interaction decays with adsorbate separation thickness [124-127].
Monte Carlo Simulations [128-130] and subsequently derived mean field theories
[129,131], showed the existence of higher island densities than those expected
by standard nucleation theory.
It is shown that several experimental works, e.g. [127,132-134], supports these
predictions, such as the deposition of $Au$ on mica substrates at a high
temperature [135-142].

In most cases, the electrical conductivity and practical
utility of discontinuous metal films, have been extensively explored [143-145].
On the other hand, media that are magnetized perpendicular to the plane of the
film, such as ultra thin cobalt films or multi layered structures [146,147], are
more stable against thermal self-erasure [144,148] than conventional
memory devices. In this context, magneto-optical memories seem particularly
promising for ultrahigh-density recording on portable disks. The roughness
and mobility of the magnetic domain walls [149,150] prevents closer packing
of the magnetic bits, and therefore presents a challenge to reaching even
higher bit densities.
Increasing information storage to high densities may evolve through extensions
of current magnetic recording technologies. But such increases in storage
density might be achieved by using other techniques such as holography, or
micro machined nano-cantilever arrays [151].
The bit-writing with local probes may be thermally assisted by a current
[152] or a laser beam that raises local temperature to the vicinity of the
Curie temperature, resulting in the formation of a reversed domain with
a rough wall. On the other hand, a strain implemented by a linear defect, in a magnetic
thin film [153], can realize smooth and stable domain walls, that can be
implemented without nano-scale patterning,
Furthermore, composite materials such as multilayer coatings and isotropic
nanocomposite coatings, having structures in the nanometer scale, can even
show properties, which can not be obtained by a single coating material
alone [154]. The most widely used coatings are $TiN$, $TiC$, $TiCN$, and
combination thereof, as well as some coatings with lubrificating properties
such as diamond-like carbon.
Monte Carlo simulations are applied to study the equilibrium magnetic
domain structure of growing ultra thin ferromagnetic films with a realistic
atomic structure [155]. Near the percolation threshold a metastable magnetic
domain structure is obtained with an average domain area ranging between the
area of individual magnetic islands and the area of the large domains
observed for thicker ferromagnetic films. This micro-domain structure is
controlled and stabilized by the non-uniform atomic nano-structure of the
ultra-thin film, causing a random interaction between magnetic islands with
varying sizes and shapes.
The investigation of the magnetic domain formation, in nano-structured
ultra-thin films, is an active field of current research. The influence of
the atomic morphology on the magnetic properties is known to be especially
strong during the initial states of the thin film growth. A small variation
of the preparation conditions may considerably change the obtained magnetic
structure. This has been shown, by resolving imaging techniques, which allows
the investigation of atomic structure as well as the magnetic domain
structure of several nanostructured systems [156,160-161].
The influence of the presence of a surface up on the bulk critical behavior
has been the focus of scientific interest during the last decade [162-164].
Since the free surface is the natural break of the translational symmetry
in any real system, critical phenomena at perfect surfaces have attracted
much interest. Furthermore, the phase diagram for the semi-infinite Ising
model is well established [162,164].  Moreover,
surface critical exponents have been estimated, both theoretically.
The general agreement is quite satisfactory [162-164]. The critical behavior
of magnetic materials with variable strength of coupling on the surface
is characterized by surface critical exponents independent of the bulk
ones. For their theoretical determination $\epsilon$-expansion [164],
series expansion [165], and Monte-Carlo simulations [166,167] were applied.
Moreover, the magnetic properties depend on several geometric features,
which depend on the conditions during the growth. thus, the connection
between experimental and theoretical results requires a detailed study
of the dependence of physical quantities on the geometry of the systems.
Some specific properties have already been considered in theoretical
works, such as the size and the lattice structure of small clusters
[168,169], the thickness of the film [170] and the roughness of one
dimensional structures [171], in particular, the role of surface
imperfections, unavoidable in real materials, on the critical
properties has been studied [172-175]. According to the Harris
criterion [176], the relevance of the perturbationis connected to the
sign of the specific heat exponent $\alpha$ in the pure system. Small
amounts of disorder do not change the nature of the  phase transition,if the corresponding pure system's specific heat exponent is negative. Currently, there appears to be widespread consensus that critical exponents of bond disordered systems in the weak disorder limit are
the same as those of the pure one [177-179].
Monte-Carlo simulations of bond disordered systems show logarithmic
corrections in the behavior of the magnetizations which persist even in the
strongly disordered regime [180,181]. However, in the $2D$ spin diluted
Ising model a continuous variation of critical exponents with the spin density has been found [182]. A Monte-Carlo study [183] shows that the specific heat does not diverge at the critical point, at least when the concentration of dilution is sufficiently large; the critical behavior of the magnetic susceptibility and the correlation length are consistent with the pure power-law behavior with the concentration dependent critical exponents. Moreover , using Monte Carlo simulations, the site-diluted Ising model in two dimensions has been studied [184] where it is found that such model belongs to the same universality class as the pure one, in spite of the strong logarithmic effects.
Besides, the surface critical behavior of the three-dimensional disordered semi-infinite Ising ferromagnetic has been studied, using Monte Carlo simulations [185]; it is found that for $J_s$ less than $J_{sc}$, the corresponding surface critical exponent is equal to the one of perfect surface [186-192].

Using the perturbative theory, Harris [193] have studied the layering transitions at $T=0$ in the presence of a transverse field. The effect of finite size on such transitions has been studied, in thin film confined between parallel plates or walls, by Nakanishi and Fisher [194] using mean field theory and by Bruno ${\it et al.}$ [195] taking into account the capillary condensation effect. The critical wetting transitions have been considered in systems with corrugated periodic walls [195] by using an effective Hamiltonian study.
By applying Monte Carlo simulations on thin Ising films with competing walls, Binder {\it et al.} [196], found that occurring phase transitions belong to the universality class of the two-dimensional Ising model and found that the transition is shifted to a temperature just below the wetting transition of a semi-infinite fluid [197,198]. Hanke ${\it et al.}$ [199] show that symmetry breaking fields give rise to nontrivial and long-ranged order parameter profiles for critical systems such as fluids, alloys, or magnets confined to wedges. The existence of the layering transitions for a film, with corrugated surface has been studied using mean field theory and Monte Carlo simulations [200]. It is found that the wetting temperature $T_w$ depends weakly on the surface corrugation degree for fixed values of the  surface magnetic field. In addition, by applying a suitable effective interface model at liquid-vapor coexistence, Rejmer ${\it et al.}$ [201] found a filling transition at which the height of the meniscus becomes macroscopically large while the planar walls of the wedge far away from its remain non-wet up to the wetting transition occurring at $T_w$. They also showed that the discontinuous filling transition is accompanied by a pre-filling line extending into the vapor phase of the bulk phase diagram and describing a transition from a small to a large, but finite, meniscus height.
Much attention has been paid to the properties of layered
structures consisting of alternating magnetic materials. The most commonly
studied magnetic multilayers are those of ferromagnetic transition metal such
as Co or Ni. Many experiments have shown that the magnetization enhancement
exists in multilayered films consisting of magnetic layers. It was found that
ferromagnetic coupling can exist between magnetic layers. From the
theoretical point of view, great interest has been paid to spin wave
excitations as well as critical phenomena [202-215]. The study of thin films is partly motivated by the development of new growth and characterization techniques, but perhaps more so by the
discovery of many exciting new properties, some quite unanticipated. These
include, more recently, the discovery of enormous values of magnetoresistance
in magnetic multilayers far exceeding those found in single layer films and
the discovery of oscillatory interlayer coupling in transition metal multilayers.
These experimental studies have motivated much theoretical work. However
these developments are applied to a large extent powered by "materials engineering"
and the ability to control and understand the growth of thin layers only a
few atoms thick. However, Many experimental studies have shown that the magnetization enhancement exists in multilayered films consisting of magnetic layers.
The development of the molecular beam epitaxy technique and its application to the growth of thin metallic films has simulated renewed interest in thin film magnetism. The fabrication of thin metallic films has led to a surge of experimental activity [216-219]. More recently, the thickness dependence of the critical temperature of thin Ising magnetic films having surface exchange enhancement has been studied by Hong [71] and by Aguilera-Granja and Moran Lopez [220] within mean field approximation, by Hai and Li [221] and Wiatrowski {\it et al.} [222] within an effective field treatment that accounts for the self-spin correlations.

On the other hand, ferroelectric films can be described by an Ising model in a transverse field [223-225], and when the film becomes very thick, its properties are those of the semi-infinite Ising system [108-110,226]. Phase transitions in Ising spin systems, driven entirely by quantum fluctuations, have been getting a lot of attention [227]. The simplest of such systems is the Ising model in a transverse field which can be exactly solved in one dimension. This model has been introduced to explain the phase transition of hydrogen-bonded ferroelectrics such as $KH_2PO_4$ [228]. Since then, this model has been applied to several physical systems like DyVO$_2$ and studied by a variety of sophisticated techniques [92,96,229,230].The technique of differential operator introduced by Kaneyoshi [231] is as simple as the mean field method and uses a
generalized but approximate Callen relation derived by Sa Barreto and
Fittipaldi [115]. The system has a finite transition temperature, which can be
decreased by increasing the transverse field to a critical value $\Omega _{c}
$. The effect of a transverse field on the critical behavior and the
magnetization curves was studied [92,96,229-233]. Using the perturbative theory, Harris {\it et al.} [193] have studied the layering transitions at $T=0$ in the presence of a transverse field.
The effect of a uniform transverse magnetic field on the layering and wetting transitions of a spin$-1/2$ Ising model in a longitudinal magnetic field and showed the existence of layering and wetting transitions above a critical transverse field $\Omega_w$, which is a function of the temperature and surface magnetic field [79] . Recently, Karevski {et al.}[82] have studied the random transverse Ising spin chain according to a distribution of the transverse field governed by a law of type: $z^{-\alpha}$; $z$ being the distance from the surface and $\alpha$ a constant).

On the other hand, The simplest of all random quantum systems is the random transverse Ising model [234,235]. However, the random systems have been known to be dominated by rare regions. This effect is particularly pronounced for random quantum systems at low or zero temperature far from critical points. Indeed, Griffiths [236] showed that the free energy is a non analytic function, because of rare regions. Having found all the derivatives being finite, Harris [237] concluded that this effect was very weak for classical systems.

The spin-1 Ising model with nearest-neighbor interactions, both bilinear and bi-quadratic, and with a crystal-field interaction was introduced by Blume, Emery, and Griffiths (BEG)[104] to describe phase separation and superfluid ordering in $^{3}He- ^{4}He$ mixtures. With vanishing bi-quadratic interactions the model is known as the Blume-Capel model [103]. Since its introduction, the spin-1 BEG model has been extended to solid-liquid-gas systems, and multi-component fluid liquid crystal mixtures [105], magnetic materials [238-240], critical behavior and multi-critical phase diagrams [241-244]. Furthermore it is the simplest model that can be used for modeling the behavior of the liquid, solid and vapor phases of a real materials. In this context it was used by Jayanthi [245] to study the surface melting of the solid phase near the triple point and by Gelb [246] to characterize the layering transitions and surface melting both near and away from the bulk coexistence lines under the effect of temperature. The effects of quantum transverse field on wetting and layering transitions of a spin-1 BEG model are also studied [79,247,248].

\section{Effects of surface potential and surface coupling on wetting and
layering transitions}
\subsection{The effect of magnetic surface field}
\subsubsection{Model}
We consider a spin-1/2 Ising model, with ferromagnetic coupling on a three dimensional cubical lattice consisting of N parallel square lattice layers at spacing a to make a "film" of finite thickness , limited by a single substrate. Each layer has A sites, labeled  i,j , and at each site is an Ising spin variable $\sigma_i$, which is in the simplest $S=1/2$ case takes the values $\pm1$. In the simplest model, only nearest-neighbor interactions are considered and the Hamiltonian is then
\begin{equation}
{\cal H}=-\sum_{p=1}^N\sum_{\langle i,j \rangle}J_{i,j}\sigma_i^p\sigma_j^p-\sum_{p=1}^{N-1}J_{p,p+1}\sum_i\sigma_i^p\sigma_i^{p+1}-\sum_{p=1}^NH_p\sum_i\sigma_{i}^p
\end{equation}
where $\sum_p$ is the  sum over layers, $\sum_{i,j}$ means a sum over distinct nearest-neighbors pairs ($i$,$j$) on the layer p and $\sum_i$ means a sum over nearest-neighbors sites between adjacent layers. While $\sum_i$  indicates a sum over all sites of the lattice. N is the number of layers. $H_p$, the magnetic field applied to the layer p. $J_{i,j}=J_p$ for ($i$,$j$) located in the layer p ( the intra-layer coupling), while $J_{p,p+1}$ is the analogous interlayer coupling.
In the uniform magnetic field  case $H_p$ is given by:
\begin{equation}
H_p=H_s\delta_{p,1}+H
\end{equation}
with $\delta_{p,1}=1$ for $p=1$ and zero elsewhere.
While in the variable magnetic field case $H_p$ is given by
\begin{equation}
H_p={H_s\over{p^3}}+H
\end{equation}
$H_s$ is the surface magnetic field, which depends on the nature of the substrate and H is the external magnetic field.
For technical reason it is practical to consider the reduced Hamiltonian as follows:
\begin{equation}
{-\beta\cal H}=-\sum_{p=1}^{N}K_p\sum_{\langle i,j \rangle}\sigma_i^p\sigma_j^p-\sum_{p=1}^{N-1}K_{p,p+1}\sum_i\sigma_i^p
\sigma_i^{p+1}-\sum_{p=1}^Nh_p\sum_i\sigma_{i}^p
\end{equation}
$\beta={1\over{K_BT}}$, $K_P=\beta J_p$ and $h_p=\beta H_p$;
$K_B$ is the Boltzmann constant and T is the absolute temperature .
\subsubsection{Real space renormalization group technique}
The Hamiltonian in the Eq.(4) is not invariant under the Migdal-Kadanoff renormalization group transformation. Then, it is more appropriate to redefine $\beta\cal H$ in terms of a sum of general $2N$-sites interactions:
$h_{2N,ij}^{(n)}(\sigma_i^1\sigma_i^2...\sigma_i^N;\sigma_j^1\sigma_j^2...\sigma_j^N)$
so that
\begin{equation}
{\beta\cal H}=-\sum_{\langle i,j \rangle}h_{2N,ij}^{(0)}(\sigma_i^1\sigma_i^2...\sigma_i^N;\sigma_j^1\sigma_j^2...\sigma_j^N)
\end{equation}
The sum is taken over nearest-neighbor sites on a square lattice for the three-dimensional cubical lattice, and over nearest-neighbor sites on a chain for the two dimensional square lattices.
$h_{2N,ij}^{(n)}(\sigma_i^1\sigma_i^2...\sigma_i^N;\sigma_j^1\sigma_j^2...\sigma_j^N)$, where $n$ is the renormalization-group transformation index  given initially by :
\begin{equation}
h_{2N,ij}^{(0)}=-\sum_{p=1}^NK_p^{(0)}\sigma_i^p\sigma_j^p-{1\over z}\sum_{p=1}^{N-1}K_{p,p+1}^{(0)}(\sigma_i^p\sigma_i^{p+1}+\sigma_j^p\sigma_j^{p+1})-{1\over z}\sum_{p=1}^Nh_p^{(0)}(\sigma_{i}^p+\sigma_{j}^p)
\end{equation}
Where $z=4$ in the three dimensional  cubical lattice case, and $z=2$ in the two-dimensional square lattice .$K_p^{(0)}=K_{p,p+1}^{(0)}=\beta J$ and $h_p^{(0)}=h_p$ are the parameters of the system before renormalization. Eq.(6)  has the simplicity of formally having reduced the N-layer system to an effective one layer system with a "2N-site" nearest-neighbor coupling. Hence, the partition function is given by:
\begin{equation}
Z=Tr_{\sigma_i}\prod_{\langle i,j \rangle}exp(-h_{2N,ij}^{(0)})
\end{equation} 
We now invoke a Migdal-Kadanoff bond-moving transformation in order to reduce  systematically the degrees of freedom in Eq. (7). We may use the displacement of bonds on square lattice, with a scale factor b=2. Then we sum over the variables associated to the crossed sites[69,211] (see Fig. 1). The recursion relation is easily found to be :
\begin{eqnarray}
A^{(n)}U^{(n)}(\sigma_i^1\sigma_i^2...\sigma_i^N;\sigma_j^1\sigma_j^2...\sigma_j^N)=\sum_{{\sigma_i}}[U^{(n-1)}(\sigma_i^1\sigma_i^2...\sigma_i^N;\sigma_t^1\sigma_t^2...\sigma_t^N)\nonumber\\U^{(n-1)}(\sigma_t^1\sigma_t^2...\sigma_t^N;\sigma_j^1\sigma_j^2...\sigma_j^N]^\epsilon
\end{eqnarray}
with $\epsilon=1$ in the two-dimensional square lattice case and $\epsilon=2$ in the three dimensional cubical lattice case; $A^{(n)}$ is the normalization constant, where we have used the reflection symmetry,
\begin{equation}
U^{(n)}(\sigma_i^1\sigma_i^2...\sigma_i^N;\sigma_j^1\sigma_j^2...\sigma_j^N)=U^{(n)}(\sigma_j^1\sigma_j^2...\sigma_j^N;\sigma_i^1\sigma_i^2...\sigma_i^N)
\end{equation}

\begin{figure}
\begin{center}
\includegraphics{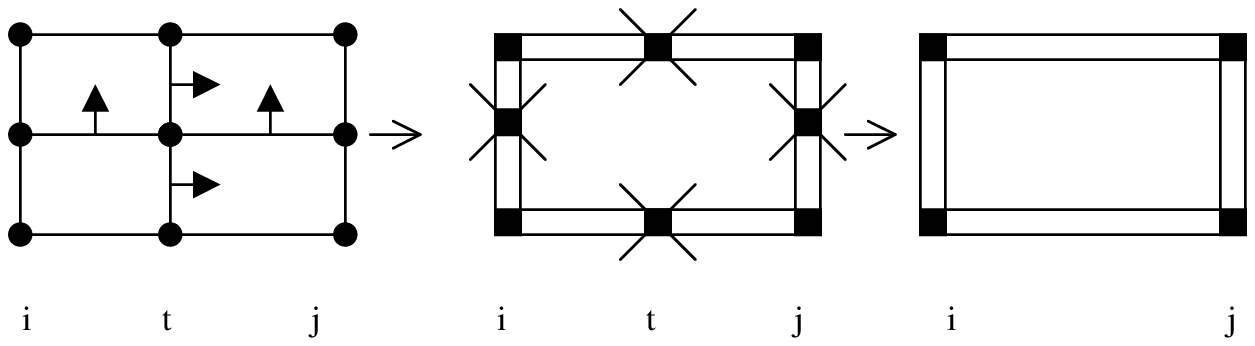}
\caption{The Migdal-Kadanoff  Model  for $b=2$.}
\label{fig1}
\end{center}
\end{figure}

Note that Eq.(8) provides one with a set of $2^{2N}$ recursion relations since each variable $\sigma_i$  can takes two values $\pm1$. The reflection symmetry (9) reduces this number to $2^{2N-1}+2^{N-1}$ , one of which is the normalization condition. In order to extract useful information from iterations of Eq.(7), we write the general 2N-site Hamiltonian in the form
\begin{equation}
-h_{2N,ij}^{(n)}(\sigma_i^1\sigma_i^2...\sigma_i^N;\sigma_j^1\sigma_j^2...
\sigma_j^N)=\sum_{r=1}^{2N}\sum_{m=1}^{f(r)}K_{m,r}^{(n)}\theta_{m,ij}
\end{equation}
where $f(r)$ is the number of different r-site Ising interactions, and
\begin{equation}
\sum_{r=1}^{2N}f(r)=2^{2N-1}+2^{N-1}-1
\end{equation}
$K_{m,r}^{(n)}$ is the interaction of type $m$ between $r$ spins,$\theta_{m,ij}$ is the function of type $m$ of $r$ coupled spins, which is symmetric by the permutation of $i$ and $j$. As an example, we give the simplest case $N=2$, where

\begin{eqnarray}
-h_{4,ij}^{(n)}(\sigma_i^1\sigma_i^2;\sigma_j^1\sigma_j^2)=K_{1,1}^{(n)}(\sigma_i^1+\sigma_j^1)+ K_{2,1}^{(n)}(\sigma_i^2+\sigma_j^2)+K_{1,2}^{(n)}(\sigma_i^1\sigma_j^1)\nonumber\\+K_{2,2}^{(n)}(\sigma_i^2\sigma_j^2)+K_{3,2}^{(n)}(\sigma_i^1\sigma_i^2+\sigma_j^1\sigma_j^2)+K_{4,2}^{(n)}(\sigma_i^1\sigma_j^2+\sigma_j^1\sigma_i^2)\nonumber\\+K_{1,3}^{(n)}(\sigma_i^1\sigma_i^2\sigma_j^2+\sigma_j^1\sigma_j^2\sigma_i^2)+K_{2,3}^{(n)}(\sigma_i^1\sigma_j^1\sigma_j^2+\sigma_j^1\sigma_i^1\sigma_i^2)+K_{1,4}^{(n)}\sigma_i^1\sigma_j^1\sigma_i^2\sigma_j^2
\end{eqnarray}
with $K_{1,1}^{(n)}$ and $K_{2,1}^{(n)}$ are respectively the renormalized reduced magnetic field; $K_{1,2}^{(n)}$ and $K_{2,2}^{(n)}$ are respectively the intra-layer renormalized reduced coupling in the first and the second layer; $K_{3,2}^{(n)}$ is the interlayer renormalized reduced coupling ;$K_{4,2}^{(n)},K_{1,3}^{(n)}$, $K_{2,3}^{(n)}$ and $K_{1,4}^{(n)}$ (initially equal to zero) are the renormalized reduced coupling between two, three and four spins, which permit to have an homogeneous system of equations. Our renormalization group calculation show in all situations that the reduced interlayer coupling iterates to infinity for the two-dimensional layers, while these parameters iterate to a fixed point (at $T=0$, this parameters iterate to infinity) in the one-dimensional layer case. The behavior of the other parameters in Eq. (10) is much more interesting in that it reflects the transition found in  the three-dimensional ferromagnet films [69,70]. The ferromagnetic layer state is characterized by the value of the renormalized reduced magnetic field, applied on this layer, in the trivial fixed point. If $h_p^{(n)}$ iterates to $+\infty$ for sufficiently large iteration n, the state of the layer p is ferromagnetic "up" and if $h_p^{(n)}$ iterates to $-\infty$, the state of the layer p is ferromagnetic "down". We remark that the result is independent of the choice of the scale factor of the real-space renormalization group.\\
\subsubsection{Transfer matrix method (TM)}
The model described by the Hamiltonian (1) is also studied using transfer
matrix method . This method consists in studying a d-dimensional strips
which are infinite in one direction and finite in the remaining ones [70].
This system with a periodic boundary condition on one finite direction
(parallel to the substrate) and the free boundary conditions on the
perpendicular direction to the substrate, is similar to a ($d-1$)
dimensional layers (see Fig. 2). The partition function $Z$ of this strip
is written under the matrix product form $(2^{NM},2^{NM})$ called the
transfer matrix, M is the width of the strips and N is the number of layers,
namely:
\begin{equation}
Z=\prod_{L=1}^{\infty}V(L,L+1)
\end{equation}
This makes the transfer, from the plane L to the plane L+1, in the infinite direction.\\

\begin{figure}
\begin{center}
\includegraphics{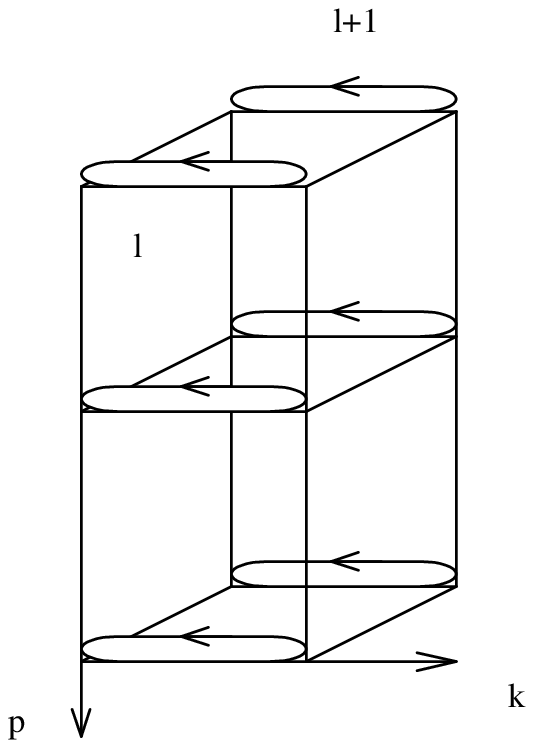}
\caption{The strip used within the finite size method, the transfer is made from the plane $l$ to the plane $l+1$. Cyclic boundary conditions are made in the direction k and free boundary conditions in the direction p (p is the layer index).}
\label{fig2}
\end{center}
\end{figure}

The elements of this matrix are given by
\begin{equation}
V(i,j)=exp(\sum_{p,k}\sigma(p,k,i)(\sigma(p+1,k,i)+\sigma(p,k+1,i)+\sigma(p,k,j))+\sum_{p}h_p\sigma(p,k,i)]/t)
\end{equation}
where  $hp={H_p/J}$; $t=T/J$ and $\sigma(p,k,i)$ is the spin variable which takes two values $\pm1$; $p=1,...�,N$; $i$ and $j$ takes $2^{NM}$ values ; $p$ is the layer index which takes $N$ values and $k$ takes $M$ values. The periodic conditions on the direction k indicate that $\sigma(N+1,k,i)=0$. The free energy per site is given by $f=-TLog(\lambda_{Max})$ where $\lambda_{Max}$ is the largest eigenvalue of the transfer matrix. In order to determine the state of each layer, it is necessary to define the magnetization per site on the layer p as follows,
\begin{equation}
m_p=-{\partial f\over \partial h_p}
\end{equation}
where $h_p$ is the reduced magnetic field applied on a layer $p$. The magnetization $m_p$ of a layer $p$ can be positive or negative. However, if $m_p$ is positive, the state of the layer $p$ is ferromagnetic "up", while if $m_p$ is negative, the state of such layer is ferromagnetic "down". The transition of this layer is given by the change of the magnetization sign.
\subsubsection{Mean Field Theory}
The mean field state equations of the Hamiltonian (1) may be obtained from the variational functional,
\begin{eqnarray}
F[m_p]=\sum_{p=1}^N T[{{1-m_p}\over 2}Log(1-m_p)+{{1+m_p}\over 2}Log(1+m_p)-Log(2)]\nonumber\\-H_pm_p-{J\over 2}m_p(m_{p-1}+4m_p+m_{p+1}\}
\end{eqnarray}
with the free boundaries conditions $m_0=m_{N+1}=0$. At the global minimum of the functional free energy(Eq.16), the magnetization per site in the layer p is $<\sigma_p>=m_p^{min}$, and the free  energy of the system is $F=F[m_p^{min}]$. The stationary condition  $\partial F/\partial m_p=0$ leads to the mean field state equations,
\begin{eqnarray}
m_p=tanh[{1\over T}(m_{p-1}+4m_p+m_{p+1}+h_p)]; p=1,...,N
\end{eqnarray}
These equations are solved numerically using iterative method. Different initial guesses can lead to different solutions of the Eq. 17; the one which makes the variational functional  free energy smallest is selected as the global minimum. At the transition of the layer p, the magnetization $m_p$ has a discontinuity for a temperature smaller than the critical-end-point temperature.
\subsubsection{Uniform surface magnetic field}
At first it is useful and instructive to find the ground state of the Hamiltonian (1) describing the phase diagrams in the $(h_s-h)$ plane at $T=0K$, this is determined analytically. We denote here after by $1^p0^q$ the configuration having p layers with positive magnetization  and q layers with negative ones. The ground state phase diagram is presented in Fig. 3. It is found that there exist two different situations depending on the value of the surface magnetic field. Indeed, for $h_s <N/(N-1)$ we have one transition $(0^N \leftrightarrow 1^N)$, at $H=-h_s/N$ in which the $N$ layers transit simultaneously and for $h_s>N/(N-1)$, we have two transitions; one at $H=-1/(N-1)$, $(1^N\leftrightarrow 10^{N-1})$ where the  surface magnetization does not changes the sign, but the remaining $(N-1)$ layers transit simultaneously , and the second $(10^{N-1}\leftrightarrow 0^N)$,at $H=a+bh_s$, where only the top layer (surface) transits (i.e  the surface magnetization changes the sign).\\

\begin{figure}
\begin{center}
\includegraphics{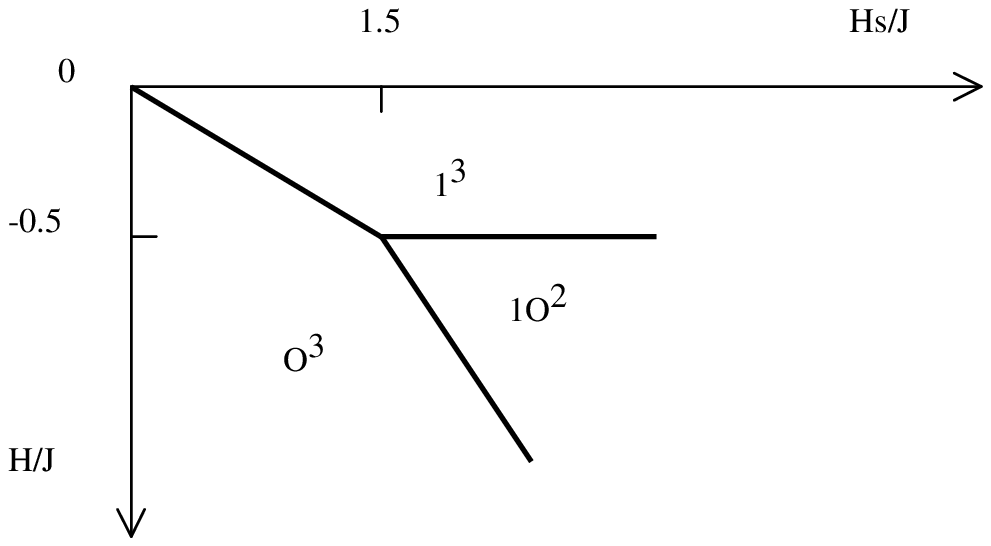}
\caption{Phase diagram at $T=0$ in the uniform field case and $N=3$. }
\label{fig3}
\end{center}
\end{figure}

At finite temperature, numerical calculations are done in the case of a thin films $N=3$. However, within RSRG and TM methods we obtain one transition $(0^N\leftrightarrow 1^N)$ for $h_s<1.5$ and $T<T_L$ (Fig. 4a), in agreement with the ground state results, and three transitions $(0^3\leftrightarrow 10^2\leftrightarrow 1^20\leftrightarrow 1^3)$ for $T_L<T<T_c$. For $h_s>1.5$, we obtain two transitions $(0^3\leftrightarrow 10^2\leftrightarrow 1^3)$ for $T<T_L$ as predicted by the ground state calculations and three transitions $(0^3\leftrightarrow 10^2\leftrightarrow 1^20\leftrightarrow 1^3)$ (Fig. 4b). $T_c$ is the transition temperature of a three layer film $(N=3)$. The layering temperature $T_L$ above which a sequence of layer transitions occurs depends strongly on the strength of the potential of substrate, i.e the value of the surface magnetic field $h_s$. Within MFT only the ground state phases appear at finite temperature, since MFT neglects correlation between spins, hence the effects of thermal fluctuations are negligible, and then does not raise the degeneration of the states at $T>0$.

\begin{figure}
\begin{center}
\includegraphics{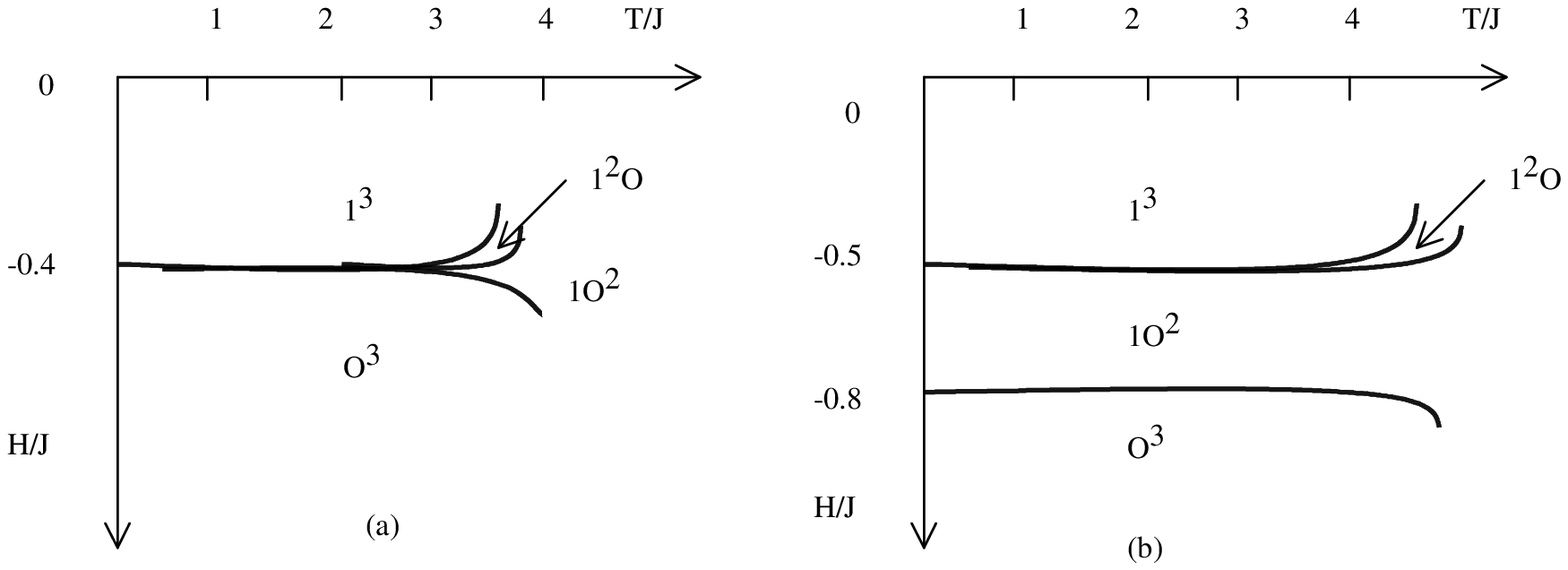}
\caption{Phase diagram in $(T-h)$ plane for $N=3$ obtained within RSRG in the uniform magnetic case ; (a) $hs=1.2 < 1.5$ ; (b)
$hs=1.8 > 1.5$ [69,70]}
\label{fig4}
\end{center}
\end{figure}

\subsubsection{Variable surface magnetic field}
In this case the problem is quite different from the uniform surface magnetic field since for sufficiently large value of $h_s$ the layering transitions occur at $T=0K$ (Fig. 5). However for $hs_q<h_s<hs_{q+1}$ with $q=1,...N$ ; we have q transitions in the $h_s-h$ plane namely $(0^N \leftrightarrow 10^{N-1}\leftrightarrow 1^20^{N-2}\leftrightarrow... 1^{q-1}0^{4-q}\leftrightarrow 1^q)$, where $h_{s_1}=0$, $h_{s_2}=1.63$, $h_{s_3}=11.36$ and $h_{s_4}=\infty $.\\

\begin{figure}
\begin{center}
\includegraphics{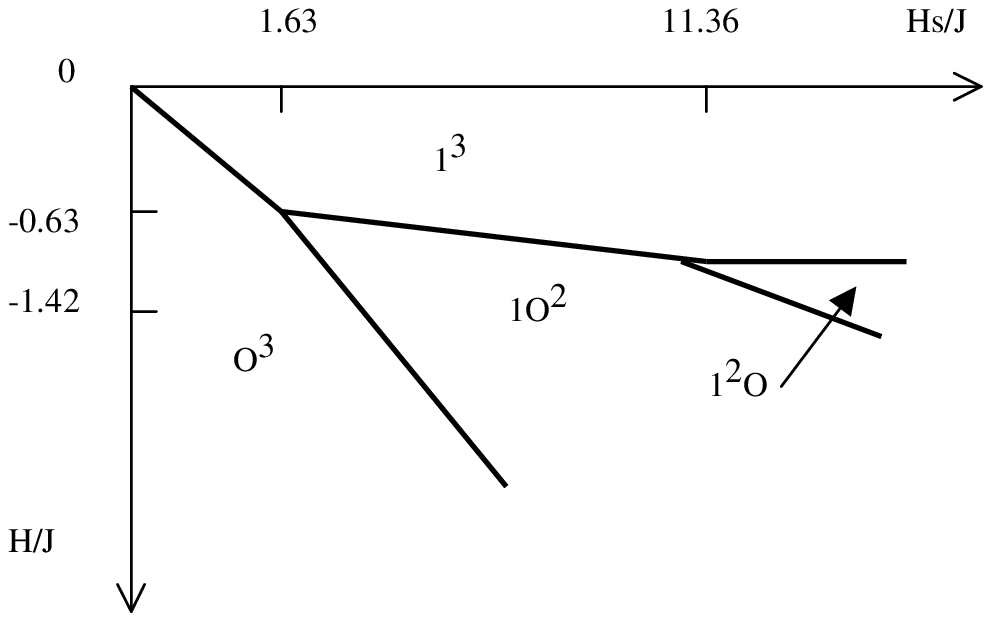}
\caption{Phase diagram at $T=0$ in the variable field case and $N=3$.}
\label{fig5}
\end{center}
\end{figure}

 Using RSRG and TM methods, three different cases are distinguished, at finite temperature:\\
i)For $h_s<h_{s_2}$ only one transition $(0^3\leftrightarrow 1^3)$ persists for $T<T_L$ and three transitions $(0^3\leftrightarrow 10^2\leftrightarrow 1^20\leftrightarrow 1^3)$ for $T_L<T<T_c$ (Fig. 6a).\\
ii)For $h_{s_2}<h_s<h_{s_3}$ we have two transitions $(0^3\leftrightarrow 10^2\leftrightarrow 1^3)$ for $T<T_L$, while for $T>T_L$, (Fig. 6b) we obtain three transitions $(0^3\leftrightarrow 10^2\leftrightarrow 1^20\leftrightarrow 1^3)$. While MFT gives rise to only two transitions $(0^3\leftrightarrow 10^2\leftrightarrow 1^3)$ for any value of $h_s$ smaller than $h_{s_3}$ even for $T>T_L$.\\
iii)For a strong potential of the substrate $(h_s>h_{s_3})$, the layering temperature $T_L$ vanishes, and then, the three methods give rise to the appearance of three transitions, $(0^3\leftrightarrow 10^2\leftrightarrow 1^20\leftrightarrow 1^3)$,  for any temperature $0<T<T_c$ (Fig. 6c)

\begin{figure}
\begin{center}
\includegraphics{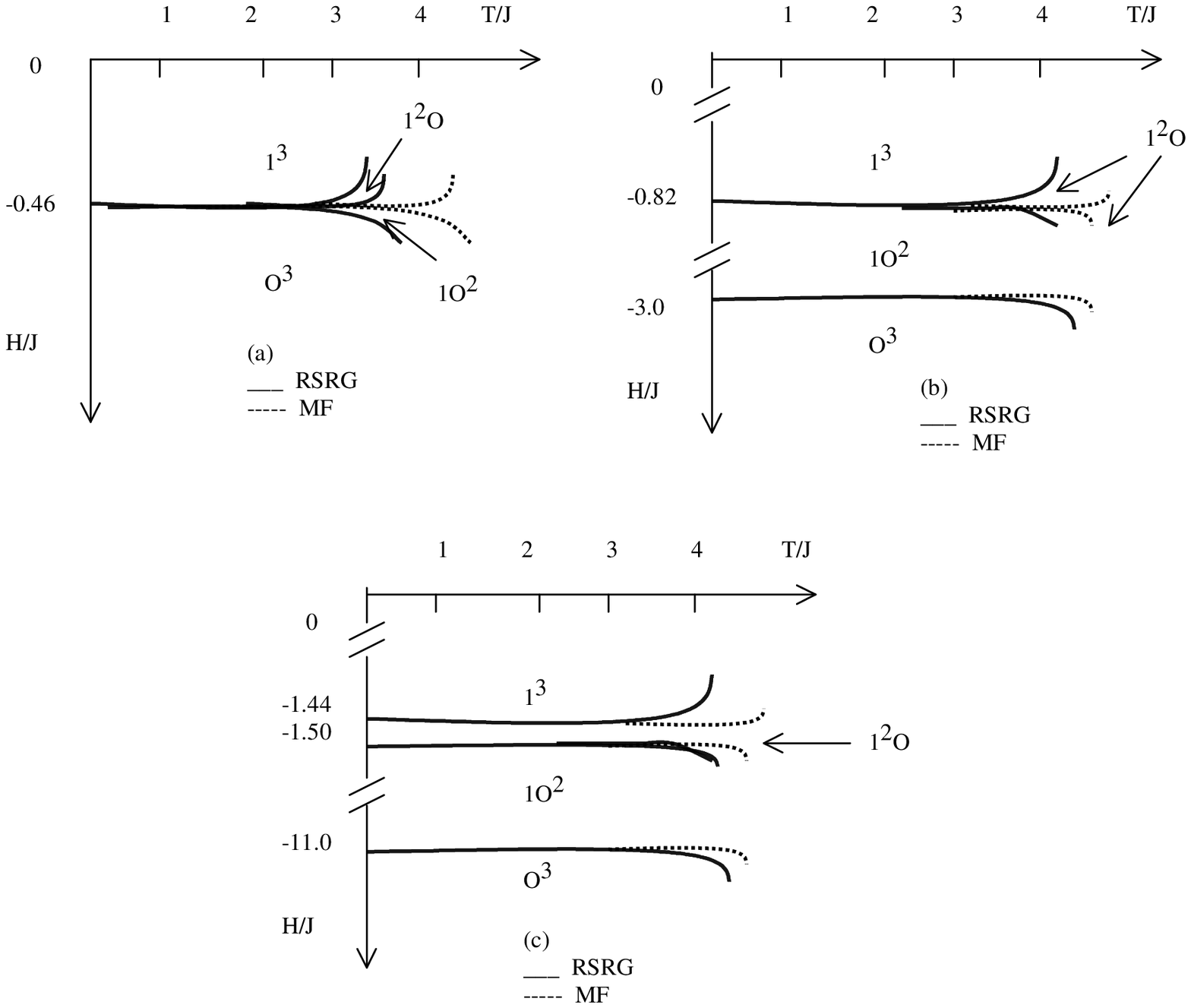}[t]
\caption{Phase diagram in $(T-h)$ plane for $N=3$ in the variable case ; (a) $hs=1.2$ ;
(b) $hs=4$ ; (c) $hs=12$. Solid lines correspond to the RSRG method and dashed lines correspond to MFT
[69]}
\label{fig6}
\end{center}
\end{figure}

\subsection{Surface coupling effects}
We consider the Hamiltonian given in Eq.1, in which, in this case, the surface and bulk couplings are different; $J_{i,j}=J_s$ and $J_{i,j}=J_b$ if $\{i,j\}$ are in the surface and the bulk respectively (Fig. 7).

\begin{figure}
\begin{center}
\includegraphics{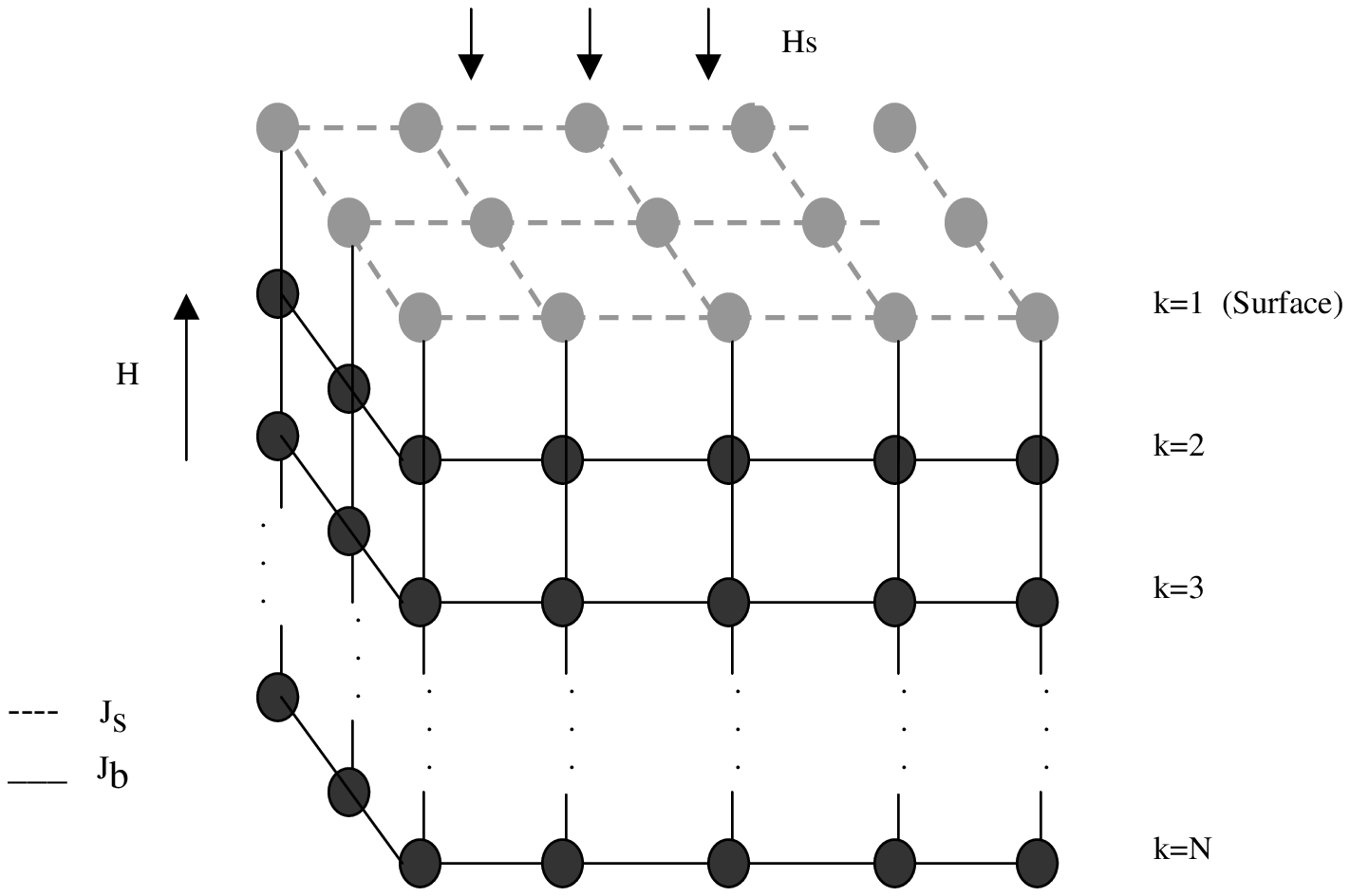}
\caption{A sketch of the geometry of the system formed with $N$ layers and subject to the surface magnetic field $H_s$ and the global magnetic field $H$. The distribution of the coupling constants : $J_s$ denoting the interaction coupling constant between the spins of the surface and $J_b$ corresponding to the interaction coupling constant between the spins of the bulk as well as between the spins of the bulk and those of the surface ; are also  presented. }
\label{fig7}
\end{center}
\end{figure}

In the following, the notation $1^{k}O^{N-k}$ means that a configuration of the system have the first $k$ layers with positive magnetizations and the remaining $N-k$ layers with negative ones. However, there exist $N+1$ possible configurations for a film formed  with $N$ square layers.
Numerical calculations are limited to $\alpha=2.0$, since the results for other values of $\alpha$ are qualitatively the same.\\
At $T=0$, and for  $(N \ge 3)$, there is only three allowed transitions.  $O^N \leftrightarrow 1^N$, $O^N \leftrightarrow 1O^{N-1}$ and $1O^{N-1} \leftrightarrow 1^N$.\\
At $T \ne 0$, Fig. 8 shows that the layering temperature, $T_{L}$, above which a succession of layering transitions occurs is greater than the "surface transition" temperature. Hence, the film thickness must be at least equal to three layers and so the results established throughout this work correspond to films with a thickness $N \ge 3$ layers.

\begin{figure}
\begin{center}
\includegraphics{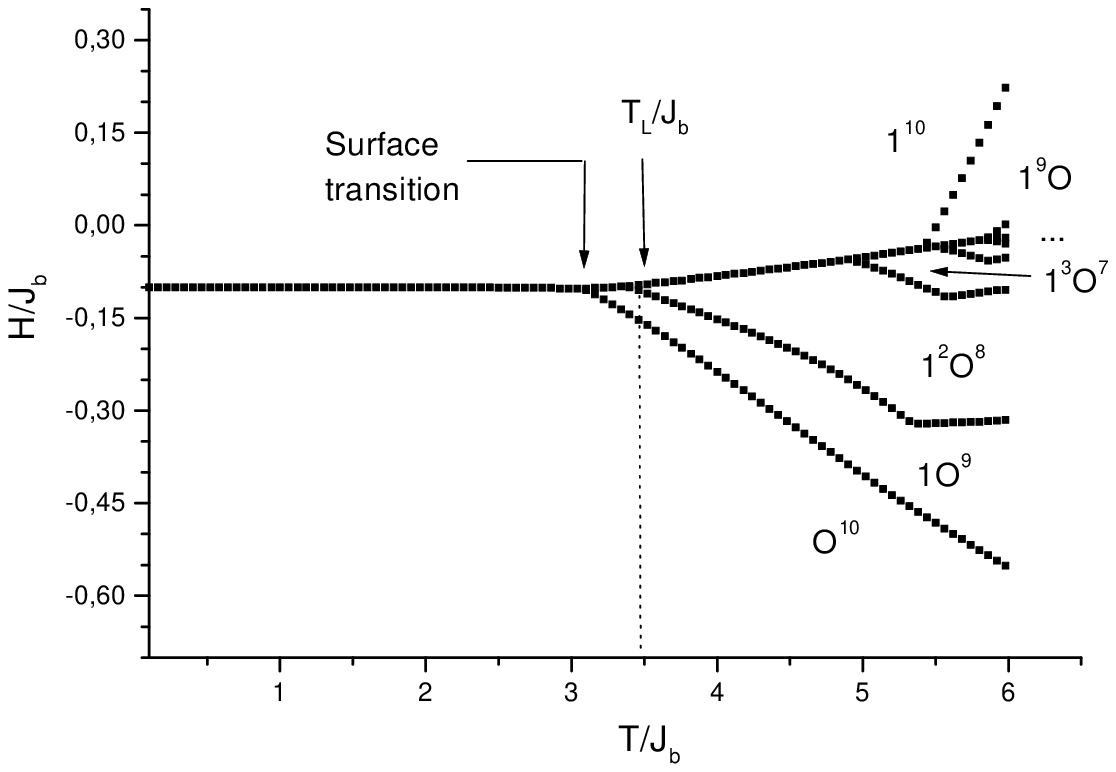}
\caption{Phase diagram, in the $(H/J_b,T/J_b)$ plane, showing the layering transitions, and the definition of the layering temperature $T_L/J_b$, which is different from the surface transition. This figure is plotted for a system size with $N=10$ layers, a magnetic surface field $H_s/J_b=1.0$ and a surface coupling constant $J_s/J_b=1.5$ [250]}
\label{fig8}
\end{center}
\end{figure}

It is understood that the wetting temperature $T_w /J_b$ is close to the limit of  $T_L/J_b$ for a sufficiently thick film $(N \rightarrow \infty)$.
The layering temperature behavior as a function of the system size is illustrated in Figs. $9a$ and $9b$ for two surface field values and several values of the surface coupling. However, for smaller values of the surface field, the layering and wetting temperatures are independent of the surface coupling for large system sizes as it is shown in Fig. 9a. While for higher values of the surface field (Fig. 9b), the wetting and layering temperatures depend strongly on the surface coupling values. Indeed, there exist three different classes of the layering temperature behaviors:\\
$(i)$$T_L$ increases with the film thickness and stabilizes at a certain fixed value. Such behavior occurs at small values of $J_s$ and any value of $h_s$.\\
$(ii)$$T_L$ increases until a certain film thickness above which it decreases exhibiting a maximum. This situation take place for medium values of both $J_s$ and $h_s$.\\
$(iii)$$T_L$ decreases continuously and stabilizes at a value close to $T_w$. This behavior occurs at higher values of $J_s$.

\begin{figure}
\begin{center}
\includegraphics{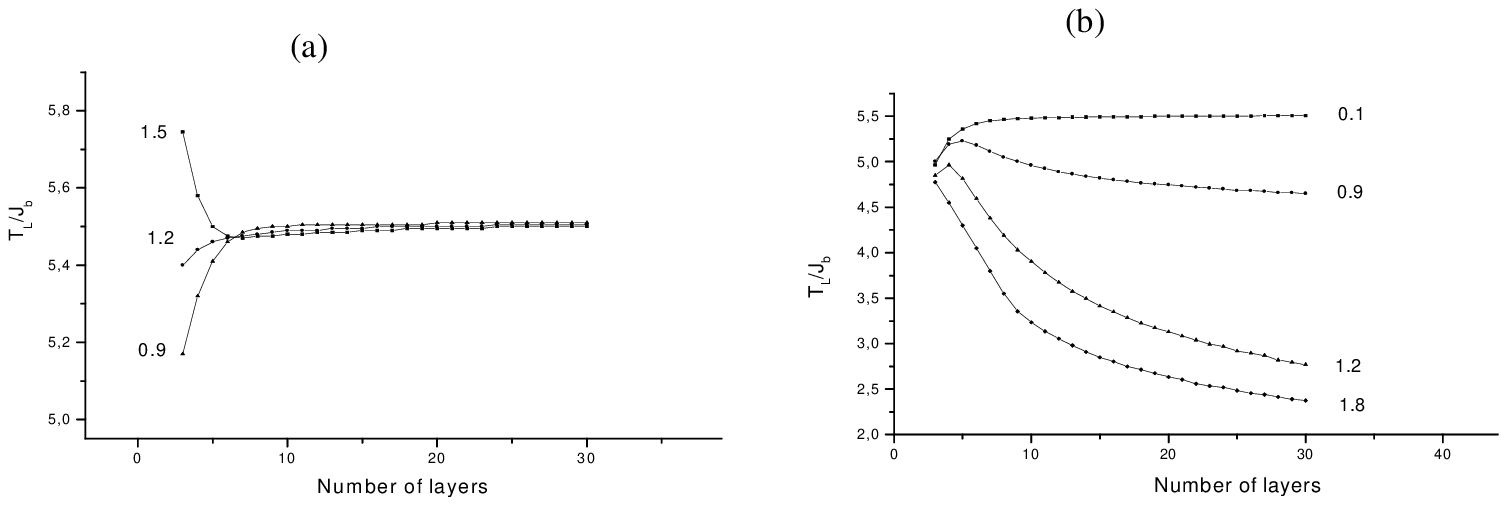}
\caption{ Layering temperature, $T_L/J_b$, behavior as a function of
the number of layers.
a) $H_s/J_b=0.1$ and b) $H_s/J_b=1.0$.
The number accompanying each curve denotes the surface coupling $J_s/J_b$
value [250] }
\label{fig9}
\end{center}
\end{figure}

Furthermore, the wetting temperature $T_w$, which is the limit value of $T_L$ for $L$ sufficiently large, is independent of $J_s$ for small values of $h_s$. While for higher values of $h_s$, it depends strongly on the $J_s$ and decreases when $h_s$ increases. Indeed, this is because the surface region becomes magnetically harder than the deeper layers, and as the film is thicker, the strength of the internal magnetic field decreases, which leads to a decrease of $T_w$. \\
We note that, the three distinct behaviors mentioned above are separated by two critical surface coupling $J_{sc1}$ and $J_{sc2}$. Indeed, $J_{sc1}$ separates the behaviors $(i)$ and $(ii)$, while $J_{sc2}$ separates the behaviors $(ii)$ and $(iii)$. The dependence of  $J_{sc1}$ and $J_{sc2}$ as a function of $h_s$ is plotted in Fig. 10.

\begin{figure}
\begin{center}
\includegraphics{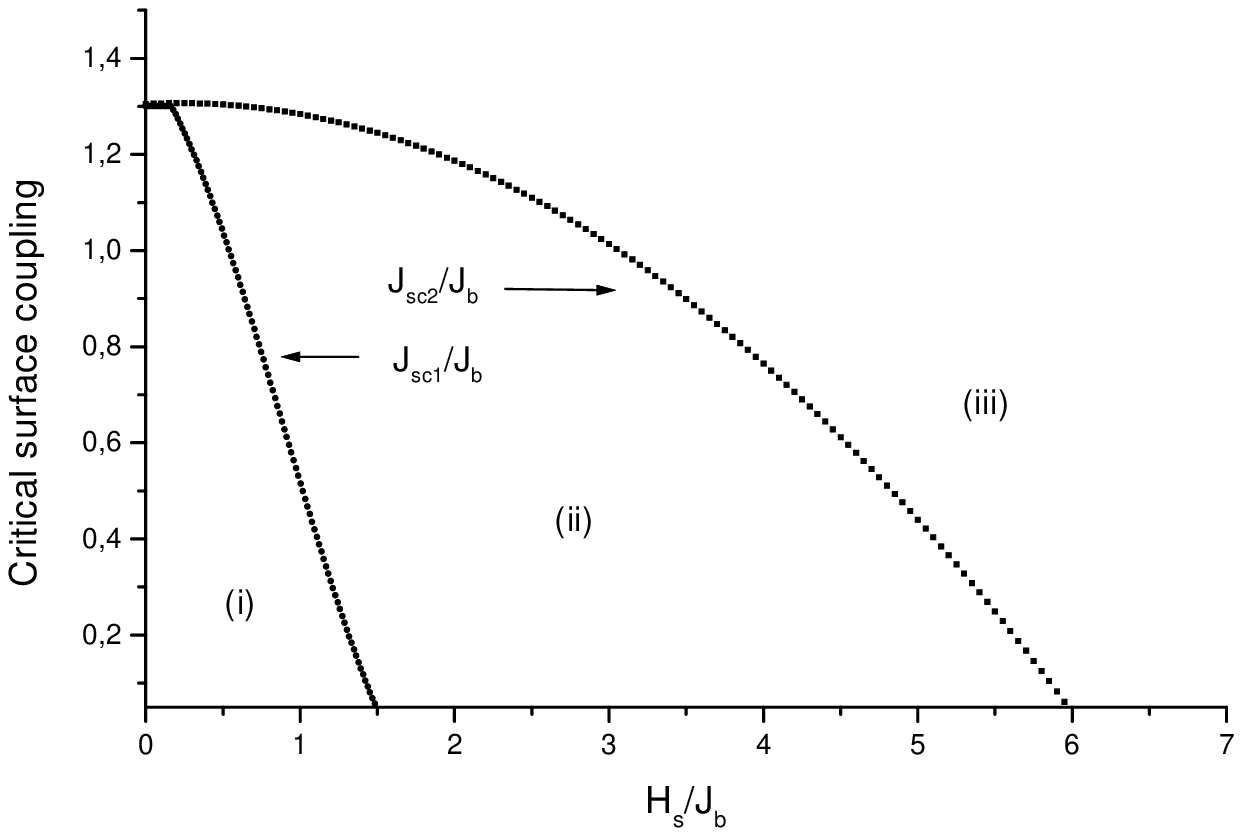}
\caption{ Critical surface coupling, $J_{sc1}/J_b$ and $J_{sc2}/J_b$, profiles as a function of the surface magnetic field $H_s/J_b$ for:
 the (USF) case.
 The behaviors $(i)$ and $(ii)$ are separated by the $J_{sc1}/J_b$ line;
 while $J_{sc2}/J_b$ separates the behaviors $(ii)$ and $(iii)$ [250] }
\label{fig10}
\end{center}
\end{figure}

$J_{sc1}$ and $J_{sc2}$ correspond to two critical surface fields $h_{sc1}$ and $h_{sc2}$ separating the three different regimes of the layering temperature $T_L$. However, for sufficiently large values of $J_s$ ($J_s/J_b \ge 1.30$), $T_L$ decreases when increasing the film thickness $N$ for any values of $h_s$. While, for $J_s < 1.3$, the temperature $T_L$ exhibits the three behaviors; (i), (ii) and (iii)  depending on the value of $h_s$. In particular for $h_s \ge h_{sc2}$, $T_L$ is always decreasing function of the film thickness.
The layering temperature behavior is plotted, in Fig. 11, for a fixed system size $N=5$ layers and several value of $h_s$. It is found that $T_L$ increases and decreases for small and higher values of $h_s$ respectively. On the other hand, the wetting temperature, Fig. 11, is not affected by the surface coupling variations for small surface field values, while for higher $h_s$, $T_w$ decreases abruptly and stabilizes at a certain limit, when increasing the surface coupling.

\begin{figure}
\begin{center}
\includegraphics{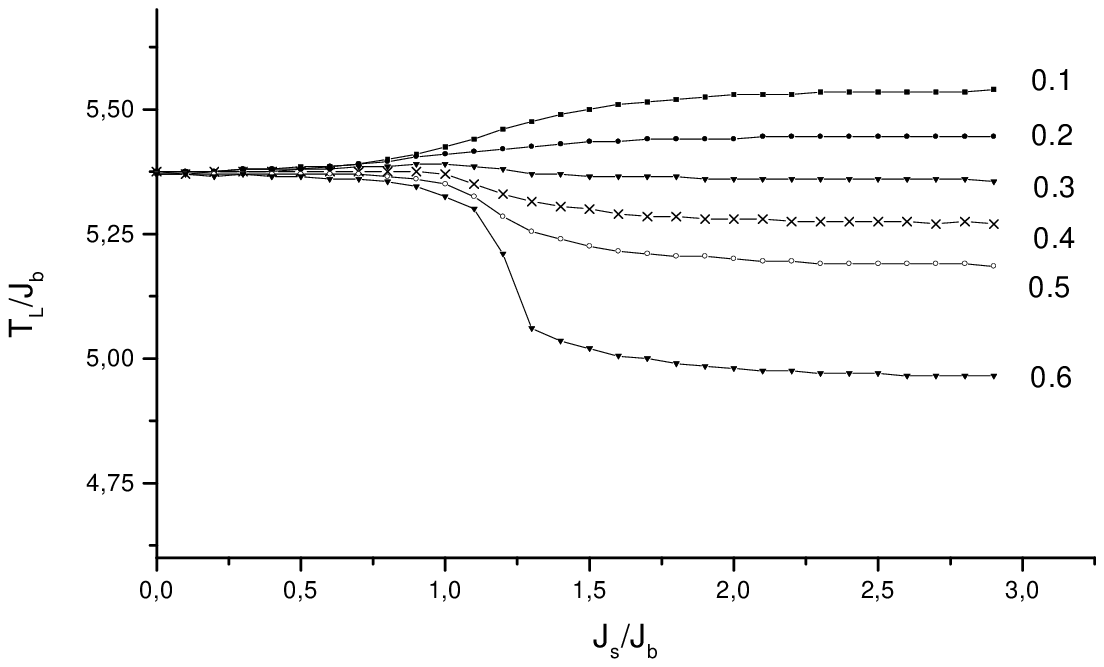}
\caption{ Layering temperature and wetting temperature behaviors, for the (USF) case, as a function of the surface coupling $J_s/J_b$ and selected values of the surface magnetic field $H_s/J_b$: $0.1$, $=0.2$, $0.3$, $=0.4$, $0.5$ and $0.6$.
$T_L/J_b$ is plotted for a fixed system size with $N=5$ layers [250]}
\label{fig11}
\end{center}
\end{figure}

\subsection{The surface crystal field effects}
The system we are studying here is formed with $N$ coupled ferromagnetic
square layers in the presence of a crystal field. The Hamiltonian governing
this system is given by
\begin{equation}
{\cal H}=-\sum_{<i,j>}J_{ij}S_{i}S_{j}+\sum_{i}\Delta_{i}(S_{i})^2
\end{equation}
where, $ S_{l}(l=i,j)=-1,0,+1$ are the spin variables. The interactions
between different spins are assumed to be constant so that $J_{ij}=J$.
All sites $i$ of a given layer $k$, $(k=1,2, ...,N)$ are subject to
a crystal field amplitude $\Delta_{i}=\Delta_{k}$ so that $\Delta_{1}>\Delta_{2}>...>\Delta_{N}$. The crystal field of a layer k is given by:
\begin{equation}
\Delta_{k}=\Delta_{s}/k^{\alpha}
\end{equation}
where $\Delta_s=\Delta_{1}$ is the crystal field acting on the surface
(first layer $k=1$) and $\alpha$ a positive constant. For a system with
$\alpha=0$, a constant crystal field is applied on each layer of the film. Whereas, for
$\alpha$ $\rightarrow$ $\infty$, the crystal field is applied only on the
first layer ($k=1$). Experimental studies of a sequence of reentrant layering transitions of
argon and krypton adsorbed on graphite substrate are given in Ref. [54], in which the algebraically decaying of the
 applied field according to a law of  type ($k^{-\alpha}$)have been used. However, the quantities computed, using the Monte Carlo simulations for each layer
$k$ containing $N_x$ spins in $x-$direction and $N_y$ spins in the
$y-$direction, are: \\
-The layer average magnetizations
\begin{equation}
m_k=(\sum_{i\epsilon k}{S_i})/(N_x  N_y)
\end{equation}
-The layer magnetic susceptibilities
\begin{equation}
\chi_{m,k}=\beta N_x  N_y  <(m_k-<m_k>)^2>
\end{equation}
-The layer quadrupolar magnetic susceptibilities expressed as
\begin{equation}
\chi_{q,k}=\beta  N_x  N_y  <(q_k-<q_k>)^2>
\end{equation}
with
\begin{equation}
q_k=(\sum_{i\epsilon k}{S_i{^2}})/(N_x  N_y).
\end{equation}
-The layer critical exponents $\gamma_{m,k}$ related to
the corresponding layer magnetic susceptibilities $\chi_{m,k}$, for a
fixed layer $k$, giving by
\begin{equation}
\gamma_{m,k}=\frac{\partial Log(\chi_{m,k})^{-1}}
{\partial Log \mid T_{c,k}/J-T/J \mid}
\end{equation}
where $T/J$ stands for the reduced absolute temperature and $T_{c,k}/J$ is the reduced
critical temperature of the layer $k$.
In the above equations $\beta=1/(k_B T)$ with $k_B$ being the Boltzmann
constant. \\
The notation $D^kO^{N-k}$ means that the first $k$ layers
from the surface are disordered while the remaining ${N-k}$ layers are ordered. The ground state phase diagram (Fig. 12) of this model
[102], shows that, there exist only $k_0$ possible layer transition which
depends on $\alpha$ and N.

\begin{figure}
\begin{center}
\includegraphics{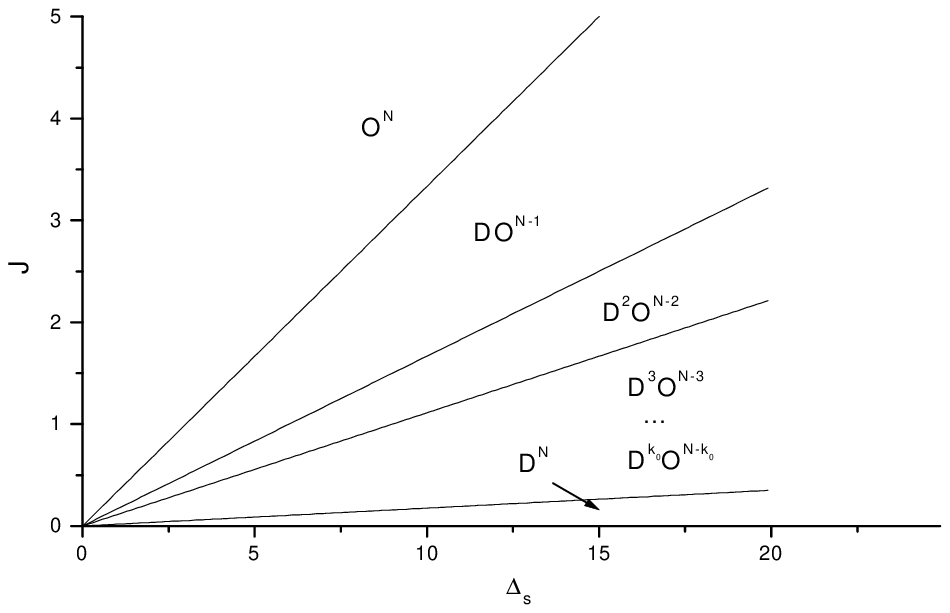}
\caption{ The ground state phase diagram in the plane $(J,\Delta_s)$ for a film formed with $N$ layers. The notations $D^kO^{N-k}$ are defined in the body text. This figure is plotted for $N=10$ layers. }
\label{fig12}
\end{center}
\end{figure}

For $T>0$ Monte Carlo simulations shows, for $N=10$, that the last layers
$k=8$, $9$ and $10$
transit simultaneously at $\Delta_s/J \approx 23.84$.
\\
Phase diagrams exhibit critical, tricritical and reentrant behaviors (Fig. 13). The latter is due, for the first $k_0$ layers, to the competition between thermal fluctuations, the crystal
field and an effective
magnetic field created by the deeper ordered layers. Indeed, when these thermal
fluctuations become sufficiently important, the magnetization of some spins,
of deeper layers, becomes nonzero (+1 or -1). This leads to the appearance
of an effective magnetic field.

\begin{figure}
\begin{center}
\includegraphics{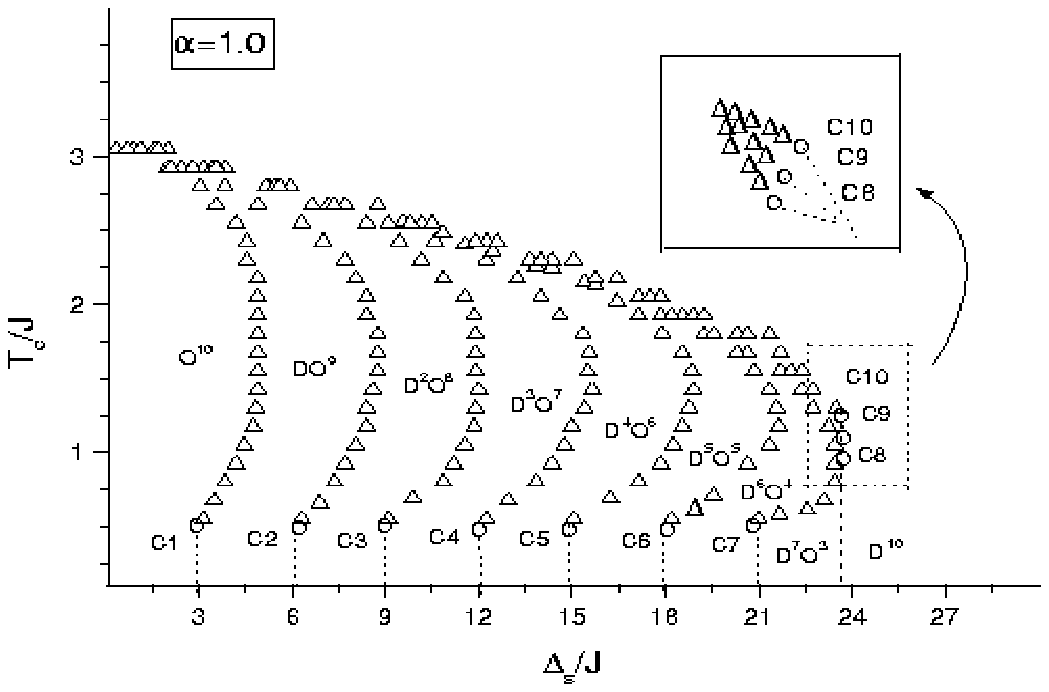}
\caption{ The critical temperature behavior as a function
of the surface crystal field $\Delta_s /J$ for $\alpha=1.0$ and a film
thickness $N=10$ layers. For each layer $k$ $(k=1,2,...,N)$, the first-order
transition line (vertical dashed line) is connected to the second-order
transition line (up-triangular points) by a tri-critical point $C_k$
(open circle) [251]}
\label{fig13}
\end{center}
\end{figure}

This magnetic field is responsible of the ordered
phase seen for the layer $k$.
This argument can also explain the absence of the
reentrant phenomena for the last $N-k_0$ layers, once the value of the
crystal field is not sufficient to overcome the effective magnetic field
created by the remaining deeper layers. It is worth to note that the reentrant phenomena is always present for the layers $k$, ($k \leq k_0$), and the corresponding tri-critical points
$C_i$ are located at a constant temperature. At the first order transition the layer quadrupolar magnetic susceptibilities $\chi_{q,k}$
, present a strong peak at the
surface crystal field transition (Fig. 14).

\begin{figure}
\begin{center}
\includegraphics{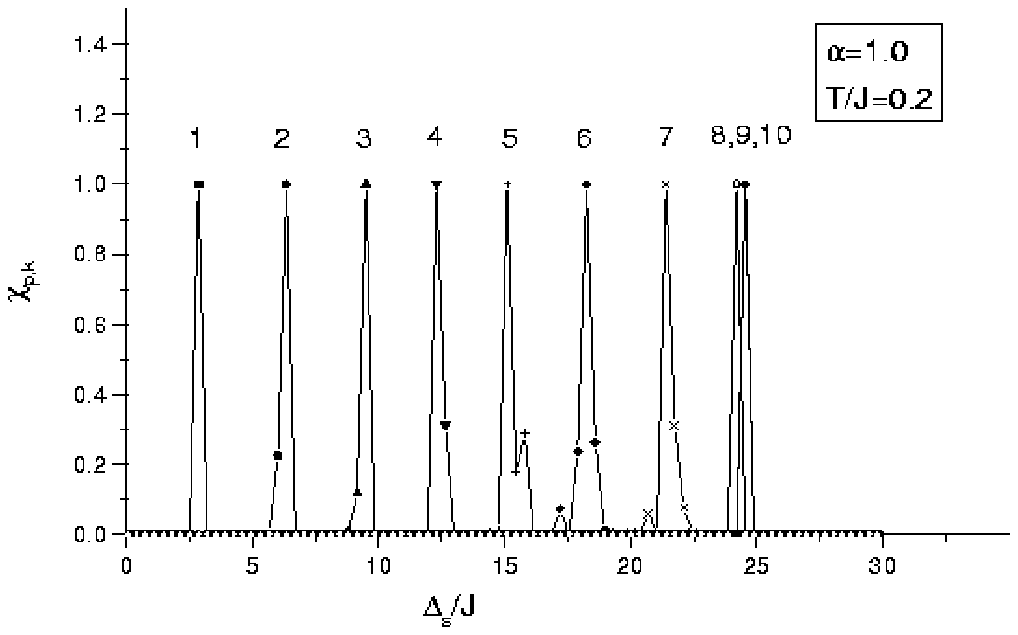}
\caption{ The behavior of the layer quadrupolar magnetic susceptibilities
$\chi_{q,k}$, showing a first-order transition, as a function of
the surface crystal field $\Delta_s /J$, for a very low temperature:
$T /J=0.2$. The number accompanying each curve denotes the layer order from the surface
to deeper layers [251]}
\label{fig14}
\end{center}
\end{figure}

While  the  magnetic susceptibilities $\chi_{m,k}$, present a strong peak at the
second order transition (Fig. 15).

\begin{figure}
\begin{center}
\includegraphics{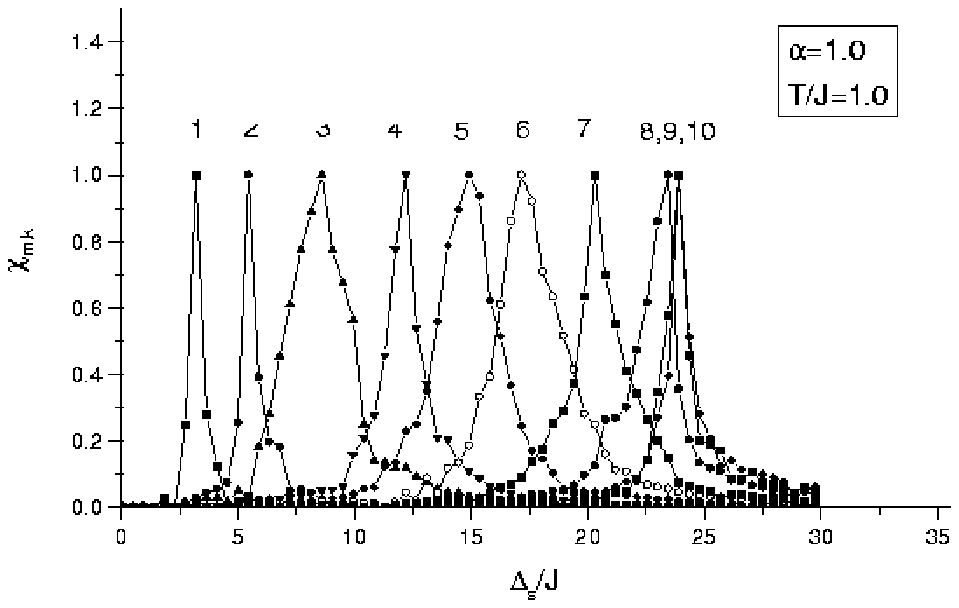}
\caption{ The dependence
of the reduced layer magnetic susceptibilities as a function of the surface
crystal field $\Delta_s /J$ for $T/J=1$ [251] }
\label{fig15}
\end{center}
\end{figure}

In the case of reentrant behavior, $\chi_{m,k}$ present two successive peaks at the transition points (Fig. 16). The critical
exponents $\gamma_{m,k}$ of magnetic susceptibility of each layer, for $N=10$.

\begin{figure}
\begin{center}
\includegraphics{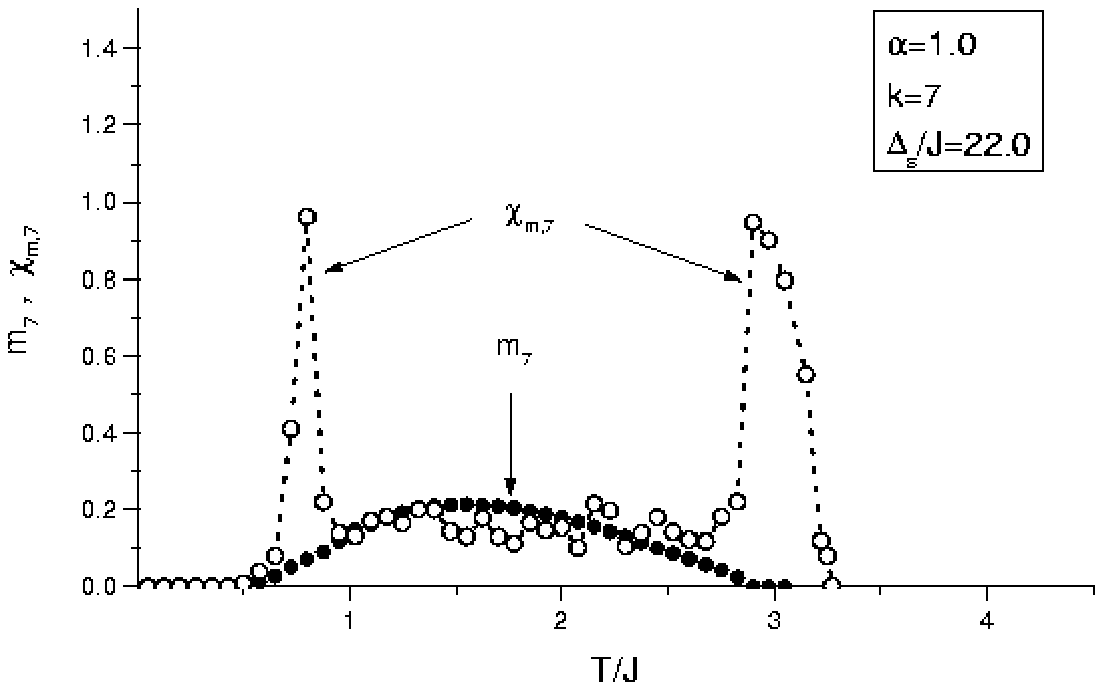}
\caption{Thermal behavior of the layer magnetization $m_7$
(line with solid circles) and the reduced layer magnetic susceptibility
$\chi_{m,7}$ (line with open circles) for a deeper layer $k=7$. for
$\Delta_s/J=22.0$ [251]. }
\label{fig16}
\end{center}
\end{figure}

It is found that
$\gamma_{m,k}$ decreases, for a fixed order $k$, with the system size
$N_x \times N_y$ and stabilizes at certain value; and due to the free
boundary conditions the critical exponents of the layers $k=1$ and $k=N$
are found to be greater than those of the internal layers $2 \le k \le N-1$. We note that the critical exponent of a 2D layer in a film is different than the one of a free $2D$-layer.Indeed the critical exponent of a layer depends on its position (Fig. 17).

\begin{figure}
\begin{center}
\includegraphics{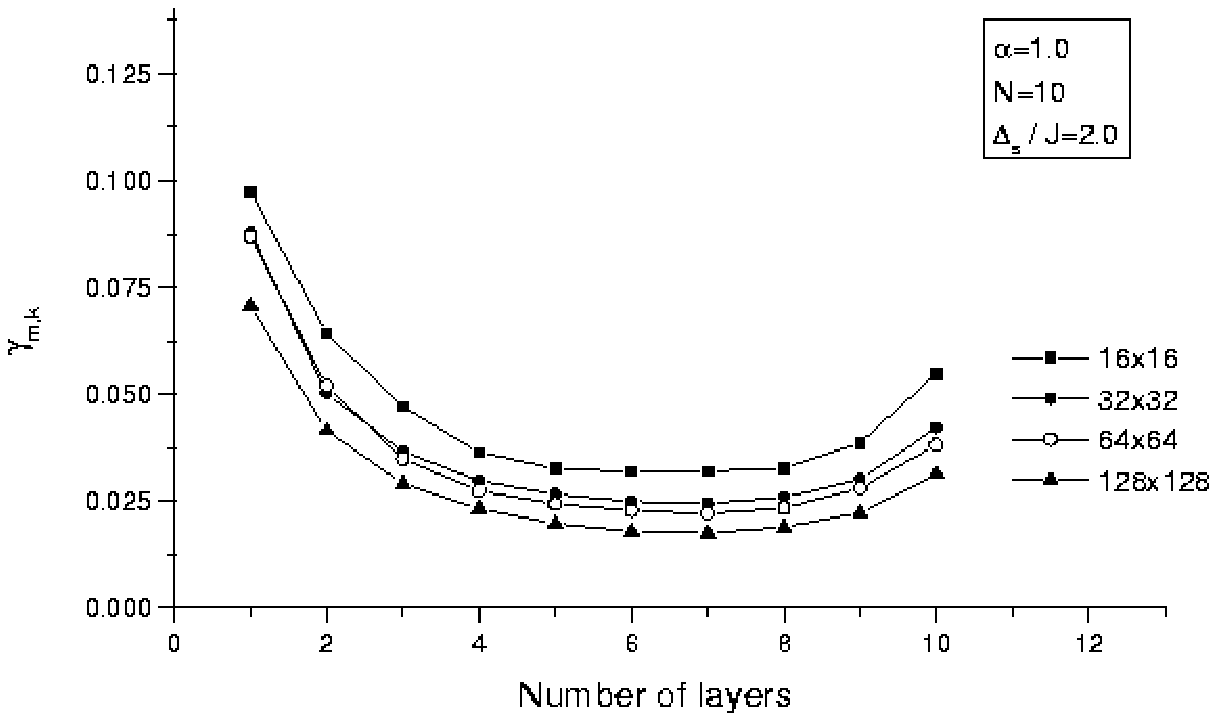}
\caption{ Layer critical exponents $\gamma_{m,k}$ for $\Delta_s/J=2.0$
as a function of the position 'k' for a film thickness $N=10$ layers, and several sizes :
$N_x \times N_y =16 \times 16$ (fill squares);
$N_x \times N_y =32 \times 32$ (fill circles);
$N_x \times N_y =64 \times 64$ (open circles);
and $N_x \times N_y =128 \times 128$ (filled up-triangulars) [251] }
\label{fig17}
\end{center}
\end{figure}

\subsection{The effect of the inhomogeneity of substrate}
\subsubsection{Model}
The effect of the inhomogeneity of the substrate on the wetting transitions is studied, using Monte Carlo simulations, in  the case of a spin-$1/2$ Ising ferromagnetic film formed by $N$ coupled square layers, in which the surface magnetic field $H_s$  acts only on alternate clusters of spins of the surface ($k=1$) indicated by the symbols $(+)$; while $H_s$ is absent on clusters of symbols $(o)$ (Fig. 18). The Hamiltonian governing this system is given by
\begin{equation}
{\cal H}=-J\sum_{<i,j>}S_{i}S_{j}-\sum_{i}(H+H_{s_i})S_{i}
\end{equation}
where, $ S_{l}(l=i,j)=-1,+1$ are the spin variables. The interaction
between different spins is assumed to be constant.
$H$ is the external magnetic field.

\begin{figure}
\begin{center}
\includegraphics{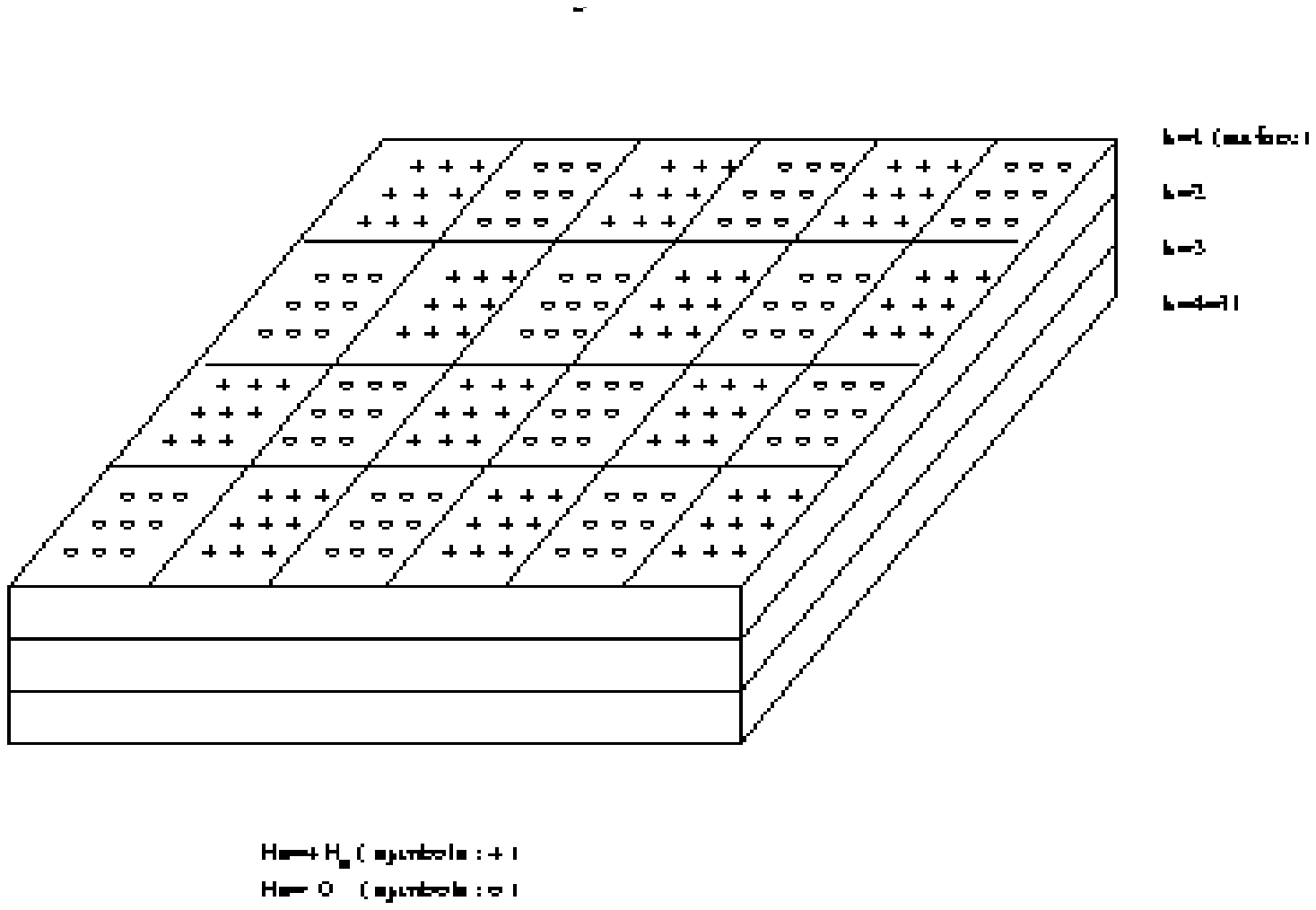}[t]
\caption{ A sketch of  the system geometry  for $N=4$ layers.
A surface magnetic field $Hs$ is acting on alternate island spins of
the surface $k=1$. $H_s$ is present on clusters  with symbols: $'+'$, and absent in clusters with symbols $'o'$. }
\label{fig18}
\end{center}
\end{figure}

The surface magnetic field $H_{s_i}$ applied on each site $i$ of the surface
$k=1$, is distributed alternatively, so that:
\begin{equation}
H_{s_i}=\left\{
	\begin{array}{ccc}
	+H_s &  \mbox{for all sites i $\epsilon$} & \mbox{clusters with
	symbols (+)} \\
	0    &  \mbox{for all sites i $\epsilon$} & \mbox{clusters with
	symbols (o)}.
	\end{array}
    \right.
\end{equation}
\subsubsection{Phase diagram and separation of holes transitions}
We mean by hole a cluster formed by positive spin surrounded by negative one. However, depending on the values of $H$, the $(T,H)$ phase diagram established in Fig. 19 shows the existence of three different phases, for a fixed surface magnetic field $H_s$. The totally wet (TW), the nonwet (NW)  and the partially wet phases.

In the  latter case, we can distinguish, in each layer, three different holes configurations namely; the (PWTD) configurations in which the layer is partially wet with a total disconnection of the holes; the (PWPD) configurations in which the layer is partially wet with a partial disconnection of the holes; the (PWTC) configurations corresponding to a partially wet layer  with a totally connected holes. Fig. 19 shows that at sufficiently low temperature, the Partial wet disappears, while
at higher temperature values, provided that $T$ is kept less than the
critical value $T_c=3.87$, the partially wet phase is found.

\begin{figure}
\begin{center}
\includegraphics{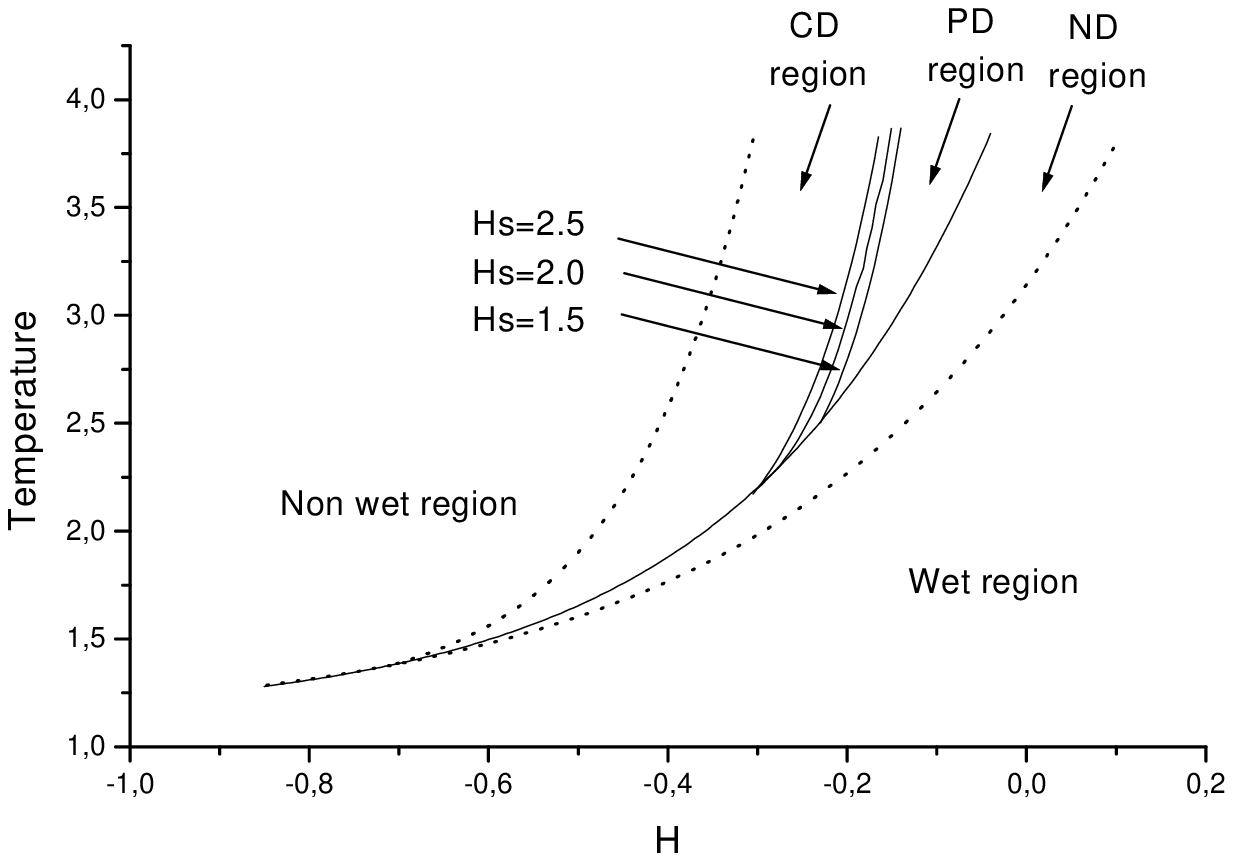}
\caption{ The $(T,H)$ phase diagram  for the layer $k=4$, for $H_s$: $1.5$;
$2.0$ and $2.5$ [252].}
\label{fig19}
\end{center}
\end{figure}

Furthermore the surface magnetic field ${H_s}$ favors the appearance of the (PWPD) phase.
In Figs. 20, we give, for the fourth layer, an example of three different partial wet phases; (PWTC), (PWPD) and (PWTD) in which the distributions of holes are different.

\begin{figure}
\begin{center}
\includegraphics[height=22cm]{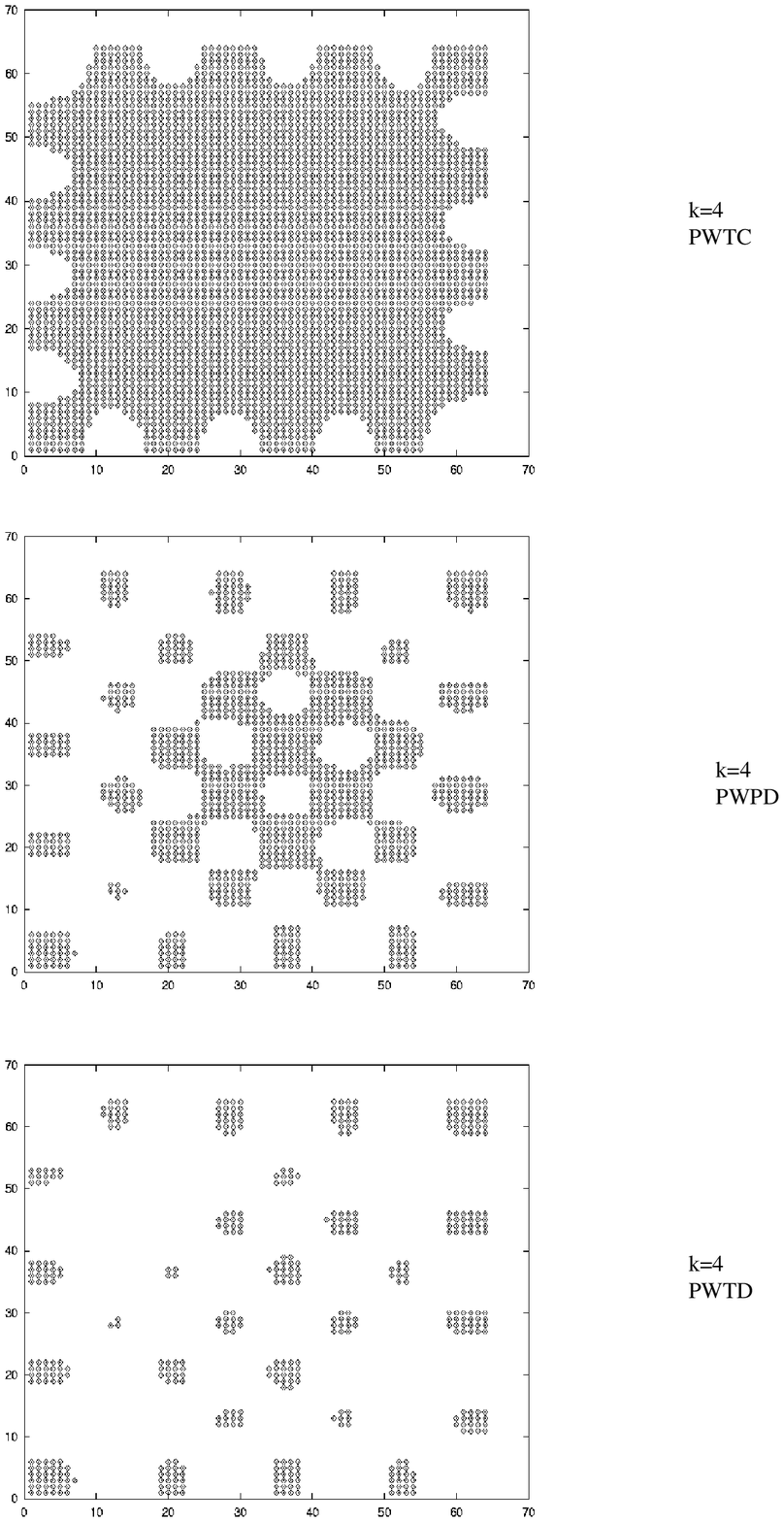}
\caption{ Island maps of positive spins for the  layer $k=4$, with  $H=-0.13$, $H=-0.15$
and $H=-0.17$. This  layer  exhibits three configurations
(PWTC), (PWTD) and (PWTD) respectively, for  $T=3.7$ and $H_s=2.0$ [252]. }
\label{fig20}
\end{center}
\end{figure}

\subsubsection{The size distribution of holes in a layer}
In the (TW), (PWTC) and (NW) phases, all spins are negative, only one big hole exist and all spins are positive respectively. However the distribution of size of holes is important in both  (PWTD) and (PWPD) phases in which there is a great number of disconnected holes with different sizes. In Fig. 21
we give the distribution of sizes of holes, in each layer ($k=1,2,3,4$) of the film, in the (PWTD) phase, for a temperature $T=3.5$, $H_s=2.0$ and $H=-0.142$. It is clear that, there exist three different peaks at three different sizes $2$, $40$ and $60$.The difference between the size frequencies decreases when increasing the value of k. Indeed, far from the surface, the effect of the surface field becomes less relevant and then the holes are equally distributed. Such result is illustrated in Fig. 21 in the $k=4$ case.

\begin{figure}
\begin{center}
\includegraphics{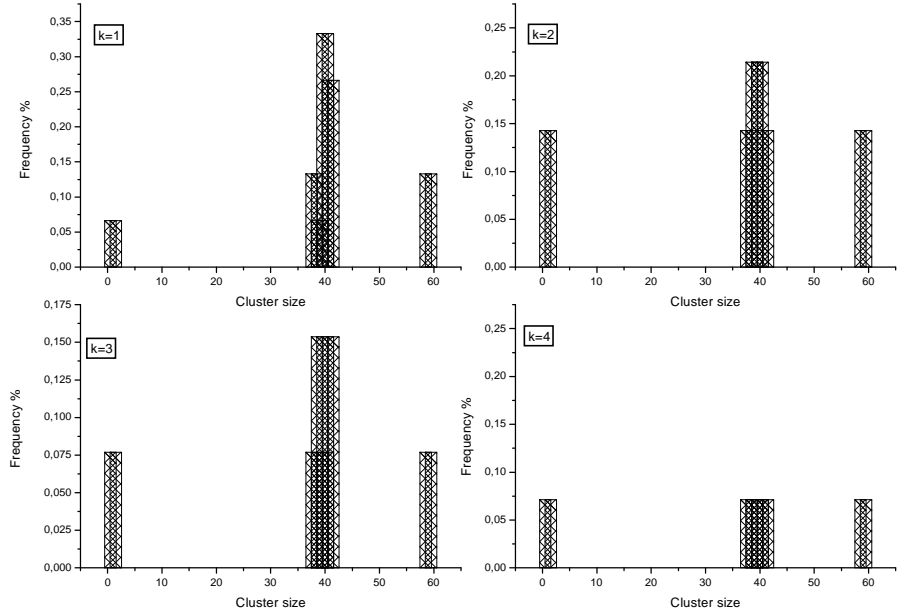}[t]
\caption{ Frequency islands size distributions for the layers $k=1$, $k=2$, $k=3$ and
$k=4$ for the configuration (PWTD) for $T=3.5$, $H_s=2.0$ and $H=-0.142$ [252]. }
\label{fig21}
\end{center}
\end{figure}

\section{Geometry effect on wetting and layering transitions}
\subsection{Corrugated surface effects}
\subsubsection{Model and method }
We consider a film with finite thickness of a three dimensional spin-$1/2$ Ising model, with M square layers, limited by two surfaces (Fig. 22). The top one is corrugated  with n steps, each step contains L spins. The top corrugated surface $St$ is under a uniform magnetic field $H_{St}$ and the bottom surface $Sb$ is a perfect plane with an applied uniform magnetic surface field $H_{Sb}$.

\begin{figure}
\begin{center}
\includegraphics{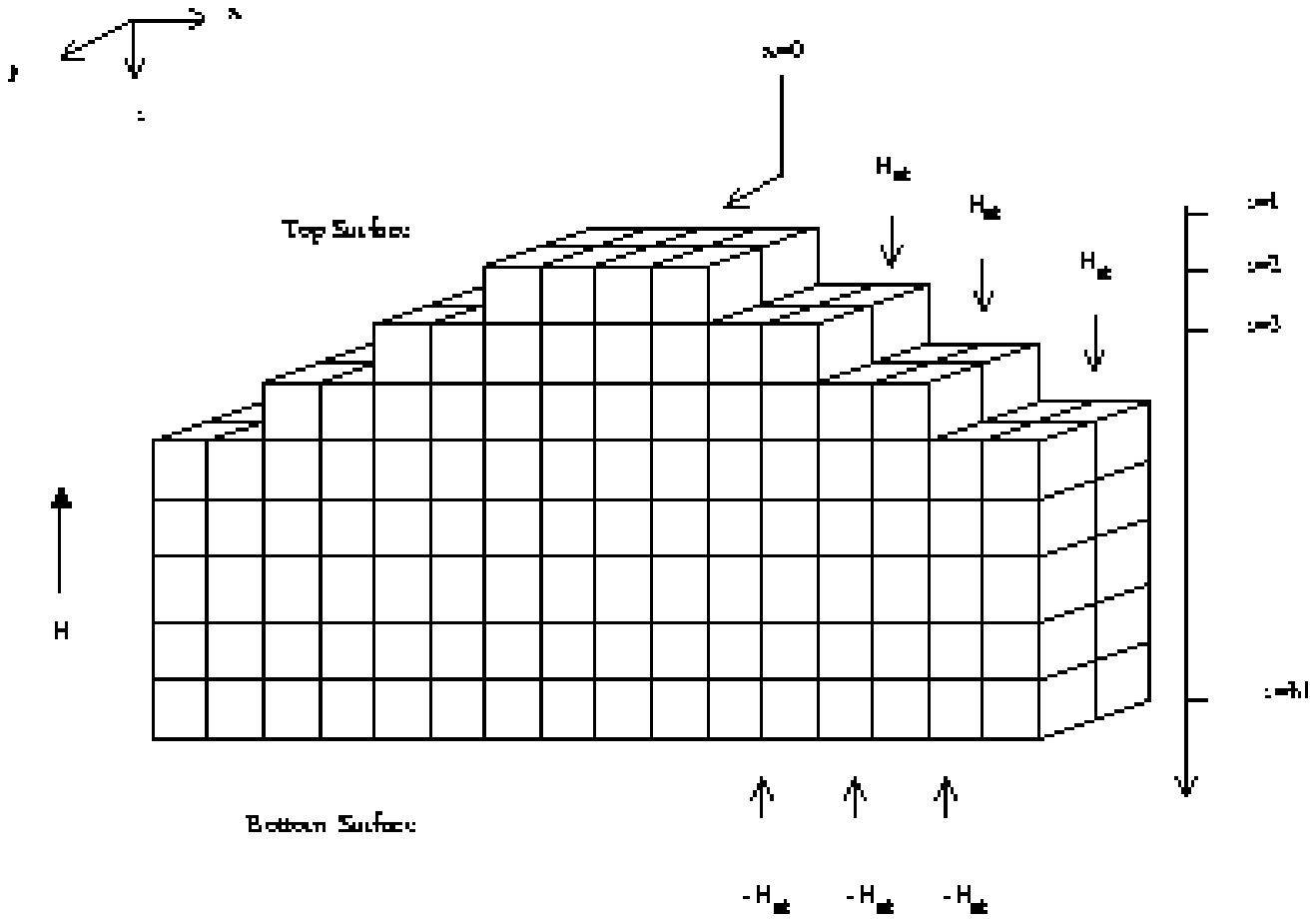}[t]
\caption{ Geometry of a corrugated surface system with $M$ layers in the $z-$direction and $n$ steps, each step contains $L$ spins in the $x-$direction. The system is infinite in the $y-$direction. A surface field $H_{St}$ is applied on the top surface $St$, whereas the bottom surface $Sb$ is under a surface field $H_{Sb}=-H_{St}$. An external field $H$ is applied to the system. }
\label{fig22}
\end{center}
\end{figure}

The Hamiltonian of the system, can be written as
\begin{equation}
{\cal H}=-\sum_{<i,j>}J_{ij}S_{i}S_{j} -\sum_{i}H_{i}S_{i}
\end{equation}
where, $S_{i}=\pm 1$ are spin$-1/2$ Ising random variables, and $J_{ij}=J$ are the exchange interactions assumed to be constant. The system is assumed to be infinite in the direction $y$, so the variable $y$ will be cancelled in all the following. $H_{i}$ is the total
longitudinal field applied on a spin located on a site 'i' of coordinates $(x,z)$.
 defined by:
\begin{equation}
H_{i}=H(x,z)=\left\{
\begin{array}{lll}
H+H_{St} & \mbox{for} & i \epsilon St \\
H & \mbox{for} &  i \epsilon Bulk  \\
H+H_{Sb} & \mbox{for} & i \epsilon Sb
\end{array}
\right.
\end{equation}
$H$ is the external magnetic field. In the following we will limit our interest to a system with $H_{Sb}=-H_{St}$. \\
The mean field equations describing the magnetization of the system is given by
\begin{equation}
m(x,z)=\tanh(\beta(2m(x,z)+m(x,z+1)+m(x,z-1)+m(x+1,z)+m(x-1,z)+H(x,z)))
\end{equation}
The total free energy of the system can be derived as :
\begin{equation}
\begin{array}{l}
F[m(x,z)]=\sum_{x,z}\{T[\frac{1-m(x,z)}{2}Log(1-m(x,z))
+\frac{1+m(x,z)}{2}Log(1+m(x,z))]   \\
-\frac{J}{2}m(x,z)[2m(x,z)+m(x,z+1)+m(x,z-1)+m(x+1,z)+m(x-1,z)] \\
-m(x,z)H(x,z)
\}
\end{array}
\end{equation}
The free boundary conditions are considered to solve the mean field equations.
\subsubsection{The corrugated surface critical behavior}
In this case the critical exponents are computed, for $H_s=0$ and $H=0$, using mean field theory.
At the vicinity of the critical temperature ${T_c}$, the spontaneous magnetization $m(z,t)$, of each layer z, vanishes with a power law
given by
\begin{equation}
m(z,t)=t^{\beta_{eff}(z,t)}
\end{equation}
Then the effective exponent $\beta_{eff}(z,t)$ can be defined as
\begin{equation}
\beta_{eff}(z,t)=dln(m(z,t))/dln(t)
\end{equation}
where $t=\vert T-T_c \vert/T_c$ is the reduced temperature and $m(z,t)$ is the average value of $m(x,z,t)$ on $x$.
For sufficiently small values of $t$, $\beta_{eff}(z,t)$ approaches the asymptotic critical exponent $\beta(z)$. The surface critical occurs when $z=1$ denoted by $\beta_1=\beta(z=1)$. The effective exponent can be approximated at discrete reduced temperatures
$t_i$ and $t_{i+1}$ as
\begin{equation}
\beta_{eff}(z,t)=ln[m(z,t_i)/m(z,t_{i+1})]/ln(t_i/t_{i+1}).
\end{equation}
However, in the vicinity of the film critical temperature $T_c$, the profile of the effective critical exponent $\beta_{eff}(z,n)$ is presented in Fig. 23.

\begin{figure}
\begin{center}
\includegraphics{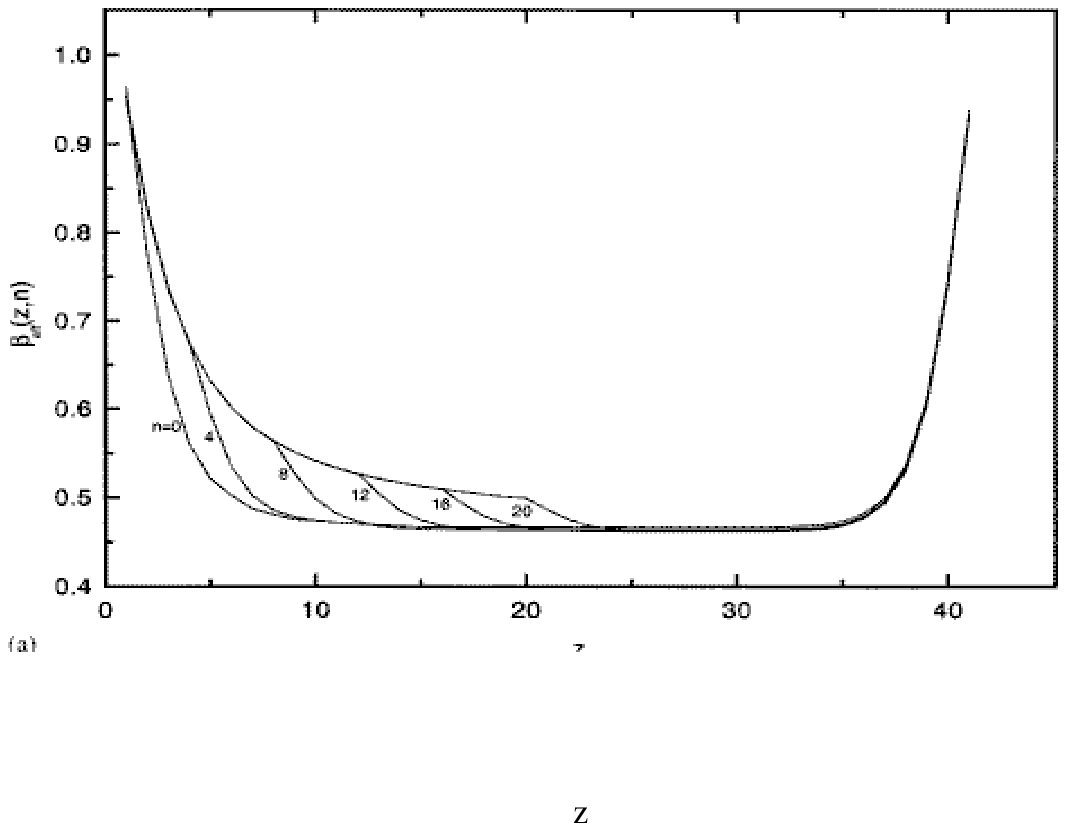}
\caption{ The dependence of  the effective exponent as a function of layer
position $z$ in the case of corrugated surface, for $L=31$, $d=60$, $n=8$,
and $J_1= J_s= J_L$. The number accompanying each curve denotes the
number of steps $n$ [198]. }
\label{fig23}
\end{center}
\end{figure}

It is clear that at the surface, the effective critical  exponent remains independent of the number of steps. While for any layer $z$ with $n_1<z<n_2$, it depends strongly on the number of steps. For a fixed number of steps, the thermal variation of the effective exponent is presented in Fig. 24.

\begin{figure}
\begin{center}
\includegraphics{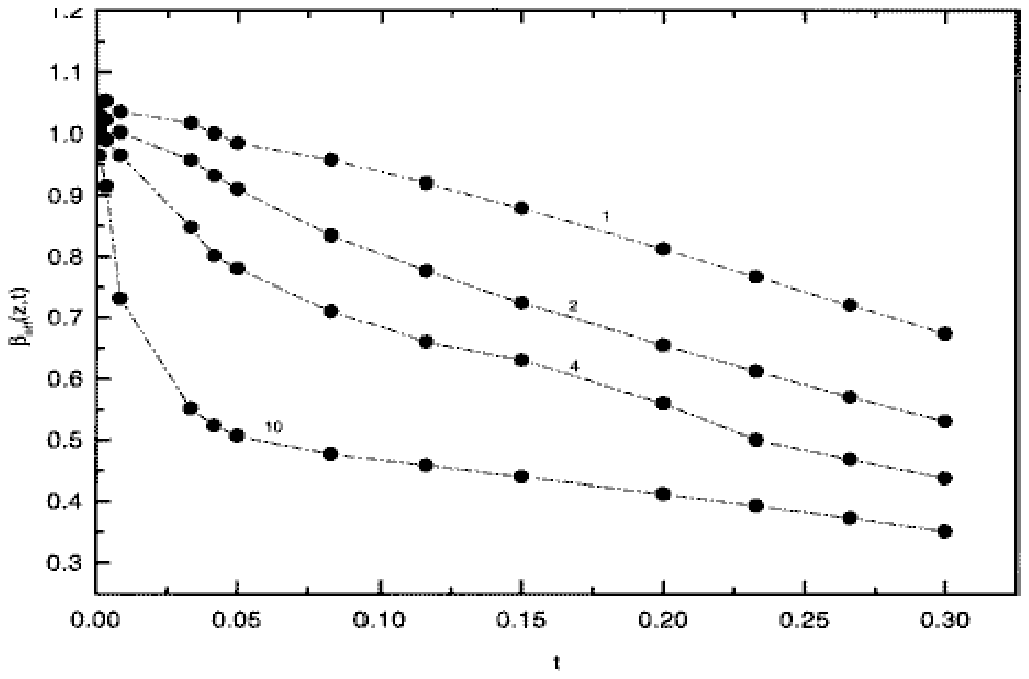}
\caption{ The dependence of  the effective exponent as a function of
reduced temperature $t$, for $L=31$, $d=60$, $n=4$, and $J_1= J_s= J_L$.
The number accompanying each curve denotes the value of the layer position
$z$ [198] }
\label{fig24}
\end{center}
\end{figure}

It is found that far from the transition temperature, $\beta_{eff}(z,n)$ decreases with increasing z, shows a crossover at $z=n$, continues to decrease until it becomes constant in the depth of the system and increases again in the vicinity of the bottom surface. If we increase the temperature to approach the critical temperature, the crossover becomes less pronounced. The dependence of the effective critical exponent as a function of temperature shows that the exponent is insensitive to the corrugation in agreement with Monte-Carlo results [175]. The effect of the number of steps on the behavior of $\beta_{eff}(z,t)$ for a given layer z, in the vicinity of the ordinary transition, is illustrated in Fig. 25.

\begin{figure}
\begin{center}
\includegraphics{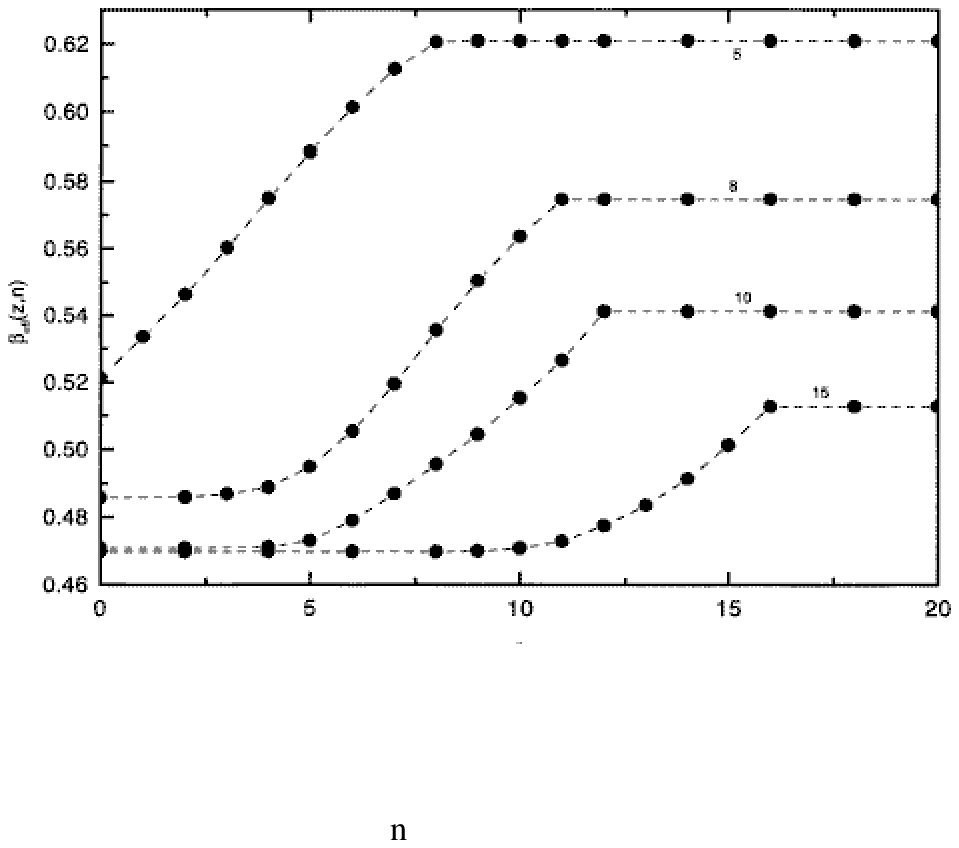}
\caption{The dependence of  the effective exponent as a function of the
number of steps $n$, for $L=31$, $d=60$, $T=0.90 \mbox{} T_c$
and $J_1= J_s= J_L$. The number accompanying each curve denotes the
 layer position $z$ [198] }
\label{fig25}
\end{center}
\end{figure}

However, when the number of steps n is sufficiently smaller than the  values of z, $\beta_{eff}(z,n)$ is insensitive of the number of steps, when n increases more, $\beta_{eff}(z,n)$ increases and once n exceeds the value of z, then $\beta_{eff}(z,n)$ becomes independent of the number of steps, i.e after a certain degree of corrugation,$\beta_{eff}(z,n)$ will be insensitive to the number of steps and then takes a saturation value, this value varies upon varying z.

\subsubsection{Wetting transition of a corrugated surface}
In the following, we denote by $(S_k, k=1,2,...)$ a configuration in which the top surface $S_t$ and the first $k-1$ layers of the bulk have positive magnetizations; while the remaining $M-k$ layers have negative ones. The ground state phase diagram is investigated exactly. In contrast to the case of perfect surface, in the case of short range substrate potential, the prewetting and layering transitions occurs at $T=0 \mbox{} K$ (Fig. 26).

\begin{figure}
\begin{center}
\includegraphics{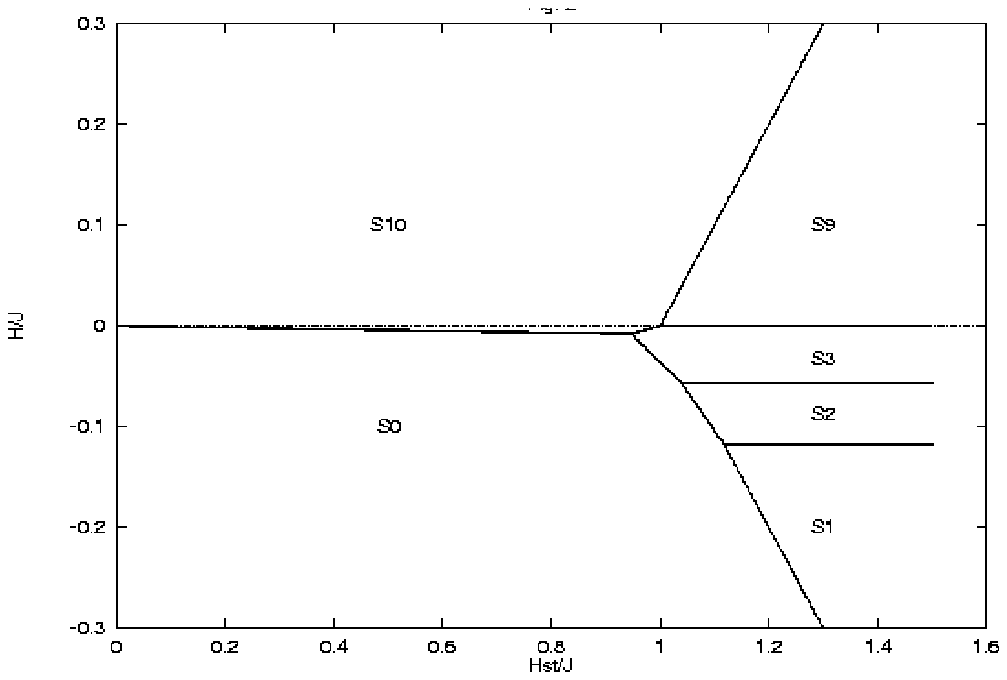}
\caption{ Phase diagram of the ground state, in the $(H/J,H_{St}/J)$ plane,
for a system with $M=10$ layers, $n=2$ steps and $L=10$ spins per step,
with the condition $H_{Sb}=-H_{St}$ [200]. }
\label{fig26}
\end{center}
\end{figure}

At finite temperature, phase diagrams obtained using Monte-Carlo simulations (Fig. 27) shows that the wetting temperature depends weakly on the surface corrugation degree but it depends strongly on lateral size of the system. Furthermore the behavior of the magnetization as a function of the external magnetic field is presented in Fig. 28a , for a fixed layer ($z=3$) and several
values of the position $x$ in the x-direction.

\begin{figure}
\begin{center}
\includegraphics{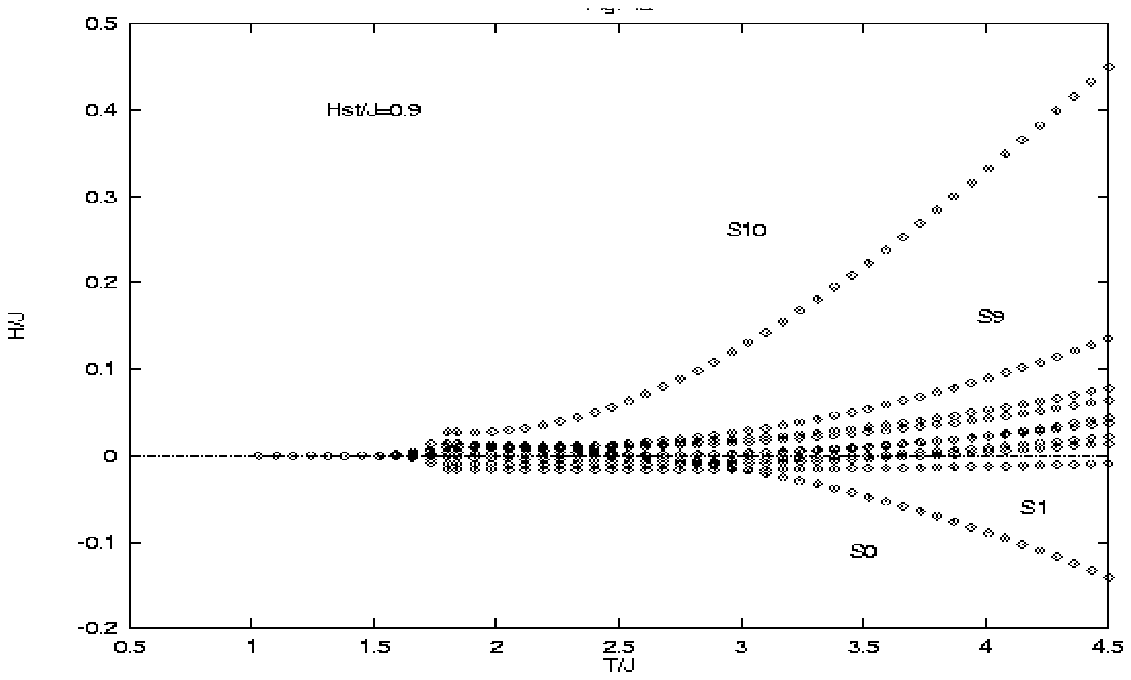}
\caption{ Layer transitions of the system, in the $(H/J,T/J)$ plane, using
Monte Carlo simulations for $H_{St}/J=0.9$ [200]. }
\label{fig27}
\end{center}
\end{figure}

It is clear that, the increasing of the external bulk field, the top surface (x=18) exhibits a first order transition from negative to positive magnetization, accompanied by the transition of the bulk site located near the surface at z=3. This is due to both thermal fluctuations and the surface effect in the x-direction. The other bulk sites keep their negative magnetizations. This is called "prelayer transition".
Such behavior was obtained neither in the perfect surface nor at $T=0K$ cases. However, by increasing $H$ the so-called layer transition occurs when the magnetizations of the remaining bulk sites of the layer $z$,($z=3$) jump from negative values to positive ones. The effect of surfaces in the z-direction, is presented in Fig. 28b in which  the local magnetization $m(x,z)$ is presented as a function of $H$ for a fixed  value of $x (x=0)$ and several layers $z$. It is clear that the bulk field, at which the transition $S_k \leftrightarrow S_{k+1}$, $(k=0,1,...,M-1)$ occurs, increases with increasing k.

\begin{figure}
\begin{center}
\includegraphics{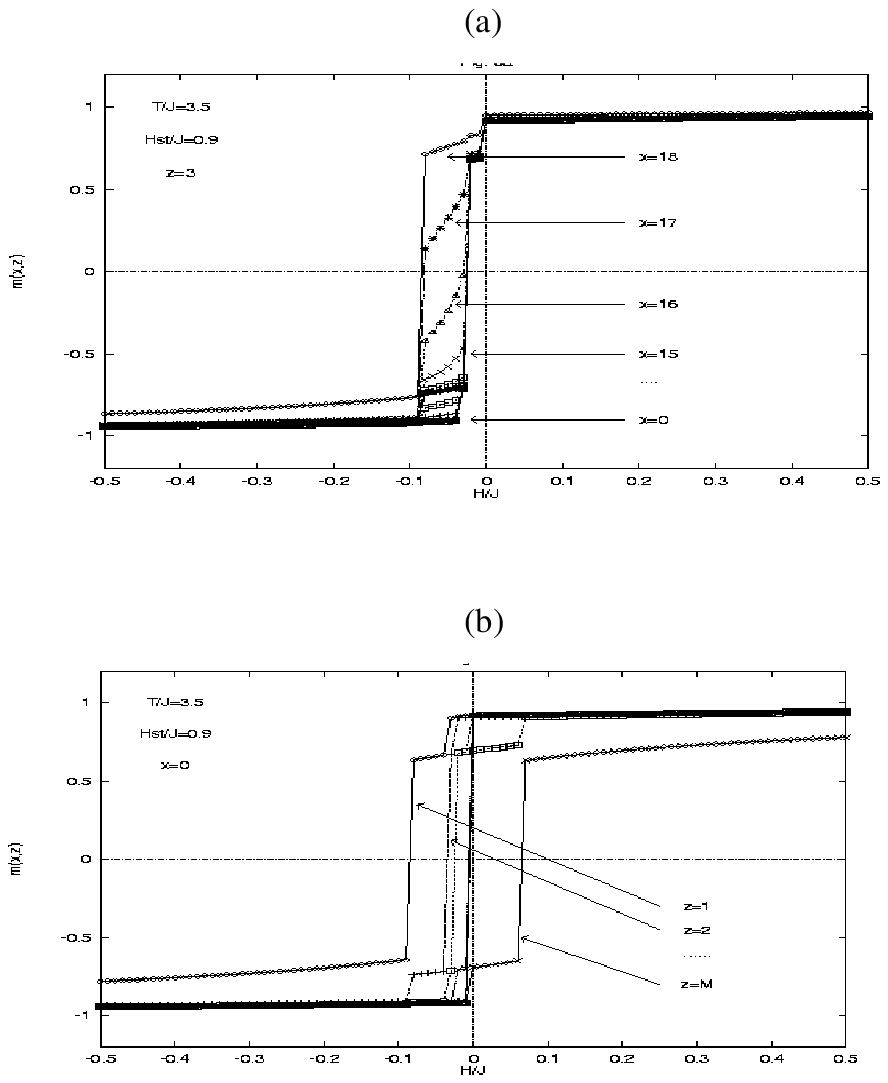}
\caption{ Magnetisation profiles as a function of the reduced bulk field
$H/J$ for $H_{St}/J=0.9$ and $T/J=3.5$. The magnetizations $m(x,z)$ are
plotted for a fixed layer $z=3$ and different positions $x$ (a), and for a
fixed position $x=0$ and different layers $z$ (b) [200].}
\label{fig28}
\end{center}
\end{figure}

Moreover, using mean field theory, it is shown that the wetting temperature depends weakly on the corrugation degree $n$. Indeed, $T_w=2.55$ for $n=2$, $T_w=2.59$ for $n=4$ for a film with thickness $N=10$. We note that the corrugate surface leads to the appearance of wetting in both longitudinal (z) and  transverse (x) directions.
\subsection{Edge wetting in a three-dimensional system}
We consider a spin-1/2 Ising model in a lattice formed with $N$ layers, each one of them is constituted by two perpendicular perfect planes and
contains $2(N-k)+1$ spin chains which are infinite in the y-direction (Fig. 29). The system is considered under the effects of an uniform surface magnetic field $H_{s}$  applied on the planes $(x,y,z=1)$ ,$(x=1,y,z)$ of the layer $k=1$ and an external magnetic field $H$ applied on all spins of the system.

\begin{figure}
\begin{center}
\includegraphics{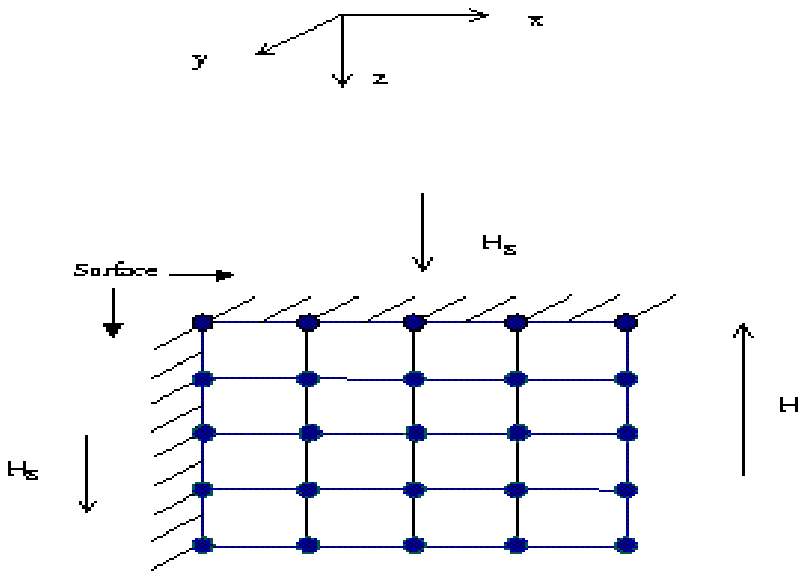}
\caption{ Geometry of the system formed with two surfaces $(x,y,z=1)$ and $(x=1,y,z)$ with $N$ spins in both the $x$ and $z-$directions. The system is infinite in the $y-$direction. A uniform surface magnetic field $H_{s}$ is applied on the planes $(z=1,x,y)$ and $(x=1,y,z)$. An external magnetic field $H$ is applied to the global system.}
\label{fig29}
\end{center}
\end{figure}

However the Hamiltonian of the system is given by
\begin{equation}
{\cal H}=-\sum_{<i,j>}J_{ij}S_{i}S_{j} -\sum_{i}H_{i}S_{i}
\end{equation}
where, $S_{i}=\pm 1$ are spin-$1/2$ Ising random variables, and $J_{ij}=J$ are the exchange interactions assumed to be constant. Since the system is invariant by translation, in the following, the coordinate y will be cancelled. The total magnetic field $H_{i}$ applied on a site 'i' of coordinates $(x,z)$ is given by:
\begin{equation}
H_{i}=H(x,z)=\left\{
\begin{array}{lll}
H+H_{s} & \mbox{for} & x=1, z=1,...,N \\
H+H_{s} & \mbox{for} & z=1, x=1,...,N \\
H & \mbox{~~~} &  elsewhere  \\
\end{array}
\right.
\end{equation}
$H$ and $H_{s}$ are, respectively, the external and surface magnetic fields
both applied in the $z$-direction. \\
The notation ($1^{p}1_{q}$) with $p=0,1,2,...,N$ and $q=0,1,3,5,...,2(N-p)-1$, will be used to denote that the first $p$ layers and the $q$ first spin chain of the layer $p+1$ are in a magnetic state "up"; while the remaining $N-(p+1)$ layers are in the state "down". In particular, the notation $1^{N}$ (resp. $O^{N}$) will be used to denote a configuration with positive magnetization for all layer spins (resp. negative magnetization for all layer spins) of the system. Fig. 30 give a ketch of different possible configurations for a system of thickness $N=4$.\\

\begin{figure}
\begin{center}
\includegraphics{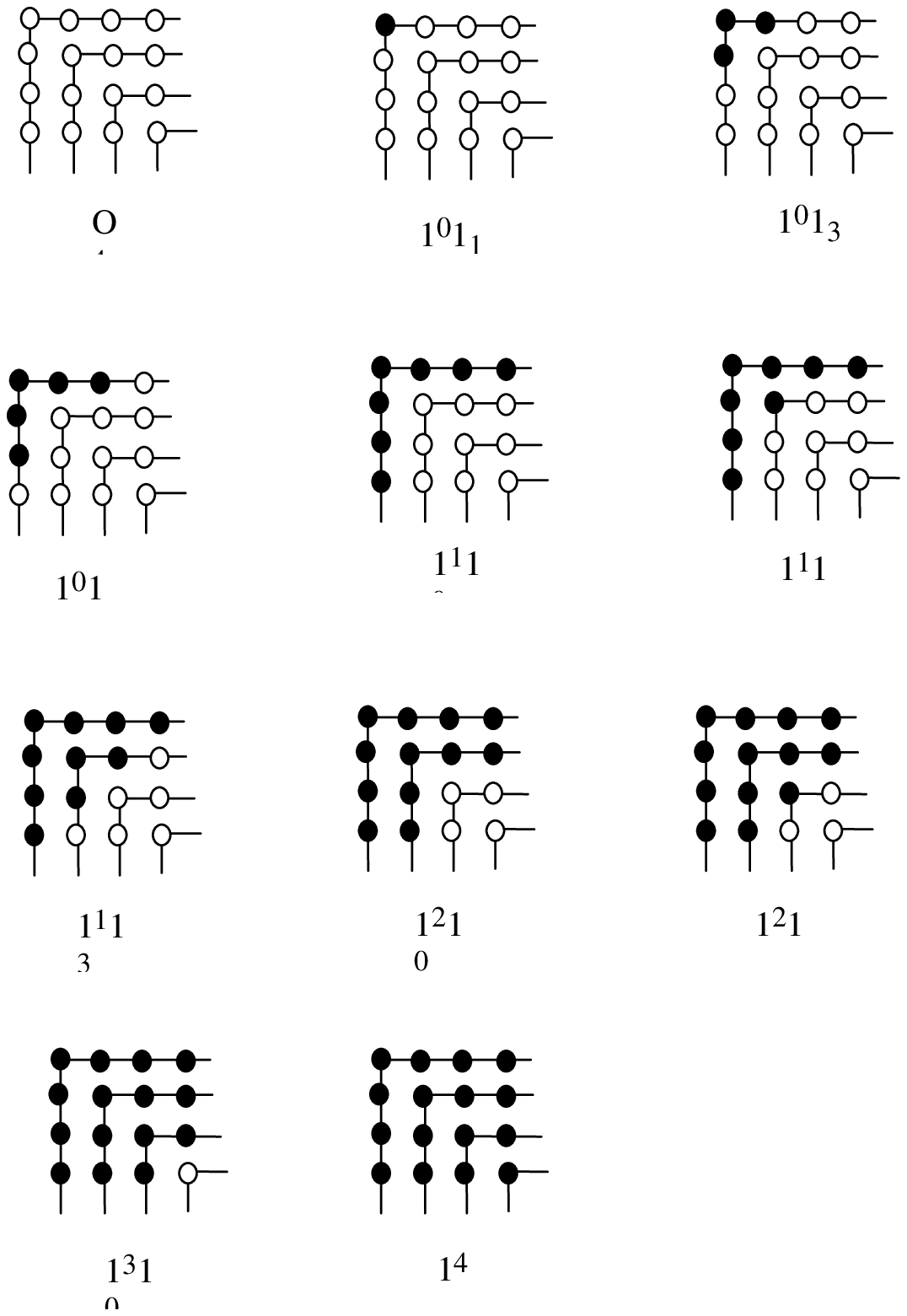}[t]
\caption{ Sketch of different possible configurations for a system with
$N=4$. Symbols ($\circ$) and ($\bullet$)  correspond to spin "down" and
spin "up" respectively. }
\label{fig30}
\end{center}
\end{figure}

Numerical results are given for two system sizes $N=4$, using Monte Carlo simulations and $N=20$ using mean field theory. However, the phase diagram obtained using the Monte Carlo (MC)simulations is represented in Fig. 31 for $H_{s}/J=1.0$.

\begin{figure}
\begin{center}
\includegraphics{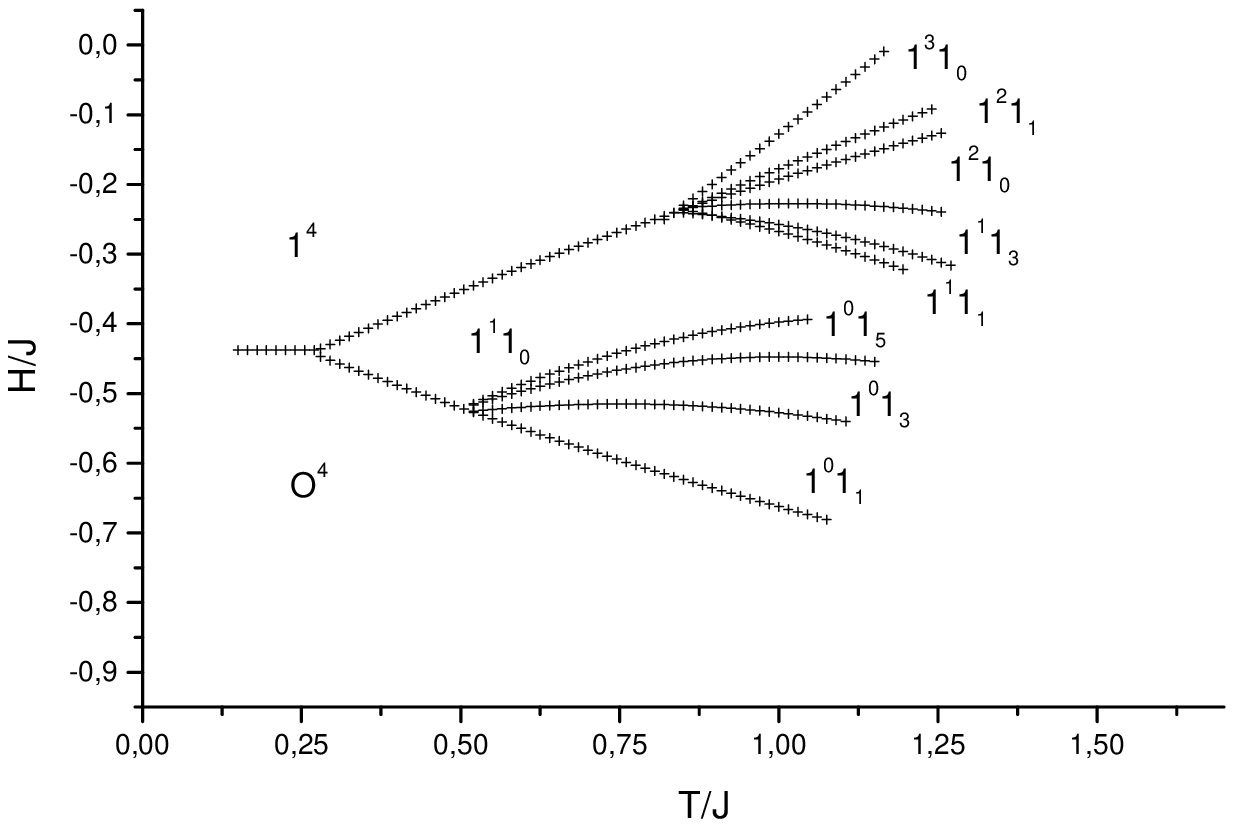}
\caption{ Phase diagram of the intra-layering and layer-by-layer
transitions, in the plane $(H/J,T/J)$, using the Monte Carlo simulations
for a system with $N=4,n_y=100$ and $H_{s}/J=1$ [40]. }
\label{fig31}
\end{center}
\end{figure}

The behavior of the local magnetizations $m(1,1)$, $m(1,2)$, $m(2,2)$ and $m(2,3)$ are given in Figs. 32a and 32b, as a function of the external magnetic field $H$. It is found that, when increasing the external field, the corners transits  before the other intra-layering transitions. Such behavior was not obtained in the perfect surface case[1-12,69,70]. However, the other layer-by-layer transitions occur when increasing the external field at specific temperature values.

\begin{figure}
\begin{center}
\includegraphics{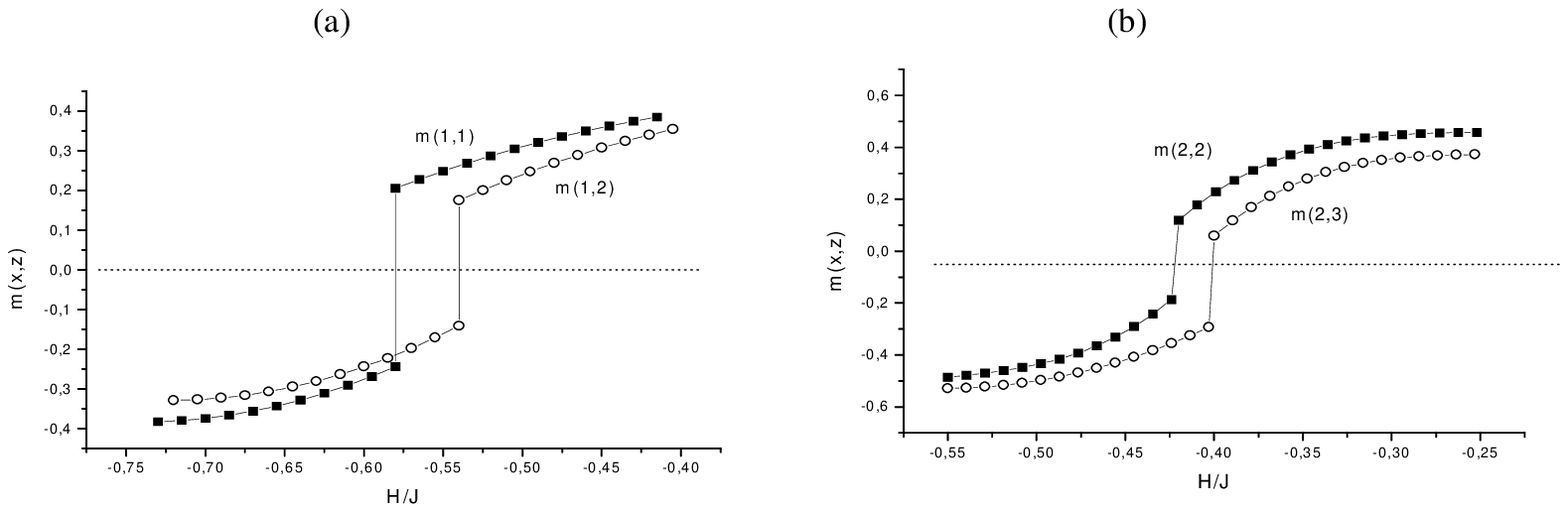}
\caption{ Magnetization profiles as a function of the reduced bulk magnetic
field $H/J$ for $N=4$ and $H_{S}/J=1.0$, of $m(1,1)$ and $m(1,2)$ at
$T/J=0.75$ (a); $m(2,2)$ and $m(2,3)$ at $T/J=1.0$ (b), using the Monte
Carlo method [40]. }
\label{fig32}
\end{center}
\end{figure}

As one can expects, the external field values needed to make arising the intra-layering and inter-layering transitions[40]  increase with increasing the order of the layer counted  from the surface $k=1$. This is qualitatively in good agreement with previous works [1-12,69,70].  it is seen that the increasing system size effect is to decrease the intra and inter layering temperature values. The bulk layering transitions are shifted to higher external magnetic field values, for a fixed surface magnetic field value. \\
The behavior of the surface, $T_{L}^{s}$, and the bulk $T_{L}^{b}$ intra-layering temperature profiles as a function of the thickness $N$, for two values of $H_{S}/J=1.0$ and $H_{S}/J=0.9$ is presented in Figs.33.

\begin{figure}
\begin{center}
\includegraphics{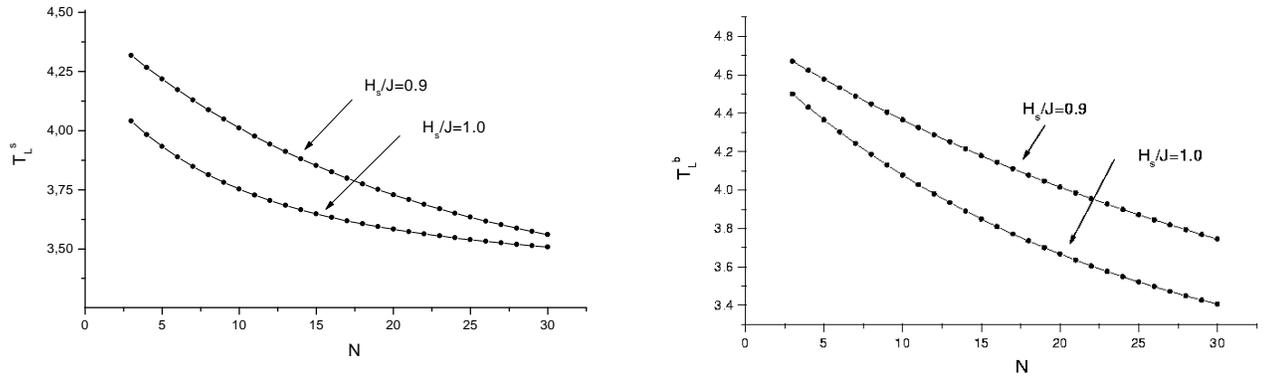}[t]
\caption{ Surface layering temperature profiles as a function of the system
size $N$ for two surface magnetic field values $H_{S}/J=1.0$ and
$H_{S}/J=0.9$, by using the mean field method [40]. }
\label{fig33}
\end{center}
\end{figure}

It is clear that $T_{L}^{s}$ as well as $T_{L}^{b}$ decreases when increasing the system size. For a fixed system size $N$, these intra-layering temperatures decrease when increasing the surface magnetic field values. Moreover, $T_{L}^{b}$ reaches the wetting temperature for sufficiently large values of the system size. These results means that the edge leads to the appearance of the intralayering transitions which are absent in the case of perfect surfaces [1-12,45,69,70] and the continuous model (Milchev {\it et al.} [40]),[47-51]; On the other hand it leads to a decay of the wetting temperature compared with the perfect surface case [1-12,45,69,70]. This is in good agreement with the results obtained in Ref. [47-51].
\section{Quantum effects on wetting and layering transitions}
\subsection{Uniform surface transverse field}
We consider N coupled ferromagnetic square layers in a transverse and longitudinal magnetic fields. The Hamiltonian of the system is given by
\begin{equation}
H=-\sum_{<i,j>}J_{ij}S_{i}^{z}S_{j}^{z}-\sum_{i}(\Omega S_{i}^{x}+H_{i}S_{i}^{z})
\end{equation}
where, \( S_{i}^{\alpha} ,(\alpha=x,z) \) are the spin-1/2 Pauli matrices, $\Omega$ is the transverse magnetic field in the x direction and \(H_{i}\) is the magnetic field in the z direction, defined by:
\begin{equation}
H_{i}=\left\{ \begin{array}{lll}
	      H+H_{s1} & \mbox{for}&  i=1  \\
	      H        & \mbox{for}&  1 < i < N \\
	      H+H_{s2} & \mbox{for}&  i=N
	      \end{array}
	\right.
\end{equation}
The surface magnetic field $H_{s1}$ is applied on the first layer $(i=1)$, the last layer $(i=N)$ is under a surface magnetic field $H_{s2}$ and H is the bulk magnetic field.\\
For a simple cubic lattice, the magnetization and the free energy for each plane are given, respectively, by the following expressions:
\begin{equation}
m_{p}=\frac{1}{\lambda_{p}}[4m_{p}+r(m_{p+1}+m_{p-1}+H)]\tanh(\beta\lambda_{p})
\end{equation}
and
\begin{equation}
 F_{p}=-\frac{1}{\beta}\log(2\cosh(\beta \lambda_{p}))+\frac{1}{2}m_{p}(4m_{p}+r(m_{p+1}+m_{p-1})
\end{equation}
with
\begin{equation}
\lambda_{p}=\sqrt{ (4m_{p}+r(m_{p+1}+m_{p-1})+H)^2+\Omega^{2}},
\end{equation}
The state equations are solved numerically, it is found that there exist  a wetting transverse field $\Omega _{w} /J$, above which a sequence of layering transitions occurs, which depends on the value of the temperature and the surface magnetic field, $h_{s_1}$. On the other hand, when $T > T_{w}$, the layering transitions exist in the absence of the transverse magnetic field, this means that the temperature effect is sufficient to produce layering transitions. $(T,\Omega)$ Phase diagram is presented in Fig. 34.

\begin{figure}
\begin{center}
\includegraphics{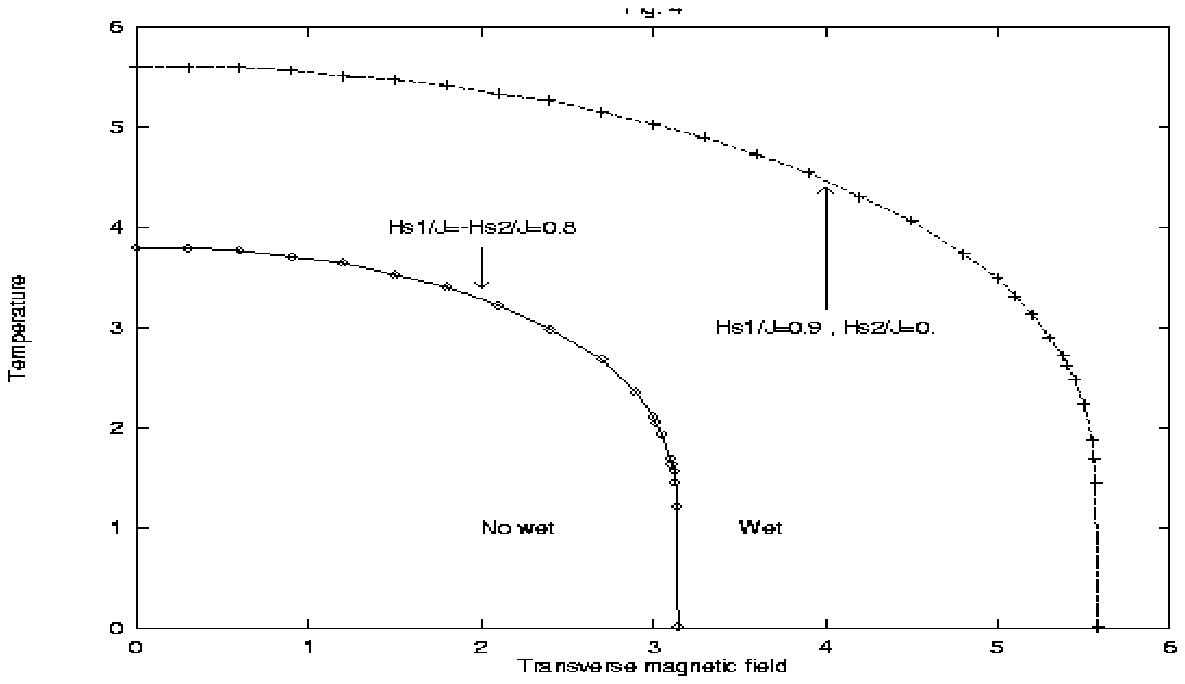}
\caption{ Phase diagram in $(T/J,\Omega/J)$ plane,
for both cases: $(H_{s1}/J=0.8,H_{s2}/J=-H_{s1}/J)$ and
$(H_{s1}/J=0.9,H_{s2}/J=0.)$, and $N=20$ [79]. }
\label{fig34}
\end{center}
\end{figure}

However, the wetting transverse magnetic field, $\Omega_{w}$, decreases with increasing temperature and/or decreasing surface magnetic field.
\subsection{Variable surface transverse field}
In this case we consider a surface variable transverse field, the transverse field of a layer z (distance from the substrate) is given by:
\begin{equation}
\Omega_{z}=\Omega_s/z^{\alpha}
\end{equation}
with $\Omega_s$ corresponding to the transverse magnetic field applied on the first layer $z=1$. Values of the parameter $\alpha$ will be discussed in the following, in particular,
$\alpha=0$ corresponds to a uniform transverse field applied over all the layers.
The notation $O^pD^{N-p}$ means that the top $p$ layers are ordered while the remaining ${N-p}$ are disordered. Phase diagrams for  $\alpha=1.0$, is presented in Figs.35. It is found that the layers with higher value of $z$ need greater magnetic transverse field values to become disordered.

\begin{figure}
\begin{center}
\includegraphics{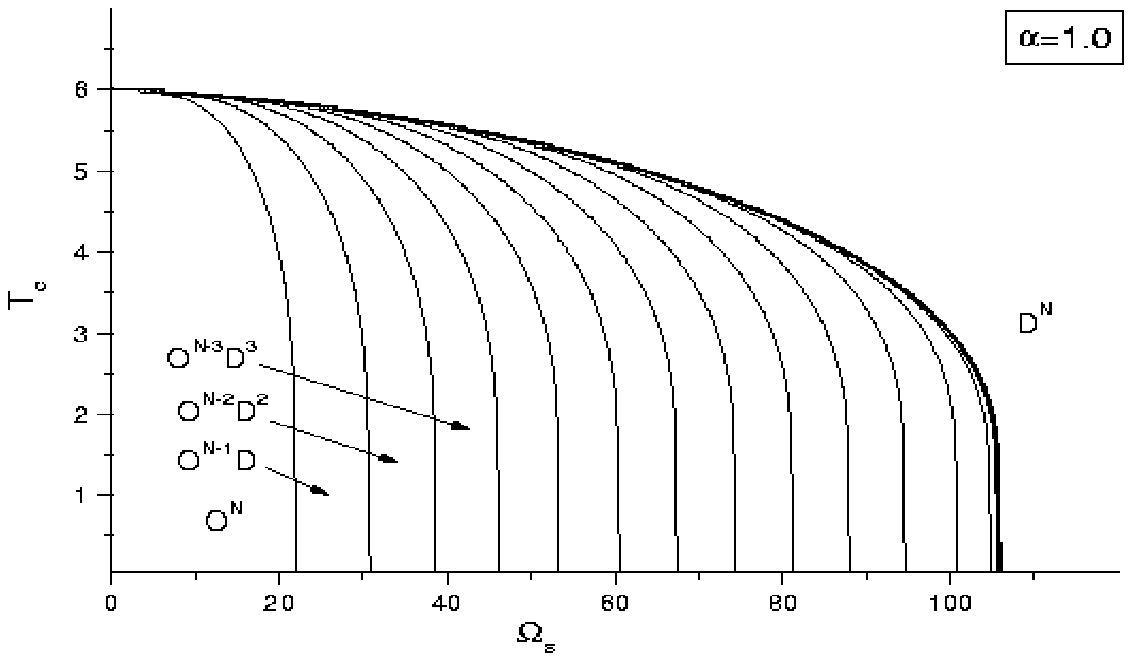}
\caption{ The dependence of the critical temperature $T_c$ on the surface
transverse magnetic field, for $\alpha=1.0$ (N=20 layers), using the mean
field theory [253]. }
\label{fig35}
\end{center}
\end{figure}

However, for a fixed temperature, the disorder of each layer occurs when increasing the transverse magnetic field so that $\Omega_{c}(z=1)<\Omega_{c}(z=2)< ... < \Omega_{c}(z=N)$.
It is found that in absence of the transverse magnetic field, $\Omega_s=0$, the critical temperature for all the layers, $1\le z \le N$, is close to:
$T_c^{MF}( {\rm bulk})=6.0$, for the mean field theory case, and
$T_c^{FC}( {\rm bulk})=5.073$ for the finite cluster method.
The corresponding thermal variation of the longitudinal and transverse magnetizations are computed using mean field theory and presented respectively in Fig. 36a and 36b.

\begin{figure}
\begin{center}
\includegraphics{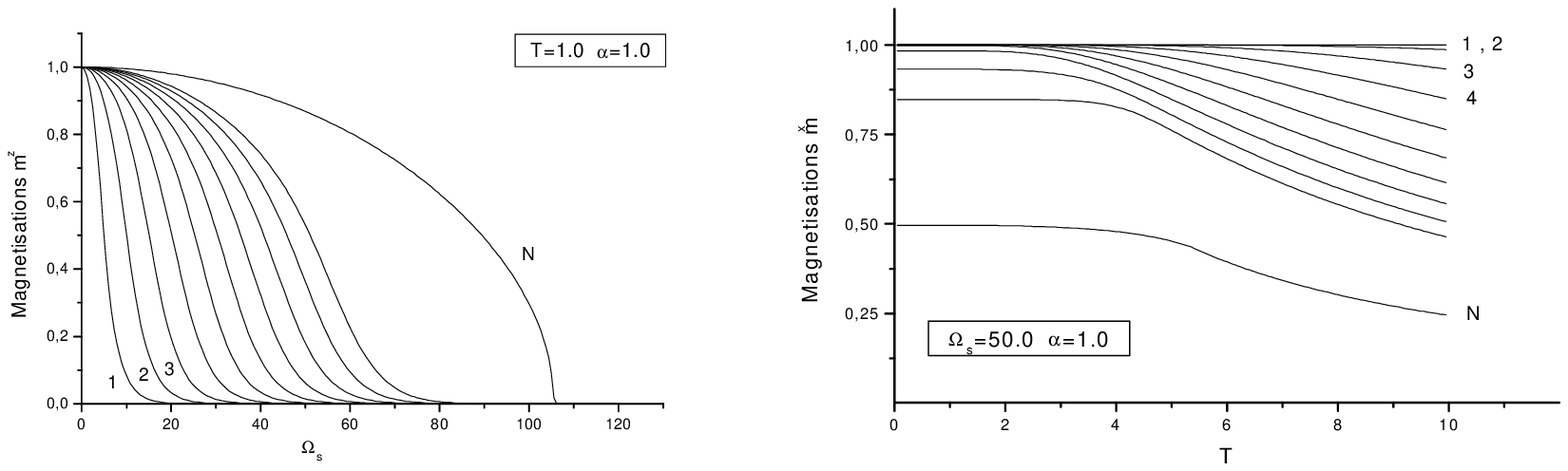}
\caption{ The dependence of the longitudinal layer magnetizations $(a)$ as a function the transverse magnetic
field for $\alpha=1.0$ and a fixed temperature: $T=1.0$ ; and the thermal
behavior of the transverse layer magnetizations $(b)$, for $\alpha=1.0$
and $\Omega_{s}=50.0$. The number accompanying
each curve denotes the value of the layer $z$ [253]. }
\label{fig36}
\end{center}
\end{figure}

The behavior of the longitudinal magnetization critical exponent  $\gamma $ at a fixed temperature $T=1.0$ is also calculated as a function of the exponent $\alpha$ . However, $\gamma$ depends on the  layer $z$, and decreases from $1$, for small values of $\alpha$, to $0.5$ for higher values of $\alpha$ (Fig. 37)

\begin{figure}
\begin{center}
\includegraphics{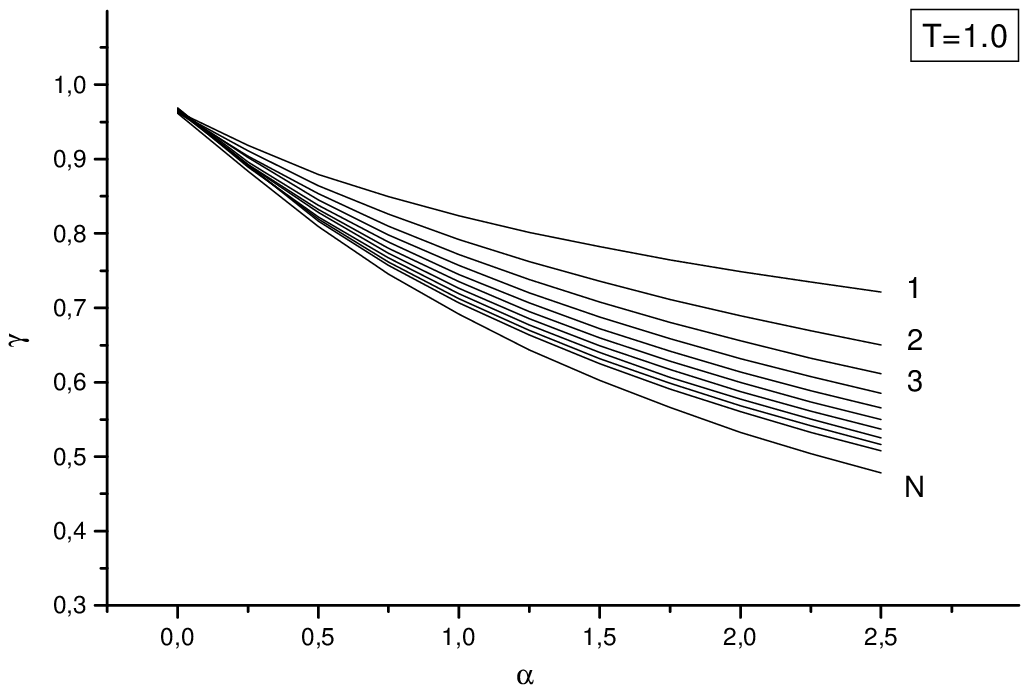}
\caption{ The dependence of the critical exponent $\gamma$ on the parameter
$\alpha$ for $T=1.0$. The number accompanying each curve denotes the value
of the layer $z$, using the mean field theory [253].}
\label{fig37}
\end{center}
\end{figure}

\subsection{Random transverse field effect on order-disorder layering transitions}
Now we consider the case of a surface random transverse field governing by a two peaks distribution law namely:
\begin{equation}
{\cal P}(\Omega_i)=p\delta(\Omega_i-\Omega_s)+(1-p)\delta(\Omega_i)
\end{equation}
where $p$ is the probability that  $\Omega_i$ acts on a site i of the surface with the value $\Omega_s$.
using finite cluster approximation within an expansion technique for spin-$1/2$ cluster identities, the longitudinal magnetizations $m_{k}^{z}$, for a plane '$k$' $(k=1,...,N)$ are well calculated. However,
On the other hand, it is shown that for a fixed $\alpha$ and very low temperatures, each layer $k$ disorders at a critical probability $p_c(k)$ corresponding to a critical surface transverse magnetic field $\Omega_{s}^{c}(k)$. Indeed, as it is shown in Fig. 38a, for $\alpha=1.0$, the film is ordered at $p \le p_c=0.9949$ even for very large values of the surface transverse magnetic field. While for $p_c < p \le p_c(k=1)=0.9950$ (Fig. 38b), the first surface becomes disordered. increasing $p$, the second layer disorders at $p_c(k=1)<p \le p_c(k=2)=0.9955$ (Fig. 38c), and so on. Finally the film is totally disordered when the applied surface transverse magnetic field is acting on all the surface sites (i.e $p=1$) (Fig. 38d).

\begin{figure}
\begin{center}
\includegraphics{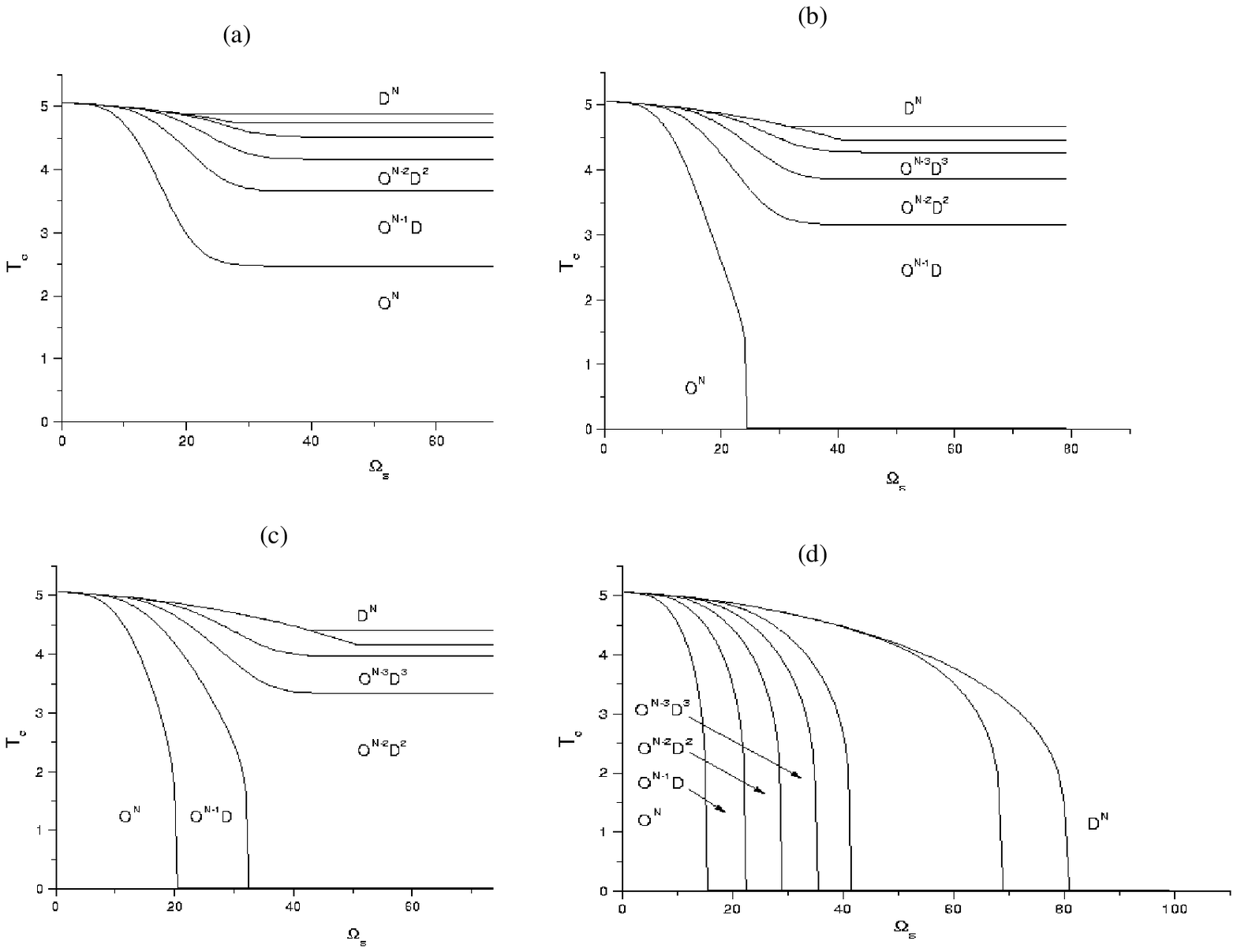}[t]
\caption{ The dependence of the critical temperature $T_c$ on the surface
transverse magnetic field, for $\alpha=1.0$ and different values of the
probability: $p(=0.9945) < p_c(=0.9949)$ (a), $p_c< p \le p_c(k=1)=0.9950$
(b), $p_c(k=1)<p \le p_c(k=2)=0.9955$ (c) and  $p=1.0$ (d) [254]. }
\label{fig38}
\end{center}
\end{figure}

It arises that a sequence of order-disorder layering transitions occurs when decreasing probability from the value $p=1.0$, corresponding to a totally ordered film at very low temperatures. We note that the critical surface transverse field $\Omega_{s}^{c}$ depends on the value of $\alpha$ and the probability $p$. Indeed it is clear from Fig. 39 that $\Omega_{s}^{c}$  increases when increasing $\alpha$ and/or decreasing $p$.
\begin{figure}
\begin{center}
\includegraphics{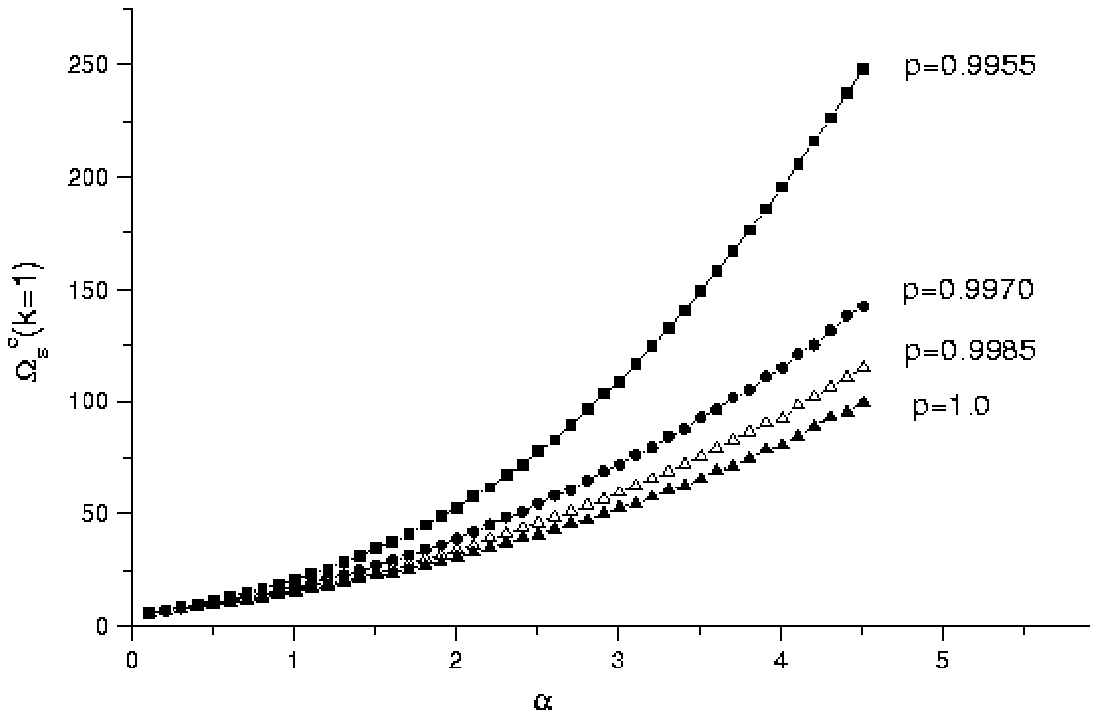}
\caption{ The dependence of the critical surface transverse magnetic field
$\Omega_{s}^{c}(k=1)$ as a function of the parameter $\alpha$ for a fixed
temperature $T=0.05$ and different probability values:
$p=0.9955 < p_c=0.9949$, $p=0.9970$, $p=0.9985$ and $p=1.0$ [247]. }
\label{fig39}
\end{center}
\end{figure}

On the other hand the profile of the surface critical transverse field (Fig. 40) and the surface critical probability (Fig. 41) shows that $\Omega_{s}^{c}$ and $p_{s}^{c}$ increases with the depth of the film.
\begin{figure}
\begin{center}
\includegraphics{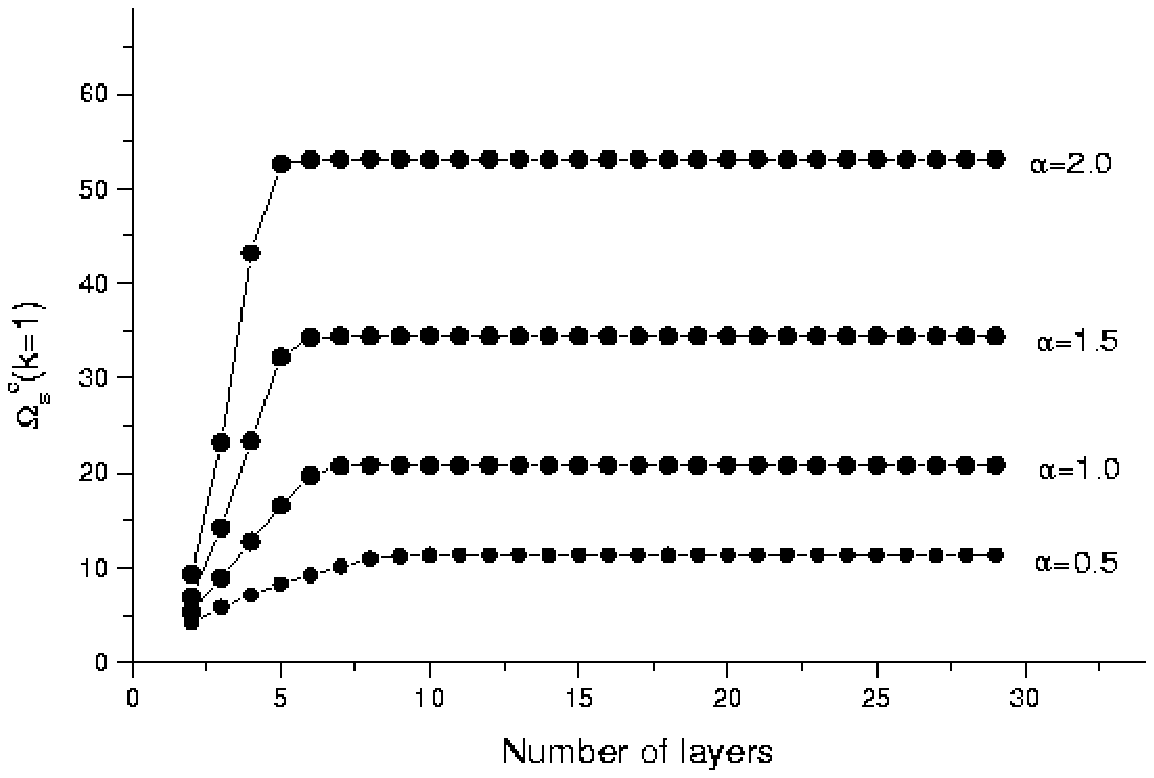}
\caption{ The dependence of the critical surface transverse magnetic
field $\Omega_{s}^{c}(k=1)$ as a function of the number of layers for
$T=0.05$ and different values of the exponent $\alpha=0.5$, $\alpha=1.0$,
$\alpha=1.5$ and $\alpha=2.0$ [247]. }
\label{fig40}
\end{center}
\end{figure}

\begin{figure}
\begin{center}
\includegraphics{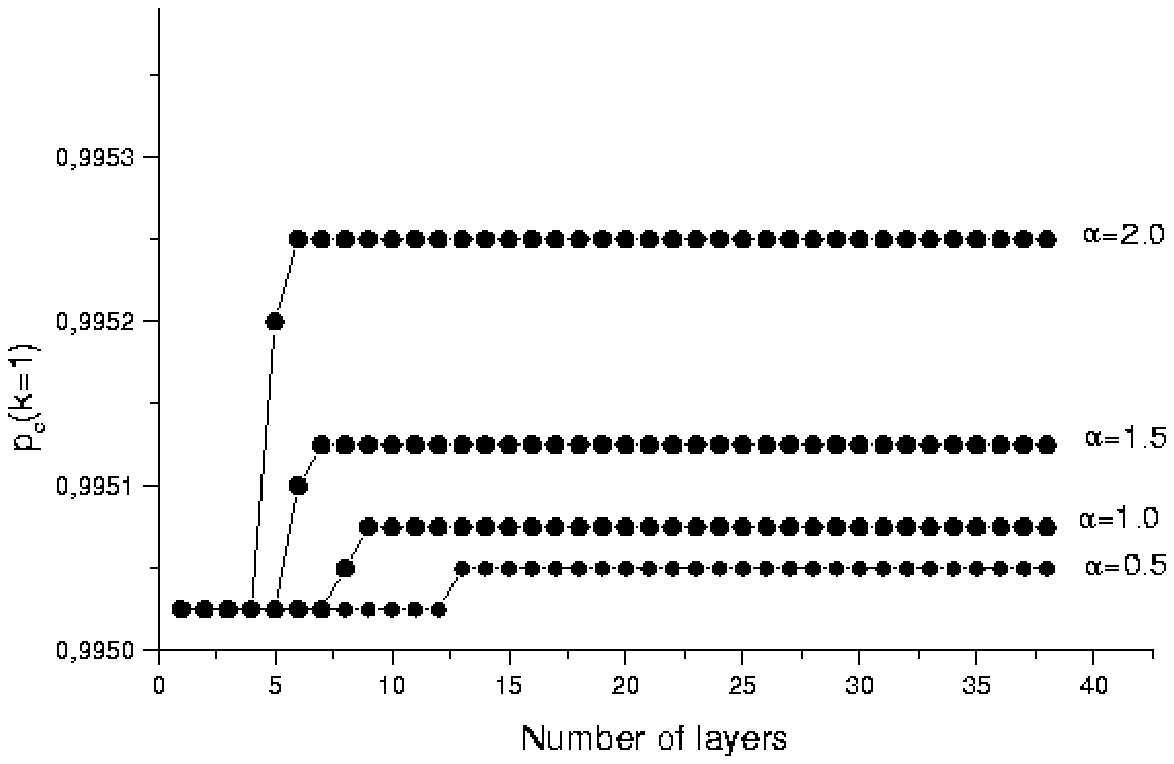}
\caption{ The dependence of the critical probability $p_{c}$ as a function
of the number of layers for $T=0.05$ and different values of the exponent
$\alpha=0.5$, $\alpha=1.0$, $\alpha=1.5$ and $\alpha=2.0$ [247]. }
\label{fig41}
\end{center}
\end{figure}

The surface critical exponent of the longitudinal magnetization is also computed (Fig. 42) as a function of $\alpha$ for several values of $p$.
It is found that
$\gamma(k)$ decreases when increasing $\alpha$ and/or decreasing $p$.
\begin{figure}
\begin{center}
\includegraphics{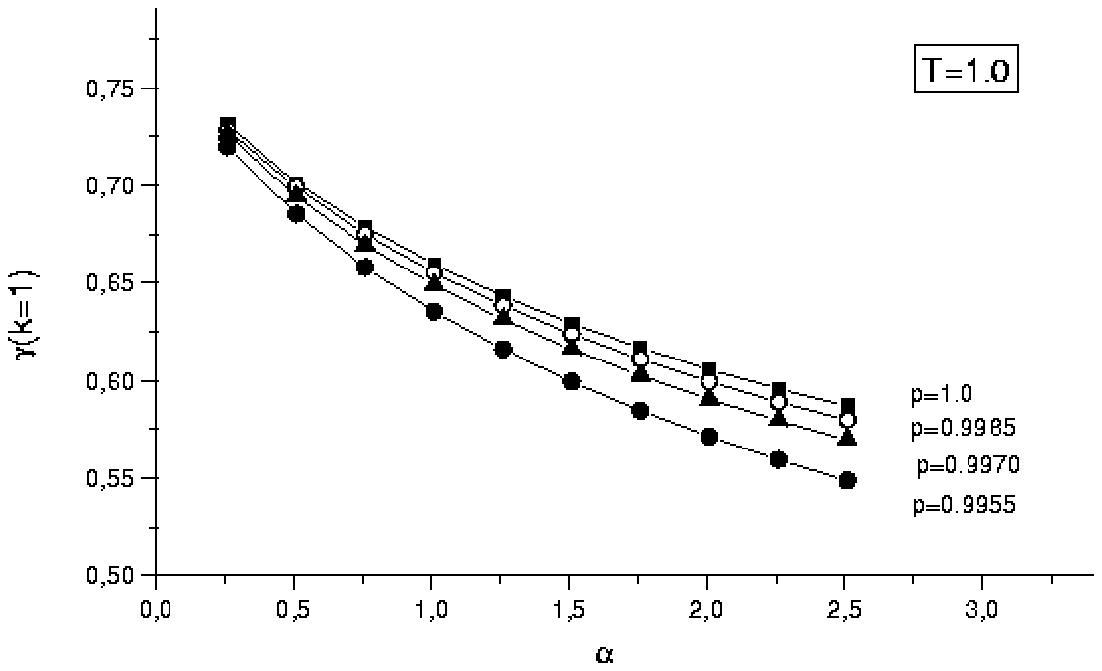}
\caption{ The dependence of the surface critical exponent $\gamma(k=1)$ on
the parameter $\alpha$ for a fixed temperature $T=1.0$ and different values
of the probability $p$ [247]. }
\label{fig42}
\end{center}
\end{figure}

\subsection{Random transverse field effect on wetting transitions}
Now we will consider the effect of random transverse field on the wetting transition of a spin-1/2 Ising  ferromagnetic film in both surface and external magnetic fields, described in Eqs.(36), (42). Using finite cluster approximation, numerical calculations are limited to a film with thickness. However, phase diagram in the space $(T, \Omega, p)$ plotted in Fig. 43, shows the existence of a critical value $p_{c}$, of the probability $p$, above which the wetting transitions disappears.

\begin{figure}
\begin{center}
\includegraphics{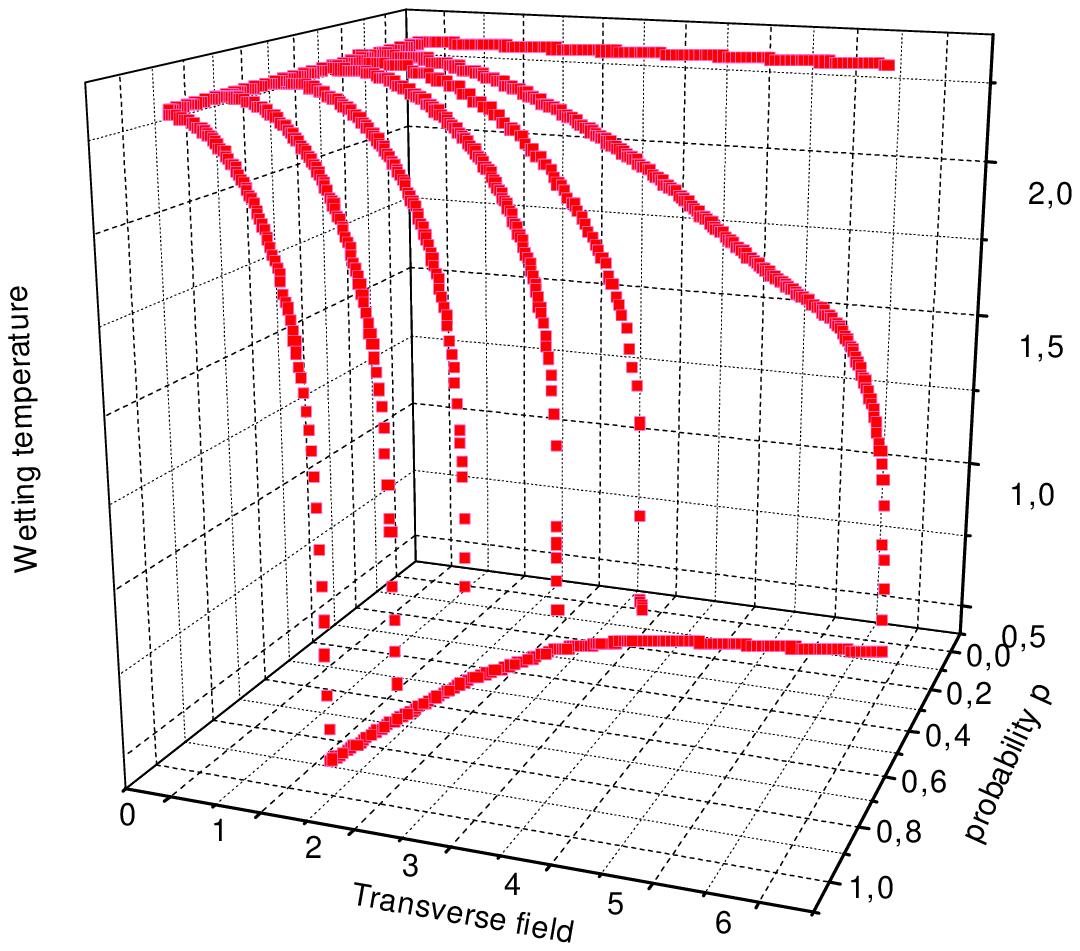}
\caption{ Phase diagrams in the $(T/J, \Omega/J ,p)$ space for
$(H_{s1}/J=0.95,H_{s2}/J=-H_{s1}/J)$ [247]. }
\label{fig43}
\end{center}
\end{figure}

The value of $p_{c}$ depends on the value of $H_{s}$. However, Fig. 44 shows that $p_c$ decreases when increasing $H_s$: indeed, $p_{c}$ decreases linearly for $H_{s}< 0.85$, and almost linearly for $H_{s} > 0.85$, with a discontinuity at $H_{s}= 0.85$.

\begin{figure}
\begin{center}
\includegraphics{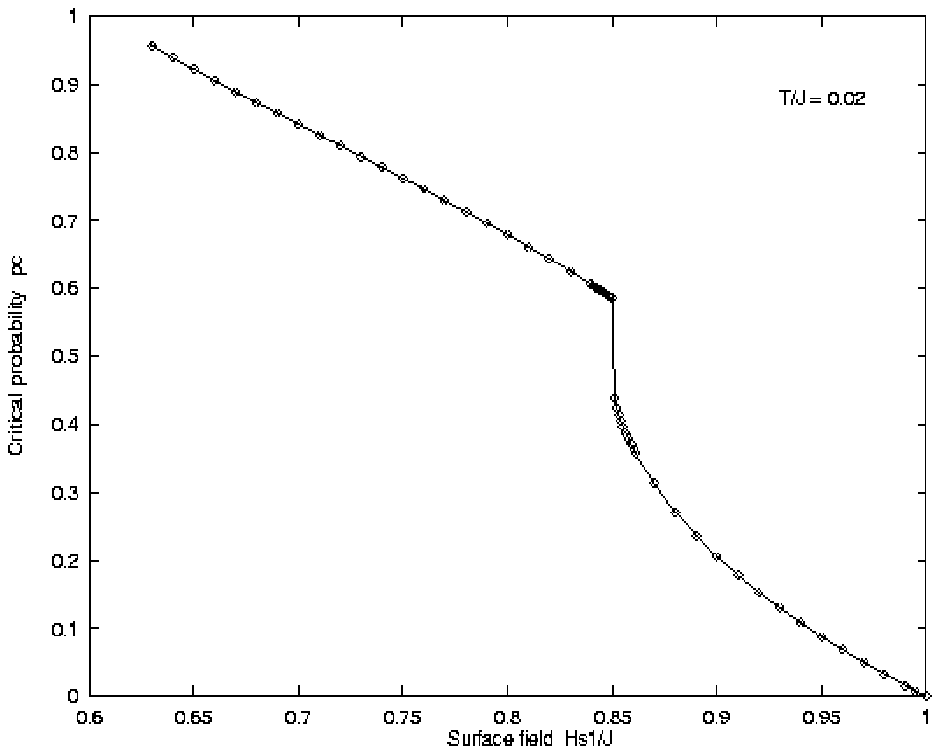}
\caption{ The critical probability $p_{c}$ dependence on the surface field
$H_{s1}/J$ for $T/J=0.02$ [247]. }
\label{fig44}
\end{center}
\end{figure}

A similar result was found by Harris [237], in the case of both  bond and site dilutions, showing that the critical transverse field at zero temperature presents a discontinuity as the concentration passes through the critical percolation concentration.

\subsection{Layering sublimation transitions of the Blume-Emery-Griffiths in a transverse field}
In this part, we consider a spin-1 BEG model on a  simple cubic lattice, governed by the Hamiltonian
\begin{equation}
H=-J\sum_{(i,j)}S_{iz}S_{jz}-K\sum_{(i,j)}S_{iz}^{2}  S_{jz}^{2}-\Delta\sum_{i}S_{iz}^{2}-\Omega\sum_{i}S_{ix}
\end{equation}
where J and K represent respectively the bilinear and bi-quadratic couplings,  $\Delta$ is the chemical potential in the lattice-gas interpretation  $[246,249]$, $\Omega$ is the transverse field, ($i$,$j$) indicates summation over nearest-neighbor sites of the simple cubic lattice, and $ S_{ix}$, $ S_{iz}$
Using the mean field theory, the free energy, the longitudinal magnetization and quadrupolar moments are given respectively by:
\begin{equation}
f=\sum_{p=1}^{N}{ -K_{B}TLog[\sum_{k=1}^{3} exp(-\beta \lambda_{k}(p))]+\frac{\rm J}{\rm 2} m_{z}(p)(m_{z}(p-1)+4m_{z}(p)+m_{z}(p+1))}
\end{equation}
$+\frac{\rm K}{\rm 2} q_{z}(p)(q_{z}(p-1)+4q_{z}(p)+q_{z}(p+1))$\\\\
\begin{equation}
{m_{z}}(p)=\frac{\rm \sum_{k=1}^{3}(\alpha_{k}^{2}(p)-\beta_{k}^{2}(p))exp(-\beta \lambda_{k}(p))}
                {\rm\sum_{k=1}^{3}exp(-\beta \lambda_{k}(p)) }
\end{equation}
\begin{equation}
{q_{z}}(p)=\frac{\rm\sum_{k=1}^{3}(\alpha_{k}^{2}(p)+\beta_{k}^{2}(p))exp(-\beta \lambda_{k}(p))}
                {\rm\sum_{k=1}^{3}exp(-\beta \lambda_{k}(p))}
\end{equation}
where\\
\begin{equation}
\lambda_{k}(p)=\frac{\rm 2}{\rm 3}(D(p)+\sqrt[3]{\varrho(p)} cos(\varphi_{k}(p)))
\end{equation}
\begin{equation}
\alpha_{k}(p)=
{\vert \Omega(\lambda_{k}(p)-D(p)+A(p)) \vert \over 2^{1/2}\sqrt {\Omega^{2}((\lambda_{k}(p)-D(p))^{2}+A^{2}(p))+((\lambda_{k}(p)-D(p))^{2}-A^{2}(p))^{2}}}
\end{equation}
\begin{equation}
\beta_{k}(p)=
{\lambda_{k}(p)-D(p)-A(p) \over\lambda_{k}(p)-D(p)+A(p)} \alpha_{k}(p)
\end{equation}
with
\begin{equation}
\varphi_{k}(p)=\frac{\rm 1}{\rm 3}arc cos(-27q(p)/2\varrho(p))+2/3(k-1)\pi
\end{equation}
\begin{equation}
\varrho(p)= \frac{\rm {3\sqrt3}}{\rm 2} \sqrt{27 q^{2}(p)+\vert4r^{3}(p)+27q^{2}(p)\vert}
\end{equation}
\begin{equation}
r(p)=-(A^{2}(p)+\Omega^{2})-D^{2}(p)/3
\end{equation}
\begin{equation}
q(p)=-D(p)(2A^{2}(p)-(2/9) D^{2}(p)-\Omega^{2})/3
\end{equation}
\begin{equation}
A(p)=-J(m_{z}(p-1)+4m_{z}(p)+m_{z}(p+1))
\end{equation}
\begin{equation}
D(p)=-\Delta-K(q_{z}(p-1)+4q_{z}(p)+q_{z}(p+1))\\
\end{equation}
With the free boundary conditions,
${m_{z}}(N+1)=
{m_{z}}(0)=
{q_{z}}(N+1)=
{q_{z}}(0)=0$.\\
Equations $(8)$ and $(9)$ are the order parameters of the system which enable us to characterize transitions. Their calculations have been performed numerically. However, in the case of several distinct solutions. The one which have the smaller free energy (19) is accepted as the physical solution. However, numerical calculations are done for $N=64$ and $K/J=3$ because at the latter value and in the absence of the transverse field, phase diagram of the spin-1 BEG model in temperature and chemical potential plane is equivalent to the solid-gas-liquid phase diagram [246].
The solid, liquid and vapor phases are characterized by a high density and non-zero magnetization,  a high density and zero magnetization and a low density and zero magnetization respectively [246]. We note that the bulk solid-vapor coexistence take place at a chemical potential $\Delta_{coe}=-12+4/N$ in the case $T=0$ and $\Omega=0$. Hence for N=64, $\Delta_{coe}=-11.9375$. Hence, we denote here after by $V^{2m}S^{N-2m}$ , $m=0,1,...,N/2$ a configuration of the system with m top layers and m bottom layers in the vapor state, while the remaining layers of the system are in the solid state; and by $L^{N}$ a configuration with N layers in the liquid state. $V^{N}$ and $S^{N}$ are respectively the bulk vapor  and  solid states respectively.\\
Phase diagram at T=Ok  (Fig. 45) shows that a sequence of layering sublimation transition $(V^{2m}S^{N-2m}\leftrightarrow V^{2m+2}S^{N-2m-2})$ appears at different values of the transverse field $\Omega_{m}$ for a fixed value of chemical potential $\Delta \rangle \ -11.9375$ (i.e in the bulk solid region); and then the bulk melting transition $(S^{N}\leftrightarrow L^{N})$ occurs at a critical bulk transverse field $\Omega_{SL}$.

\begin{figure}
\begin{center}
\includegraphics{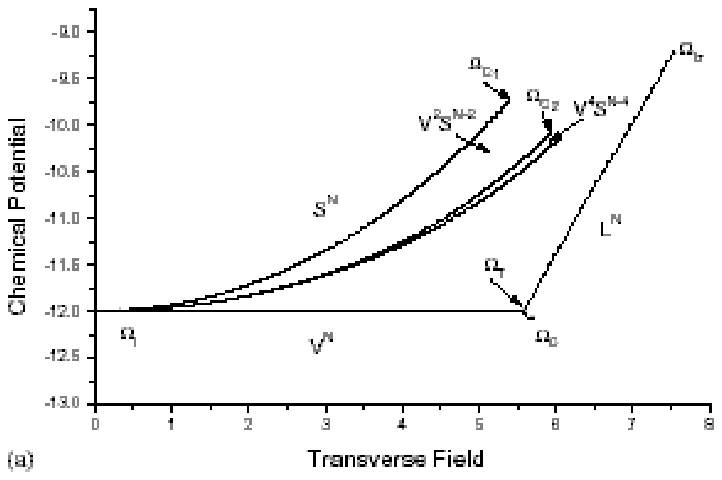}
\caption{ Phase diagram in the $(\Omega, \Delta)$ plane for $T=0$ [248]. }
\label{fig45}
\end{center}
\end{figure}

While for $\Delta \langle \Delta_{coe}$ and $\Omega \langle \Omega_{c}$ there is also the bulk vapor-liquid transition ($V^{N}\leftrightarrow L^{N}$) which take place at a critical value of a transverse field $\Omega_{VL}$, while the solid-liquid-gas bulk coexistence occurs at the triple transverse field $\Omega_{T}$ and $\Delta= \Delta_{coe}$. Such result is qualitatively similar to the one obtained by Gelb [246] under the effect of temperature. In Fig. 46, the bulk density exhibits a discontinuity at the transition low density (Vapor phase) to high density (liquid phase) which occurs at  $\Omega_{VL}=5.51$ for $\Delta=-12.002$.

\begin{figure}
\begin{center}
\includegraphics{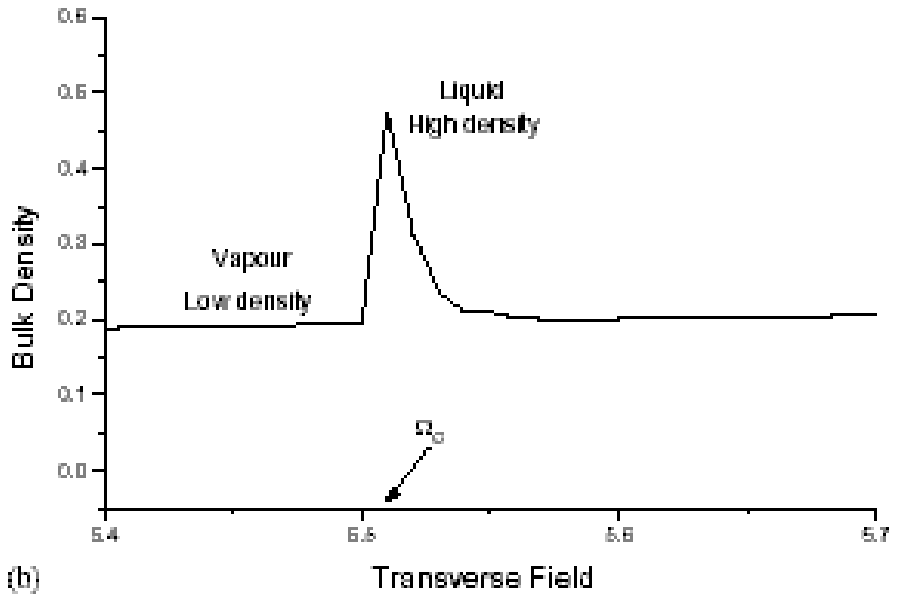}
\caption{ The dependence of the bulk density of the transverse field for $ \Delta=12.002$ and $T=0$.
The vapor liquid transition is found at   $\Omega_{VL}=5.51$ [248]. }
\label{fig46}
\end{center}
\end{figure}

The behavior of density and magnetization, as a function of the transverse field $\Omega$, are given, for $\Delta \rangle \Delta_{coe}$, in Fig. 47a  and 47b respectively.
It is clear that the transition of a layer p (from solid state to quasi-vapor state) is characterized by the jump of both density  and magnetization.
\begin{figure}
\begin{center}
\includegraphics{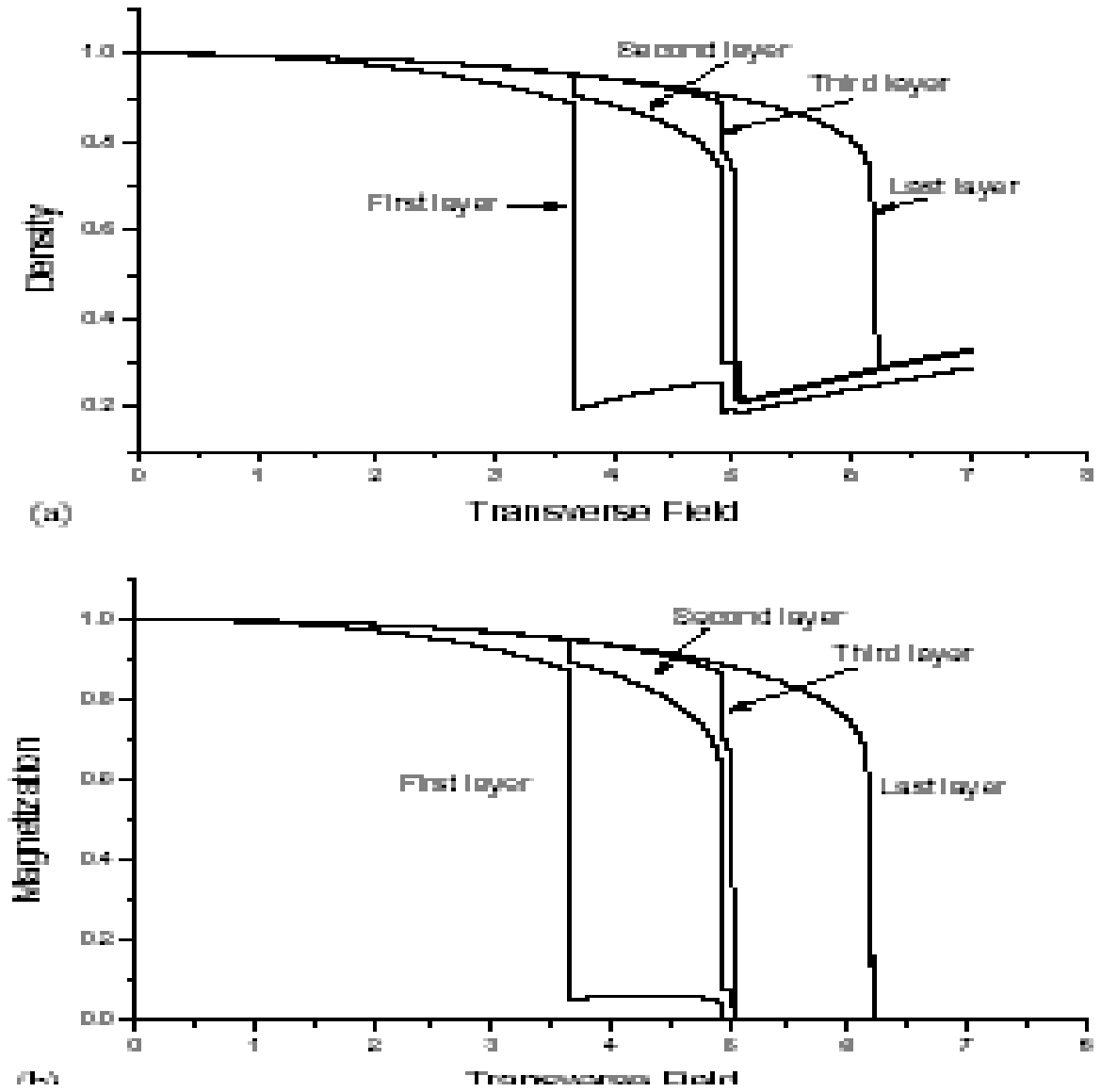}[t]
\caption{ (a) Profile of density of the top three and last layers as a
function transverse field  for $ \Delta=11$ and $T=0$. (b):  Profile of
magnetization of the first three and last layers  as a function of
transverse field for $\Delta=-11$ and $T=0K$ [248]. }
\label{fig47}
\end{center}
\end{figure}

At $T>0$, the $(T, \Omega)$ phase diagrams presented in Fig. 48 show the existence of a sequence of layering sublimation transition for each value of temperature and fixed value of chemical potential, and a solid-liquid bulk transition occurs at a transverse field $\Omega_{SL}$ which depends on the value of chemical potential.

\begin{figure}
\begin{center}
\includegraphics{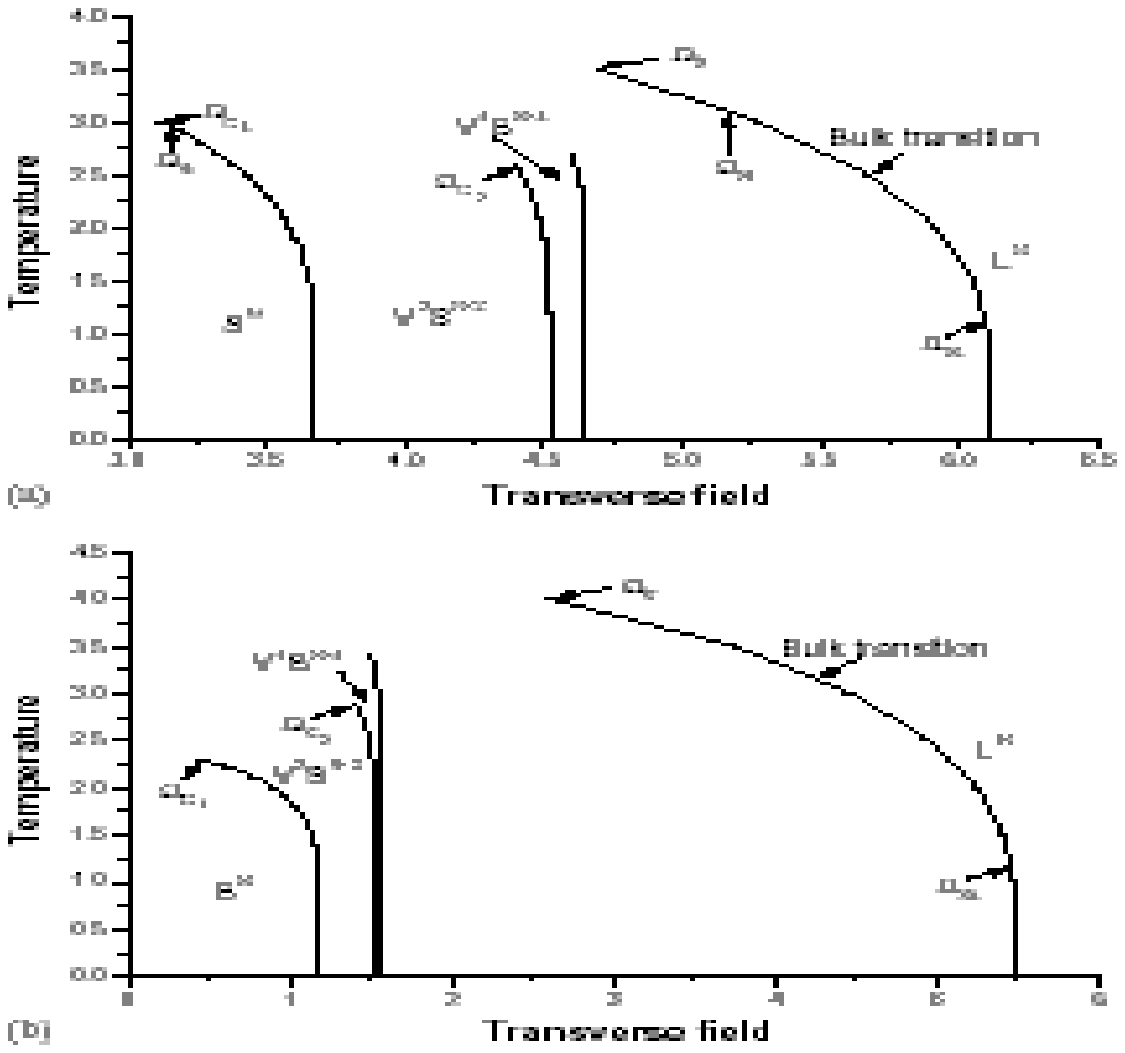}[t]
\caption{ The phase diagram in the $(\Omega,T)$ plane
(a): For $\Delta=-11.$
(b): For $\Delta=-11.9$ [248]}
\label{fig48}
\end{center}
\end{figure}

However $\Omega_{SL}$ increases with increasing chemical potential. For example for $T=1$ and $\Delta=-11$, $\Omega_{SL}=6.01$ (Fig. 48a) while it takes the value 5.30 for T=1 and $\Delta=-11.9$ (Fig. 48b). On the other hand the transition line of a layer p terminates by an end  point of coordinate $(T_ {c_{p}},\Omega_{c_{p}})$. these endpoints  converge towards a roughening temperature $T_{R}$ and roughening transverse field  $\Omega_{R}$ above which all interfaces are rough on a microscopic scale,  Hence  $T_{R}=3.07$ and $\Omega_{R}=5.18$ in the case of Fig. 48a. Besides, in this figure the critical end point of the first layer is greater than the  second layer end point while in Fig. 48b the critical end point of the first layer is smaller than the second layer one. This means that, depending on the value of the chemical potential, there exist a critical temperature $T_{s}$ and critical transverse field $\Omega_{s}$ at which the surface transition occurs while the other layers transitions cannot take place for any value of transverse field.

\section{References}
\begin{enumerate}
\item[{[1]}] R. Pandit, M. Schick and M. Wortis, Phys. Rev. B {\bf 26}, 8115 (1982).
\item[{[2]}] R. Pandit and M. Wortis, Phys. Rev. B {\bf 25}, 3226  (1982).
\item[{[3]}] M. P. Nightingale, W. F. Saam and M. Schick, Phys. Rev. B {\bf 30},3830 (1984).
\item[{[4]}] C. Ebner, C. Rottman and M. Wortis, Phys. Rev. B {\bf 28},4186  (1983).
\item[{[5]}] C. Ebner and W. F. Saam, Phys. Rev. Lett. {\bf 58},587 (1987).
\item[{[6]}] C. Ebner and W. F. Saam, Phys. Rev. B {\bf 35},1822 (1987).
\item[{[7]}] C. Ebner, W. F. Saam and A. K. Sen, Phys. Rev. B {\bf 32},1558 (1987).
\item[{[8]}] A. Patrykiejew A., D. P. Landau and K. Binder, Surf. Sci. {\bf 238}, 317 (1990).
\item[{[9]}] S. Ramesh and J. D. Maynard, Phys. Rev. Lett. {\bf 49},47 (1982).
\item[{[10]}] S. Ramesh, Q. Zhang, G. Torso and J. D. Maynard, Phys. Rev. Lett. {\bf 52},2375 (1984).
\item[{[11]}] M. Sutton, S. G. J. Mochrie  and R. J. Birgeneou, Phys. Rev. Lett. {\bf 51},407 (1983); S. G. J. Mochrie, M. Sutton, R. J. Birgeneou, D. E. Moncton and P. M. Horn, Phys. Rev. B {\bf 30},263 (1984).
\item[{[12]}] S. K. Satija, L. Passel, J. Eckart, W. Ellenson and H. Patterson, Phys. Rev. Lett. {\bf 51},411 (1983).
\item[{[13]}] A.F.G. Wyatt  and J. Klier, Phys. Rev. Lett. {\bf 85}, 2769 (2000).
\item[{[14]}] Bonn et al., Phys. Rev. Lett. {\bf 87}, 279601 (2001). S. M. M. Ramos and E. Charlaix, Phys. Rev.E {\bf 67}, 031604 (2003).
\item[{[15]}] M. P. Valignat, S. Villette, J. Li, R. Barberi, R. Bartolino, E. Dubois-Violette and A. M. cazabat, Phys. Rev. Lett. {\bf 77}, 1994 (1996).
\item[{[16]}] K. Ragil, J. Meunier, D. Broseta, J. O. Indekeu and D. Bonn, Phys. Rev. Lett. {\bf 77}, 1532 (1996).  
\item[{[17]}] Ch. Bahr, C. J. Booth, D. Fliegner and J. W. Goodby Phys. Rev. Lett. {\bf 77}, 1083 (1996);
R. Lucht, Ch. Bahr and G. Heppke, J. Phys. Chem. B {\bf 102}, 6861 (1998). 
\item[{[18]}] M. Geoghengan, R. A. L. Jones, D. S. Sivia, J. Penfold and A. S. Clough, Phys. Rev. E {\bf 53}, 825 (1996). 
\item[{[19]}]  E. Bertrand, H. Dobbs, D. Broseta, J. Indekeu, D. Bonn and J. Meunier, Phys. Rev. Lett. {\bf 85}, 1282 (2000).
\item[{[20]}] B. Nickel, W. Donner, and H. Dosch, Phys. Rev. Lett. {\bf 85}, 134 (2000).  
\item[{[21]}]  E. Montevecchi and R. Blossey, Phys. Rev. Lett. {\bf 22}, 4743 (2000).  
\item[{[22]}] Ross et al., Phys. Rev. Lett. {\bf 87}, 176103 (2001).  
\item[{[23]}] Tostmann et al., Phys. Rev. lett. {\bf 84}, 4385 (2000).  
\item[{[24]}] Huber et al., Phys. Rev. Lett. {\bf 89}, 035502 (2002).  
\item[{[25]}] Huber et al. Phys. Rev. B {\bf 68}, 085409 (2003).  
\item[{[26]}] T. Ueno, S. Balibar, T. Mizusaki, F. Caupin, and E. Rolley, Phys. Rev. Lett {\bf 90}, 116102 (2003).  
\item[{[27]}] J. Crassous, E. Charlaix and J. Loubet, Phys. Rev. Lett. {\bf 78}, 2425 (1997).  
\item[{[28]}] D. Bonn, E. Bertrand and J. Meunier, Phys. Rev. Lett. {\bf 84}, 4661 (2000).  
\item[{[29]}] F. Mugele and M. Salmeron, Phys. Rev. Lett. {\bf 25}, 5796 (2000).  
\item[{[30]}] K. G. Sukhatme, J. E. Rutledge and P. Taborek, Phys. Rev. Lett. {\bf 80}, 129 (1998).  
\item[{[31]}] N. Shhidzadeh, D. Bonn, K. Ragil, D. Broseta and J. Meunier, Phys. Rev. Lett. {\bf 80}, 3992 (1998).  
\item[{[32]}] M. J. de Oliveira  and R. B. Griffiths , Surf. Sci. {\bf 71}, 687 (1978).
\item[{[33]}] T. Jil, Phys. Rev. E {\bf 52}, 772 (1995).  
\item[{[34]}] A. V. Dobrynin, A. Deshkovski, and M. Rubinstein,
 Phys. Rev. Lett. {\bf 84}, 3101 (2000).  
\item[{[35]}] G. G. Pereira, and J.-S Wang, Phys. Rev. E {\bf 54}, 3040 (1996).  
\item[{[36]}] S. Curtarolo, G. Stan, M. W. Cole, M. J. Bojan
and W. A. Steele , Phys. Rev. E {\bf 59}, 4402 (1999).  
\item[{[37]}] W. Shi, and J. K. Johnson, Phys. Rev. B {\bf 68}, 125401 (2003).
\item[{[38]}] M. Kruk, A. Patrykiejew, and S. Sokolowski, Surf. Sci. {\bf 340}, 179 (1995).  
\item[{[39]}] A. D. Stoycheva, and S. J. Singer,
 Phys. Rev. Lett. {\bf 84}, 4657 (2000).  
\item[{[40]}] L. Bahmad, A, Benyoussef, and H. Ez-Zahraouy, Phys. Rev. E
 {\bf 66}, 056117 (2002); A. Milchev, M. Muller, and D. P. Landau,
 Phys. Rev. Lett. {\bf 90}, 136101-1 (2003).  
\item[{[41]}] A. Patrykiejew, L. Salamacha, S. Sokolowski, and O. Pizio,
Phys. Rev. E {\bf 67}, 061603 (2003).  
\item[{[42]}] G. Oshanin, J. De Coninck, A. M. Cazabat, and M. Moreau, Phys. Rev. E
 {\bf 58}, R20 (1998).  
\item[{[43]}] T. Getta, and S. Dietrich, Phys. Rev. E {\bf 57}, 655 (1998).  
\item[{[44]}] E. Montevecchi, and J. O. Indeku, Phys. Rev. B {\bf 62}, 14359 (2000).  
\item[{[45]}] M. Muller, E. V. Albano, and K. Binder, Phys. Rev. E {\bf 62}, 5281 (2000).  
\item[{[46]}] L. Szybisz, Phys. Rev. B {\bf 62}, 3986 (2000).  
\item[{[47]}] K. Rejmer, S. Dietrich, and M. Napiorkowski, Phys. Rev. E {\bf 60}, 4027 (1999).
\item[{[48]}] C. Bauer, and S. Dietrich, Phys. Rev. E {\bf 60}, 6919 (1999).  
\item[{[49]}] C. Bauer, T. Bieker, and S. Dietrich, Phys. Rev. E {\bf 62}, 5324 (2000).  
\item[{[50]}] E. Rabani, D. R. Reichman, P. L. Geisser and L. E. Brus, Nature {\bf 426}, 175 (2003), C. Bauer, and S. Dietrich, Phys. Rev. E {\bf 61}, 1664 (2000). 
\item[{[51]}] L. Harnau, and S. Dietrich, Phys. Rev. E {\bf 66}, 051702 (2002).  
\item[{[52]}] G. Palasantzas, Phys. Rev. E {\bf 66}, 021604 (2002).
\item[{[53]}] G. Palasantzas, Phys. Rev. B {\bf 68}, 035412 (2003).  
\item[{[54]}] P. B. Weichman, P. Day, and D. Goodstein, Phys. Rev. Lett. {\bf 74}, 418 (1995).  
\item[{[55]}] A. Prasad, and P. B. Weichman, Phys. Rev. B {\bf 57}, 4900 (1998).  
\item[{[56]}] C. J. Boulter, and A. O. Parry, Phys. Lett. {\bf 74}, 3403 (1995).
\item[{[57]}] D. B. Abraham, L. Fontes, C. M. Newman, and M.S.T. Piza, Phys. Rev. E 52, R1257 (1995) 
\item[{[58]}] P. Lenz, and R. Lipowsky, Phys. Rev. Lett. {\bf 80}, 1920 (1998).  
\item[{[59]}] F. P. A. Cortat, and S. J. Miklavcic, Phys. Rev. E {\bf 68}, 052601 (2003).  
\item[{[60]}] E. Bonaccurso, H.-J. Butt, Phys. Rev. Lett. {\bf 90}, 144501 (2003).  
\item[{[61]}] M. Boninsegi, M. W. Cole, and F. Toigo,
 Phys. Rev. Lett. {\bf 83}, 2002 (1999).  
\item[{[62]}] J. A. Phillips, D. Ross, P. Taborek, and J. E. Rutledge,
 Phys. Rev. B {\bf 58}, 3361 (1998).  
\item[{[63]}] A. Hanke, Phys. Rev. Lett. {\bf 84}, 2180 (2000).  
\item[{[64]}] D. B. Abraham, and A. Maciolek,
 Phys. Rev. Lett. {\bf 89}, 286101-1 (2002).
\item[{[65]}] W. Jin, and J. Koplik, Phys. Rev. Lett. {\bf 78}, 1520 (1997).  
\item[{[66]}] K. Grabowski, A. Patrykiejew, S. Sokolowski, E. V. Albano, and
A. de Virgiliis, Surf. Sci. {\bf 448}, 11 (2000).  
\item[{[67]}] C. Ebner and W. F. Saam, Phys. Rev. A {\bf 22}, 2776 (1980);
     ibid, Phys. Rev. A {\bf 23},1925 (1981);
     ibid, Phys. Rev. B {\bf 28},2890 (1983).
\item[{[68]}] D. A. Huse , Phys. Rev. B {\bf 30},1371 (1984).
\item[{[69]}]  A. Benyoussef and H. Ez-Zahraouy, Physica A, {\bf 206}, 196
(1994).
\item[{[70]}]  A. Benyoussef and H. Ez-Zahraouy, J. Phys. {\it I } France
{\bf 4}, 393 (1994).
\item[{[71]}] Q. Hong  Phys. Rev. B {\bf 41}, 9621  (1990);   ibid, Phys. Rev. B {\bf 46}, 3207  (1992).
\item[{[72]}] A. J. Freeman, J. Magn. Magn. Mater, {\bf 15-18}, 1070 (1980).
\item[{[73]}] K. Binder in {\it Phase Transitions and Critical Phenomena}, edited by C. Domb and J. L. Lebowits (Academic, New York, 1983), Vol. 8.
\item[{[74]}] R. Richter, J. G. Gay and J. R. Smith, Phys. Rev. Lett. {\bf 54}, 2704 (1985).
\item[{[75]}] R. H. Victora and L. M. Falicov, Phys. Rev. B {\bf 31}, 7335 (1985).
\item[{[76]}] C. Rau, C. Schneider, G. Xiang and K. Jamison, Phys. Rev. Lett. {\bf 57}, 3221 (1986).
\item[{[77]}] D. Pescia, G. Zampieri, G. L. Bona, R. F. Willis and F. Meier, Phys. Rev. Lett. {\bf 58}, 9 (1987).
\item[{[78]}] C. Rau, G. Xiang and C. Liu, Phys. Rev. Lett. A {\bf 135}, 227 (1989).
\item[{[79]}]  L. Bahmad, A. Benyoussef, A. Boubekri, and H. Ez-Zahraouy,
Phys. Stat. Sol. (b) {\bf 215}, 1091 (1999).
\item[{[80]}] L. Bahmad, A. Benyoussef and H. Ez-Zahraouy,
M. J. Condensed Matter  {\bf 4}, 84 (2001).
\item[{[81]}] F. Aguilera-Granja and J. L. Moran-Lopez, Phys. Rev. B {\bf 31}, 7146 (1985).
\item[{[82]}] D. Karevski, R. Juhasz, L. Turban and F. Igloi, Phys. Rev. B {\bf 60}, 4195 (1999).
\item[{[83]}] A. Benyoussef, N. Boccara and M. Saber, J. Phys. C {\bf 19}, 1983 (1986).
\item[{[84]}] A. Benyoussef, N. Boccara and M. Bouziani, Phys. Rev. B {\bf 34}, 7775 (1986).
\item[{[85]}] X. P. Jiang and M.R. Giri, J. Phys. C {\bf 21}, 995 (1988).
\item[{[86]}] I. Tamura, J. Phys. Soc. Jpn {\bf 51}, 3607 (1982).
\item[{[87]}] C. Buzano and Pelizzola, Physica A {\bf 195}, 197 (1993).
\item[{[88]}] J.B. Collins, P.A. Rikvold and E. T. Gawlinski, Phys. Rev. B {\bf 38}, 6741 (1988).
\item[{[89]}] Y. Saito, J. Chem. Phys. {\bf 74}, 713 (1981).
\item[{[90]}] Y. L. Wang and K. Rauchwarger, Phys.Lett. A {\bf 59}, 73 (1976).
\item[{[91]}] J.D. Kimel, P. A. Rikvold and Y. L. Wang, Phys. Rev. B {\bf 45}, 7237 (1992).
\item[{[92]}] A. Benyoussef and H. Ez-Zahraouy, J. Phys.: Cond. Matt.  {\bf 6}, 3411 (1994).
\item[{[93]}] R. B. Stinchcombe, J. Phys. C {\bf 6}, 2459 (1973).
\item[{[94]}] J. L. Zhong, J. Liangli and C. Z. Yang, Phys. Stat. Sol. (b) {\bf 160}, 329 (1990).
\item[{[95]}] U. V. Ulyanov, O. B. Zalavskii, Phys. Rep. {\bf 216}, 179 (1992)
\item[{[96]}] A. Benyoussef, H. Ez-Zahraouy and M. Saber, Physica A {\bf 198}, 593 (1993)
\item[{[97]}] R. Geer, T. Stoebe, C. C. Huang, R. Pindak, J. W. Goodby, M. Cheng, J. T. Ho and S. W. Hui, {\it Nature} {\bf 355}, 152 (1992).
\item[{[98]}] T. Stoebe, R. Geer, C. C. Huang and J. W. Goodby, Phys. Rev. Lett. {\bf 69}, 2090 (1992).
\item[{[99]}] A. J. Jin, M. Veum, T. Stoebe, C. F. Chou, J. T. Ho, S. W. Hui, V. Surendranath and C. C. Huang, Phys. Rev. Lett. {\bf 74}, 4863 (1995);
Phys. Rev. E {\bf 53}, 3639 (1996).
\item[{[100]}] D. L. Lin, J. T. Ou, Long-Pei Shi, X. R. Wang and A. J. Jin, Europhys. Lett. {\bf 50}, 615 (2000).
\item[{[101]}] H. S. Youn and G.B. Hess, Phys. Rev. Lett. {\bf 64}, 918 (1990).
\item[{[102]}] L. Bahmad, A. Benyoussef and H. Ez-Zahraouy, J. Magn. Magn. Mat. {\bf 251}, 115 (2002).
\item[{[103]}] M. Blume, Phys. Rev. {\bf 141}, 517 (1966); H. W. Capel, Physica {\bf 32}, 966 (1966).
\item[{[104]}] M. Blume, V. J. Emery and R. B. Griffiths, Phys. Rev. A {\bf 4}, 1071 (1971).
\item[{[105]}] J. Lajzerowicz and J. Sivardi{\'e}re, Phys. Rev. A {\bf 11}, 2079 (1975); J. Sivardi{\'e}re and J. Lajzerowicz, {\it ibid} {\bf 11}, 2090 (1975).
\item[{[106]}] K. E. Newman and J. D. Dow, Phys. Rev. B {\bf 27}, 7495 (1983).
\item[{[107]}] S. A. Kivelson, V. J. Emery and H. Q. Lin, Phys. Rev. B {\bf 42}, 6523 (1990).
\item[{[108]}] E. F. Sarmento, I. Tamura, L. E. M. C. de Oliveira and T. Kaneyoshi, J. Phys. C {\bf 17}, 3195 (1984).
\item[{[109]}] E. F. Sarmento and T. Kaneyoshi, Phys. Stat. Sol. b {\bf 160}, 337 (1990).
\item[{[110]}] Q. Jiang and Z. Y. Li, Phys. Rev. B {\bf 43}, 6198 (1991).
\item[{[111]}] N. M. Jennet and D. Dingley, J. Magn. Magn. Mater. {\bf 93}, 472 (1991).
\item[{[112]}] A. S. Edelstein {\it et al.}, Solid State Commun. {\bf 76}, 1379 (1990).
\item[{[113]}] R. Krishnan, H. O. Gupta, H. Lassri, C. Sella and J. Kaaboutchi, J. Appl. Phys. {\bf 70}, 6421 (1991).
\item[{[114]}] A. Z. Maksymowicz, Phys. Stat. Sol. b {\bf 122}, 519 (1984).
\item[{[115]}] F. C. Sa Barreto and I. P. Fittipaldi, Physica A {\bf 129}, 360 (1985).
\item[{[116]}] T. Kaneyoshi Phys. Rev.  B {\bf 39}, 557 (1989).
\item[{[117]}] A. Bobak, Phys. Stat. Sol. b {\bf 109}, 161 (1982).
\item[{[118]}] N. Cherkaoui Eddeqaqi, H. Ez-Zahraouy and A. Bouzid, Phys. Stat. Sol. (b) {\bf 213}, 421 (1999).
\item[{[119]}] N. Cherkaoui Eddeqaqi, H. Ez-Zahraouy and A. Bouzid, J. Magn. Magn. Mat. {\bf 207}, 209 (1999).
\item[{[120]}] R. Hull and J. C. Bean (Eds) Semicond. semicont. {\bf 56}
(1999).
\item[{[121]}] E. Kasper and K. Lyutovich in '{\it Properties of Silicon Germanium
and $SiGe$: Carbon}, EMIS Datareviews Series {\bf 24} (INSPEC, IEE, London,
2000).
\item[{[122]}] G. D. M. Dilliway, D. M. Bagnall, N. E. B. Cowern and C. Jeynes,
J. Mat. Sci.: Mat. in Elect. {\bf 14}, 323 (2003)
\item[{[123]}] K. A. Fichthorn and M. L. Merrick, Phys. Rev. B {\bf 68},
041404 (R) (2003).
\item[{[124]}] K. Lau and W. Kohn Surf. Sci. {\bf 65}, 607, (1977).
\item[{[125]}] P. Hyldgaard and M. Persson J. Phys.: Condens. Matter
{\bf 12}, L13, (2000).
\item[{[126]}] K. A. Fichthorn and M. Scheffler, Phys. Rev. Lett.
{\bf 84}, 5371, (2000).
\item[{[127]}] N. Knorr, H. Brune, M. Epple, A. Hirstein, M. A. Shneider and
K. Kern, Phys. Rev. B {\bf 65}, 115420, (2002).
\item[{[128]}] S. Ovesson, A. Bogicevic, G. Wahnstrom, and B. I. Lundqvist,
Phys. Rev. B {\bf 64}, 115423, (2001).
\item[{[129]}] S. Ovesson, Phys. Rev. Lett. {\bf 88}, 116102, (2002).
\item[{[130]}] K. A. Fichthorn, M. L. Merrick, and M. Scheffler, Appl. Phys.
A: Mater. Sci. Process. {\bf 75}, 17, (2002)
\item[{[131]}] J. A. Venables and W. Kohn Phys. Rev. B {\bf 66}, 195404, (2002)
\item[{[132]}] H. Brune, K. Bromann, K. Kern, J. Jacobsen, P. Stolze,
K. Jacobsen, and J. Norskov, Phys. Rev. B {\bf 52}, R14 380, (1995)
\item[{[133]}] J. V. Barth, H. Brune, B Fischer, J. Weckesser, and
M. Scheffler, Comput. Phys. Commun. {\bf 107}, 187, (1997)
\item[{[134]}] X. Y. Zheng, Y. Ding, L. A. Bottomley, D. P. Alison, and R. J.
Warmack, J. Vac. Sci. Technol. B {\bf 13}(3), 187, (1995)
\item[{[135]}] K. Reichelt and H. O. Lutz, J. Cryst. Growth  {\bf 10}, 103,
(1971)
\item[{[136]}] C. E. D. Chidsey, D. N. Loiacono, T. Sleator, and S. Nakahara,
Surf. Sci.  {\bf 200}, 45 (1988)
\item[{[137]}] A. Putnam, B. L. Blackford, M. H. Jericho, and M. O. Watanabe
Surf. Sci.  {\bf 217}, 276 (1989)
\item[{[138]}] J. A. DeRose, T. Thundat, L. A. Nagahara, and S. M. Lindsay,
Surf. Sci.  {\bf 256}, 102 (1991)
\item[{[139]}] R. Koch, D. Winau, A. Fuhrmann, and K. H. Reider, Vacuum
{\bf 43}, 521 (1992)
\item[{[140]}] M. Hegner, P Wagner, and G. Semenza, Surf. Sci.  {\bf 291},
39 (1993)
\item[{[141]}] J. E. Morris and T. J. Coutts, Thin Solid Films {\bf 47}, 3
(1977)
\item[{[142]}] P. G. Borziak, Y. A. Kulyupin, S. A. Nepijko, and V. G.
Shamonya, Thin Solid Films {\bf 76}, 359 (1981)
\item[{[143]}] {\it Magnetic Recording: The First Hundred Years.} Eds.
Daniel, E.D. Mee, C.D. and Clark, M. H. (IEEE Press, New York, 1998).
\item[{[144]}] D. A. Thompson and J. S. Best, IBM J. Res. Develop. {\bf 44},
311 (2000)
\item[{[145]}] A. Hubert and R. Schafer, {\it Magnetic Domains.}
(Springer, New York, 1998).
\item[{[146]}] R. D. Kirby, J. X. Shen, R. J. Hardy, and D.J. Sellmyer,
Phys. Rev. B {\bf 65}, R10 810, (1994).
\item[{[147]}] R. Allenspach, M. Stampanoni, and A. Bischof, Phys. Rev.
Lett. {\bf 65}, 3344 (1990)
\item[{[148]}] D. Weller and A. Moser, IEEE Trans. Magn. {\bf 35}, 4423,
(1999)
\item[{[149]}] S. K. Han, S.-C. Yu, and K. V. Rao, J. Appl. Phys. {\bf 79},
4260 (1996)
\item[{[150]}] S. Lemerle, J. Ferre, C. Chappert, V. Mathet, T. Giamarchi,
and P. Le Doussal, Phys. Rev. Lett. {\bf 80}, 849 (1998)
\item[{[151]}] J. F. Heanue, M. C. Bashaw, and L. Hasselink, {\it Science} {\bf 265},
749 (1994)
\item[{[152]}] J. Nakamura, M. Miyamoto, S. Hosaka, and H. Koyanagi,
J. Appl. Phys. {\bf 77}, 779 (1995)
\item[{[153]}] L. Krusin-Elbaum, T. Shibauchi, B. Argyle, L. Gignac, and D.
Weller {\it Nature} {\bf 410}, 444 (2001)
\item[{[154]}] R. Hauert and J. Patscheider, Adv. Eng. Mater. {\bf 2} No. 5,
247 (2000)
\item[{[155]}] R. Brinzanik, P. J. Jensen, and K. H. Bennemann, J. Magn. Magn.
Mater. {\bf 238} 258 (2002)
\item[{[156]}] R. Allenspach, J. Magn. Magn. Mater. {\bf 129}, 160 (1994)
\item[{[157]}] H. P. Oepen, M. Speckmann, Y. Millev, and J. Kirschner
, Phys. Rev. B {\bf 55}, 2752 (1997)
\item[{[158]}] E. Mentz, A. Bauer, D. Weiss, and G. Kaindl, Matter. Res. Soc.
Symp. {\bf 475}, 431 (1997)
\item[{[159]}] W. Kuch, J. Gilles, S. S. Kang, S. Imada, S. Suga, and J.
Kirschner, Phys. Rev. B {\bf 62}, 3824 (2000)
\item[{[160]}] A. Kubetzka, O. Pietzsch, M. Bode, and R. Wiesendanger,
, Phys. Rev. B {\bf 63}, 140407(R) (2001)
\item[{[161]}] A. Benyoussef, and H. Ez-Zahraouy, Physica Scripta {\bf 57}, 398 (1998); H. Ez-Zahraouy and A. Kassou-Ou-Ali, Phys. Rev. B {\bf 69}, 0266 (2004)
\item[{[162]}] K. Binder, in C, Domb, J. L. Lebwitz (Eds.), Phase Transitions and
Critical Phenomena, vol.{\bf 8} Academic Press, London, (1983), p. 1
\item[{[163]}] H. Dosh, Critical Phenomena at Surfaces and Interfaces,
Springer tracts in Modern physics, vol.{\bf 126} Springer, Heidelberg, (1992),
\item[{[164]}] W. H. Diehl, J. L. Lebwitz (Eds.), Phase Transitions and
Critical Phenomena, vol.{\bf 10} Academic Press, London, (1986).
\item[{[165]}] K. binder, P. C. Hohenberg, Phys. Rev. B {\bf 6}, 3461 (1972).
\item[{[166]}] D. P. Landau, K. binder, Phys. Rev. B {\bf 41}, 4633 (1990).
\item[{[167]}] M. Kikushi, Y. Okabe, Prog. Theor. Phys. {\bf 73}, 32 (1985).
\item[{[168]}] J. Merikoski, J. Timonen, M. Manninen, P. Jenna, Phys. Rev. Lett.
{\bf 66}, 938 (1991).
\item[{[169]}] B. V. Reddy, S. N. Khamma, Phys. Rev. B {\bf 45}, 10103 (1992).
\item[{[170]}] P. Schibi, S. Sienbentritt, K. H. Rieder, Phys. Lett. A {\bf 216}, 20 (1996).
\item[{[171]}] F. D. A.  Aarao Reis, Phys. Rev. B {\bf 55}, 11084 (1997).
\item[{[172]}] T. Kaneyoshi, Introduction to Surface Magnetism, CRC Press,
Boca Raton, 1991.  
\item[{[173]}] H. W. Diehl, A. Nusser, Z. Phys. B {\bf 79}, 69 (1990)
\item[{[174]}] W. Selke, F. Szalma, P. Lajko, F. Igloi, J. Stat. Phys.
{\bf 89}, 1079 (1997)
\item[{[175]}] M. Pleimling, J. Phys. A: Math. Gen. {\bf 37}, R79-R115 (2004), M. Pleimling, W. Selke, Eur. Phys. J. B{\bf 1}, 385 (1998)
\item[{[176]}] A. B. Harris, J. Phys. C {\bf 7}, 1671 (1974)
\item[{[177]}] Vik. S. Dotsenko, VI. S. Dotsenko, Adv. Phys. {\bf 32}, 129 (1983)
\item[{[178]}] R. Shankar, Phys. Rev. Lett.  {\bf 58}, 2466 (1987)
\item[{[179]}] A. W. W. Luding, Phys. Rev. Lett.  {\bf 61}, 2388 (1988)
\item[{[180]}] B. N. Shalnev, Sov. Phys. Solid state {\bf 26}, 1811 (1984)
\item[{[181]}] V. B. Andreichenko, VI. S. Dotsenko, W. Selke, J. -S. Wang,
Nucl. Phys. B {\bf 344}, 531 (1990)
\item[{[182]}] R. Kuhn, Phys. Rev. Lett.  {\bf 73}, 2368 (1994)
\item[{[183]}] J. K. Kim, A. Patrasciolu, Phys. Rev. Lett.  {\bf 72},
2785 (1994)
\item[{[184]}] H. G. Ballesteros, L. A. Fernandez, V. Martin-Mayor, A.
Munoz-Suduje, G. Parisi, J. J. ruiz Lorentzo, J. Phys. A  {\bf 30},
8379 (1997).  
\item[{[185]}] H. W. Diehl, Eur. Phys. J. B {\bf 1}, 401 (1998)
\item[{[186]}] N. Boccara, Phys. Lett. A {\bf 94}, 185 (1983)
\item[{[187]}] A. Benyoussef, N. Boccara, J. Phys. C {\bf 16}, 1143 (1983).
\item[{[188]}] K. K. Mon, M. P. Nightingale, Phys. Rev. B {\bf 37}, 3815 (1988).
\item[{[189]}] P. Tomezak, E. F. Sarmento, A. F. Siqueira, A. R. Ferchemin, Phys. Stat.
Sol. B {\bf 142}, 551 (1987).
\item[{[190]}]A. Benyoussef, H. Ez-Zahraouy, Phys. Rev. B {\bf 52}, 4245 (1995)
\item[{[191]}] H. Dosh, Appl. Phys. A {\bf 61}, 475 (1995)
\item[{[192]}] K. Binder, P. C. Hohenberg, B {\bf 9}, 2194 (1974)
\item[{[193]}] A. B. Harris, C. Micheletti and J. Yeomans, J. Stat. Phys. {\bf 84}, 323 (1996)
\item[{[194]}] H. Nakanishi and M. E. Fisher, J. Chem. Phys. {\bf 78},3279  (1983)
\item[{[195]}] P. S. Swain and A. O. Parry, Eur. Phys. J.  {\bf B 4}, 459 (1998);
E. Bruno, U. Marini, B. Marconi and R. Evans, Physica A {\bf 141A}, 187 (1987)
\item[{[196]}] K. Binder, D. P. Landau and A. M. Ferrenberg, Phys. Rev. Lett. {\bf 74}, 298 (1995)
\item[{[197]}] K. Binder, D. P. Landau and A. M. Ferrenberg, Phys. Rev. {\bf E 51}, 2823 (1995)
\item[{[198]}] M. Bengrine, A. Benyoussef, H. Ez-Zahraouy and F. Mhirech, Physica {\bf A 268}, 149 (1999)
\item[{[199]}] A. Hanke, M. Krech, F. Schlesener and S. Dietrich, Phys. Rev. E {\bf 60}, 5163 (1999)
\item[{[200]}]  L. Bahmad, A. Benyoussef and H. Ez-Zahraouy, Physica  {\bf A 303}, 525 (2002)
\item[{[201]}] K. Rejmer, S. Dietrich and M. Napiorkowski, Phys. Rev. E {\bf 60}, 4027 (1999)
\item[{[202]}] S. Dietrich and M. Schick, Phys. Rev B {\bf 31},4718 (1985)
\item[{[203]}] S. J. Kennedy and S. J. Walker, Phys. Rev. B {\bf 30},1498 (1984)
\item[{[204]}] P. Wagner and K. Binder, Surf. Sci. {\bf 175},421 (1986)
\item[{[205]}] K. Binder and D. P. Landau, Phys. Rev. B {\bf 37}, 1745 (1988)
\item[{[206]}] R. Kariotis  and J. J. Prentis, J. Phys. A{\bf 19}, L 455 (1986)
\item[{[207]}] R. Kariotis, B. Yang and H. Suhl, Surf. Sci. {\bf 173},283 (1986)
\item[{[208]}] J. M. Luck, S. Leibter and B. Derrida, J. Phys. France {\bf 44}, 1135 (1983)
\item[{[209]}] K. K. Mon and W. F. Saam, Phys. Rev. B {\bf 23}, 5824 (1981)
\item[{[210]}] W. F. Saam, Surf. Sci. {\bf 125},253 (1983)
\item[{[211]}] L. P. Kadanoff, Ann. Phys. (NY) {\bf 100}, 359 (1976)
\item[{[212]}] M. P. Nightingale, Physica{\bf 83a}, 561 (1979)
\item[{[213]}] M. P. Nightingale, J. Appl. Phys. {\bf 53}, 7927 (1982)
\item[{[214]}] H. J. Hermann, J. Phys. Lett. {\bf 100A}, 256 (1984)
\item[{[215]}] K. Binder and D. P. Landau, J. Appl. Phys. {\bf 57}, 3306 (1985)
\item[{[216]}] M. Przybylski and U. Gradmann, Phys. Rev. Lett. {\bf 59}, 1152 (1987)
\item[{[217]}] W. Durr, M. Taborelli, O. Paul, R. Germar, W. Gudat, D. Pescia and M. Landolt, Phys. Rev. Lett. {\bf 62}, 206 (1989)
\item[{[218]}] Z. Q. Qiu, S. H. Mayer, C. J. Gutierrez, H. Tang and J. C. Walker, Phys. Rev. Lett. {\bf 63}, 1649 (1989)
\item[{[219]}] D. P. Pappas, K. P. Kamper, B. P. Miller, H. Hopster, D. E. Fowler, C. R. Brundle, A. C. Luntz and Z.-X. Shen, Phys. Rev. Lett. {\bf 66}, 504 (1991)
\item[{[220]}] F. Aguilera-Granja and J. L. Moran Lopez, Sol. Stat. Commun. {\bf 74}, 155 (1990).
\item[{[221]}] T. Hai and Z. Y. Li, Phys. Stat. Sol. b {\bf 156}, 641 (1989)
\item[{[222]}] G. Wiatrowski, J. Mielnicki and T. Balcerzak, Phys. Stat. Sol. (b) {\bf 164}, 299 (1991).
\item[{[223]}] E. F. Sarmento and J. W. Tuker, J. Mag. Mag. Mater. {\bf 118}, 133 (1993).
\item[{[224]}] C. L. Wang, S. R. P. Smith and D. R. Tilley, J. Phys.: Condens. Matter {\bf 6}, 9633 (1994).
\item[{[225]}] C. L. Wang, W. L. Zhong and P. L. Zhang, J. Phys.: Condens. Matter {\bf 3}, 4743 (1992); Solid State Commun. {\bf 101}, 807 (1997).
\item[{[226]}] K. Swiderczak, A. Zagorski and J. Krol, Phys. Stat. Sol. b {\bf 134}, 161 (1986).
\item[{[227]}] B. K. Chakrabarti, A. Dutta and P. Sen, Quantum Ising Phases and Transitions in Transverse Ising Models, Lecture Notes in Physics, Vol. M$41$, Springer-Verlag, Heidelberg (1996).
\item[{[228]}] P. G. de Gennes, Solid State Commun.  {\bf 1}, 132 (1963).
\item[{[229]}]  A. Benyoussef and H. Ez-Zahraouy, Phys. Stat. Sol. (b) {\bf 179}, 521 (1993).
\item[{[230]}]  A. Bassir, C.E. Bassir, A. Benyoussef, A. Klumper and J. Zittart, Physica A, {\bf 253}, 473 (1998).
\item[{[231]}]  T. Kaneyoshi, Phys. Rev. {\bf B 33}, 526 (1986).
\item[{[232]}]  T. Kaneyoshi, E. F. Sarmento and I. P. Fittipaldi, Phys. Stat.
Sol. (b) {\bf 150}, 261 (1988).
\item[{[233]}]  T. Kaneyoshi, E. F. Sarmento and I. P. Fittipaldi, Phys. Rev.
{\bf B 38}, 2649 (1988).
\item[{[234]}] T. F. Cassol, W. Figueiredo and J.A. Plascak, Phys. Lett. {\bf A 160}, 518 (1991).
\item[{[235]}] Y. Q. Wang and Z.Y. Li, J. Phys.: Cond. Matt. {\bf 6}, 10067 (1994).
\item[{[236]}] R. B. Griffiths, Phys. Rev. Lett. {\bf 23}, 17 (1969).
\item[{[237]}] A. B. Harris, Phys. Rev. {\bf B 12}, 203 (1975).
\item[{[238]}] A. Hinterman, F. Rys, Helv. Phys. Acta {\bf 42}, 608 (1969)
\item[{[239]}] J. Bernasconi, F. Rys, Phys. Rev. {\bf B 4}, 3045 (1971)
\item[{[240]}] H. H.Chen, P.M.Levy, Phys. Rev. {\bf B 7}, 4267 (1973)
\item[{[241]}] A. N. Berker, M. Wortis, Phys. Rev. {\bf B 14}, 4946 (1976)
\item[{[242]}] R. R. Netz, A.N. Berker, Phys. Rev. {\bf B 47}, 1519 (1993)
\item[{[243]}] A. Falikov, A.N. Berker, Phys. Rev. Lett.{\bf B 76}, 4380 (1996)
\item[{[244]}] A. Maritan, M.Cieplak, M.R.Swift and F.Toigo, Phys. Rev. Lett.{\bf 69}, 221 (1992)
\item[{[245]}] C. S. Jayanthi, Phys. Rev. {\bf B 44}, 427 (1991)
\item[{[246]}] L. D. Gelb, Phys. Rev. {\bf B 50}, 11146 (1994)
\item[{[247]}] L. Bahmad, A. Benyoussef, A. El Kenz, and H. Ez-Zahraouy, M. J. Condensed Matter {\bf A 3}, 23 (2000)
\item[{[248]}] A. Benyoussef, H. Ez-Zahraouy, H. Mahboub, and M. J. Ouazzani, Physica {\bf A 326},220 (2003)
\item[{[249]}] A. N. Berker, S. Ostlund, F. A. Putnam, Phys. Rev. {\bf B 17},3650 (1978)
\item[{[250]}] L. Bahmad, A. Benyoussef, and H. Ez-Zahraouy, Surf. Sci. {\bf 536}, 114 (2003)
\item[{[251]}] L. Bahmad, A. Benyoussef, and H. Ez-Zahraouy, Surf. Sci. {\bf 552}, 1 (2004)
\item[{[252]}] L. Bahmad, A. Benyoussef, and H. Ez-Zahraouy, Submitted to Phys. Rev. E
\item[{[253]}] L. Bahmad, A. Benyoussef, and H. Ez-Zahraouy, J. Magn.
Magn. Mater. {\bf 238}, 115 (2002)
\item[{[254]}] L. Bahmad, A. Benyoussef, and H. Ez-Zahraouy, Chin. J. Phys. {\bf 40}, 537 (2002)

\end{enumerate}
\end{document}